%% file: main.tex
\documentclass[a4paper,11pt]{article}
\usepackage{jheppub} 
\usepackage[T1]{fontenc} 
\usepackage{amsmath,amsthm}
\usepackage{amssymb, amsfonts}
\usepackage{bbm}
\usepackage{graphicx}
\usepackage{dsfont}
\usepackage{relsize}
\usepackage{stmaryrd}
\usepackage{lmodern}
\usepackage{slantsc}
\usepackage{scalefnt}
\usepackage{wasysym}
\usepackage{xspace}
\usepackage{boxedminipage}
\usepackage{ifpdf}
\usepackage{multirow, hhline}
\usepackage{slashed}
\usepackage{xifthen}
\usepackage{tensor}
\usepackage{array, nicematrix}
\usepackage{tabu}
\usepackage{pbox}
\usepackage{multirow}
\usepackage{tikz}
\usepackage{overpic}
\usepackage{color}
\usepackage{diagbox}
\usepackage{makecell}
\usetikzlibrary{arrows, matrix, calc, scopes, decorations.markings}
\usepackage[mathcal]{euscript}
\usepackage{booktabs}
\usepackage{autobreak, subfig, diagbox, mathrsfs}
\usepackage{mathbbol}
\usepackage{tcolorbox}
\usepackage{hyperref}
\allowdisplaybreaks[4]

\newcommand{\ii}{{\mathbb{i}}}

\newcommand{\NN}{\mathbb{N}}
\newcommand{\Z}{\mathbb{Z}}
\newcommand{\A}{\mathcal{A}}
\newcommand{\Hil}{\mathcal{H}}
\newcommand{\G}{\mathcal{G}}
\newcommand{\T}{\mathcal{T}}
\newcommand{\Bimod}{{\mathsf{Bimod}}}
\newcommand{\Fibo}{{\mathsf{Fibo}}}
\newcommand{\C}{\mathbb{C}}
\newcommand{\V}{\mathcal{V}}
\newcommand{\D}{\mathcal{D}}
\newcommand{\Proj}{\mathcal{P}}
\newcommand{\Fus}{\mathscr{F}}
\newcommand{\F}{\mathcal{F}}
\newcommand{\RR}{\mathbb{R}}
\newcommand{\CC}{\mathbb{C}}
\newcommand{\idm}{{\mathds{1}}}

\newcommand{\df}{{\mathrm{d}}}

\newcommand{\dk}[2][1]{{\ifthenelse{\equal{#1}{1}}{\frac{\df{#2}}{2\pi}}{\frac{\df^{#1}{#2}}{(2\pi)^{#1}}}}}

\newcommand{\id}{{\rm id}}

\newcommand{\ket}[1]{{\left\vert{#1}\right\rangle}}

\newcommand{\mat}[1]{{\begin{pmatrix}#1\end{pmatrix}}}

\newcommand{\eq}[1]{\begin{align*}#1\end{align*}}
\newcommand{\eqn}[2][0]{\ifthenelse{\equal{#1}{0}}{\begin{equation}\begin{aligned}#2\end{aligned}\end{equation}}{\begin{equation}\begin{aligned}#2\end{aligned}\label{#1}\end{equation}}}

\usetikzlibrary{shapes.geometric, snakes}
\tikzset{>=latex}
\usetikzlibrary{arrows, matrix, calc, scopes, decorations.markings}
\usetikzlibrary{decorations.pathmorphing, positioning}
\tikzset{snake it/.style={decorate, decoration={snake,amplitude=0.2mm,segment length=1mm}}}
\tikzset{->-/.style={decoration={
			 markings,
			 mark=at position .5*\pgfdecoratedpathlength+2pt with {\arrow{>}}},postaction={decorate}}}

\tikzset{-<-/.style={decoration={
			 markings,
			 mark=at position .5*\pgfdecoratedpathlength+2pt with {\arrow{<}}},postaction={decorate}}}

\newtabulinestyle{dash=off 2pt}

{\null\,\vcenter\bgroup\scriptsize
\arraycolsep=.13885em
\hbox\bgroup$\array{@{}#1@{}}}
{\endarray$\egroup\egroup\,\null}
\input{input.tex}

\begin{document}

\title{Noninvertible gauge invariance in (2+1)d Topological Orders: A String-Net Model Realization}

\date{\today}
\author[a]{Yu Zhao}
\author[a,b,c]{Yidun Wan\footnote{Corresponding author}}
\affiliation[a]{State Key Laboratory of Surface Physics, Center for Astronomy and Astrophysics, Department of Physics, Center for Field Theory and Particle Physics, and Institute for Nanoelectronic devices and Quantum Computing, Fudan University, 2005 Songhu Road, Shanghai 200433, China}
\affiliation[b]{Shanghai Research Center for Quantum Sciences, 99 Xiupu Road, Shanghai 201315, China}
\affiliation[c]{Hefei National Laboratory, Hefei 230088, China}
\emailAdd{yuzhao20@fudan.edu.cn, ydwan@fudan.edu.cn}

\abstract{We develop a systematic framework for understanding symmetries in topological phases in \(2+1\) dimensions using the string-net model, encompassing both gauge invariances that preserve anyon types and global symmetries permuting anyon types, including both invertible symmetries describable by groups and noninvertible symmetries described by categories. As an archetypal example, we reveal the first noninvertible categorical gauge invariance of topological orders in \(2+1\) dimensions: the Fibonacci gauge invariance of the doubled Fibonacci topological order, described by the Fibonacci fusion \(2\)-category. Our approach involves two steps: first, classifying and establishing dualities between different string-net models describing the same topological order; and second, constructing symmetry transformations within the same string-net model when the dual models have isomorphic input data, achieved by composing duality maps with isomorphisms of degrees of freedom between the dual models.}

\maketitle

\flushbottom

\section{Introduction}\label{sec:intro}

Symmetry is a central concept in modern physics, traditionally described by groups. Topological orders in \(2+1\) dimensions, whose low-energy effective descriptions are topological gauge field theories, are novel phases of matter that go beyond the conventional Landau-Ginzburg paradigm, revealing more interesting symmetry structures. 

Nevertheless, the gauge symmetries\footnote{A historical misnomer. It should be more appropriately called gauge invariance or gauge redundancy because it is mathematical redundancy where different states in the Hilbert space describe the same physical state in the theory. For the $(2+1)$D topological orders, the criterion for determining whether a transformation that preserves the Hamiltonian is a gauge redundancy or a global symmetry in $(2+1)d$ topological orders is whether it changes the types of anyonic excitations---the physical topological observables. The other types of global symmetries, such as symmetry fractionalizations\cite{Wang2012, Essin2012, Xu2013, Barkeshli2014c, SongHermele2014, cheng2017, Ning2018, li2024, zhao2024} or soft symmetries\cite{davydov2014, kobayashi2025}, are not considered in this paper.} in topological orders are often vague. In cases where the gauge symmetries are described by groups, e.g., in the Dijkgraaf-Witten topological gauge field theory\cite{Dijkgraaf1990, Fuchs2014, Kapustin2014, Cong2017a} or its lattice Hamiltonian model---the (twisted) quantum double model\cite{Bravyi1998, Kitaev2003a, Beigi2011, Hu2012a, Bullivant2017, Cong2017, Wang2020, Wang2020b, Hu2020}, the gauge structure has been well understood as gauge groups, where the gauge transformations correspond to nontrivial conjugation actions on the lattice model's local basic degrees of freedom that are elements in the input group of the theory. In more general cases, such as the Turaev-Viro topological field theory\cite{Turaev1992, Turaev1994, Fuchs2014, Bhardwaj2016, Cui2016} or its lattice Hamiltonian model---the string-net (Levin-Wen) model\cite{Levin2004, Hung2012, Hu2012, Kitaev2012, schulz2013, Lan2014b, Lin2014, Hu2017, Hu2018, Bridgeman2020, Wang2022, zhao2022}, the gauge invariance is not known in general. In such cases, the basic degrees of freedom take values in the simple objects of the input fusion category, making the gauge structure not describable by groups but rather by categories\cite{polishchuk1998, Hung2013, HungWan2013a, Baratin2014, Gaiotto2014, Bhardwaj2017, Ji2019, ji2020, Kong2020, bartsch2023, bhardwaj2023}, invoking the concept of noninvertible symmetry\cite{Ji2019, ji2020, bhardwaj2023, bhardwaj2022, choi2022, choi2023, choi2024, bartsch2024}, which has greatly expanded and deepened our understanding of symmetry in physics. Noninvertible symmetry has been extensively explored in \(1+1\) dimensions\cite{ji2021, jacobsen2023, fechisin2023, xi2023, li2024}. In \(2+1\) dimensions, however, noninvertible symmetries remain largely an open problem, despite studies in certain \((2+1)\)-dimensional systems\cite{choi2024a, mana2024, inamura2024} and attempts in studying the applications of noninvertible symmetry in quantum field theory and M-theory \cite{kaidi2022, apruzzi2023, van2023, chen2023a, sela2024}.

In this paper, we tackle this problem by explicitly and systematically formulating the symmetry transformations of \((2+1)\)-dimensional topological orders as operators of the string-net model. We show that such symmetry transformations can be either gauge invariances that preserve anyon types or global symmetries permuting anyon species, and can be invertible (describable by groups) or noninvertible (described by fusion \(2\)-categories\cite{douglas2018, decoppet2021, johnson2021, decoppet2023, henriques2023}). As a key result, our construction reveals the first noninvertible categorical gauge invariance of topological orders in \(2+1\) dimensions:

\emph{The noninvertible gauge invariance (to be called the Fibonacci gauge invariance) of the doubled Fibonacci topological order described by the Fibonacci fusion \(2\)-category.}

Our general construction is rather involved and thus to be detailed in the appendices. In the main text, we shall only sketch our approach and expound on two archetypal examples to illustrate our general construction. The first example is the well-known $em$ exchange global symmetry of the $\Z_2$ toric code string-net model, and second example is the noval categorical Fibonacci gauge invariance of the Fibonacci string-net model. Besides, we also provide a clear criterion determining whether a symmetry transformation generates a global symmetry or a gauge invariance of the topological order.

Our framework offers a powerful tool for analysing the symmetry structure of topological orders. Employing it, we have recently implemented generic bosonic anyon condensation in the string-net models\cite{zhao2025}. Earlier studies realised only fluxon and simple-current condensations---two special subclasses of anyon condensation---within the string-net model\cite{zhao2022,lin2023,Bais2009a,Eliens2013,Hung2013,Wan2017,Burnell2018,Hu2021}. By applying appropriate symmetry transformations, we can perform all types of bosonic anyon condensations in familiar forms within the transformed model. Moreover, our framwork enables the explicit construction and subsequent gauging of symmetry-enriched topological (SET) phases, naturally accommodating non-Abelian global symmetries and symmetry-fractionalization charges of non-Abelian anyons. A detailed exposition of these results will be presented elsewhere.

\section{Sketch of Our Approach}\label{sec:sketch}

\begin{figure}\centering
\includegraphics[width=14cm]{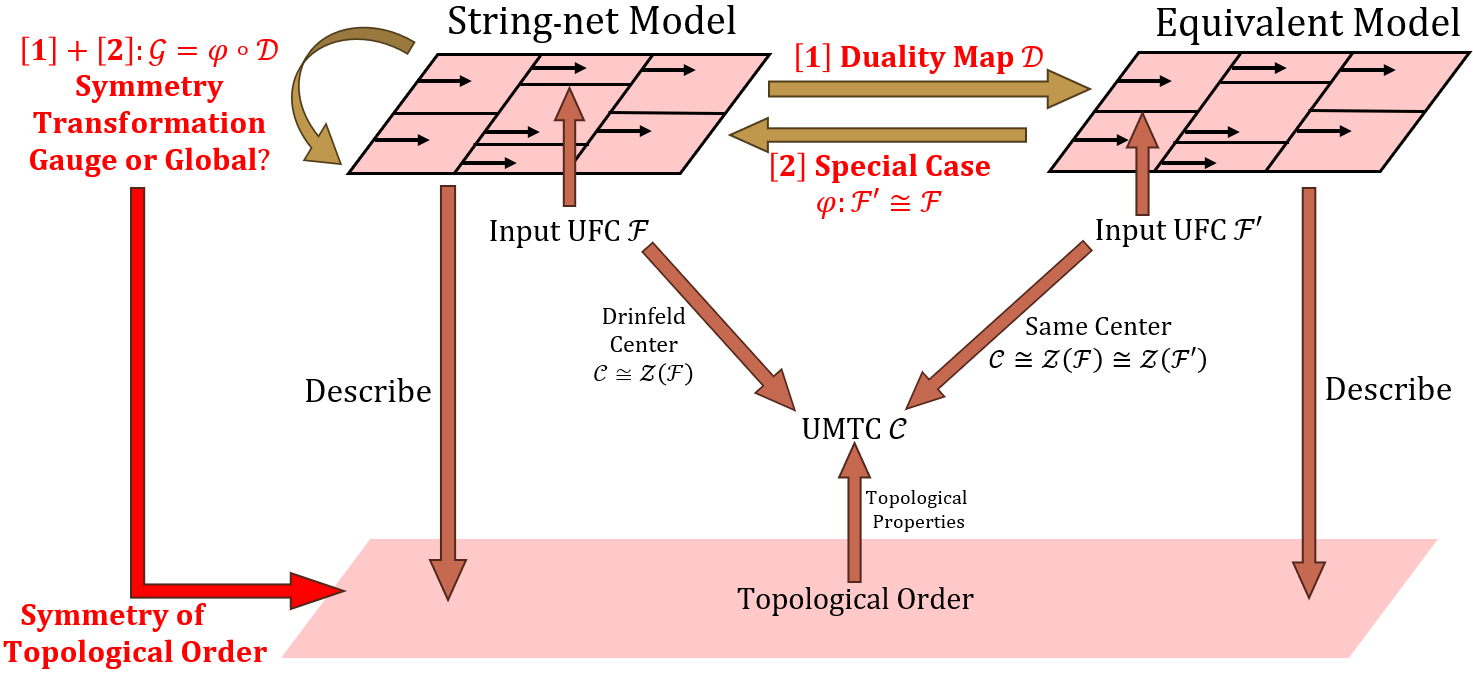}
\caption{Sketch of the construction: [1] Duality map \(\D\) between distinct but equivalent models describing the same topological order; [2] Isomorphism \(\varphi\) between the input UFCs of the equivalent models; Combining [1] and [2] yields a symmetry transformation of the string-net model, as well as the topologgical order.}
\label{fig:sketch}
\end{figure}

As the low-energy effective theories of topological orders, topological quantum field theory (TQFT) is believed to be a topological gauge theory. For the Dijkgraaf-Witten type of TQFT, the gauge invariance and transformation are manifest and clear. For more general topological orders, whose low-energy effective theory is the Turaev-Viro type of TQFT, however, their gauge symmetries and transformations are in general obscure. 

To reveal the gauge structure of topological orders in general, we first note that topological orders are primarily observed in strongly correlated electron systems\cite{Laughlin1983, TaoWu1984, Jain1989a, Wen1990b, Wen1990c, Tang2011, Clarke2013, Yang2019}, where anyons are collective excitations of the electrons. Such a system has a physical Hilbert space consisting of the fundamental degrees of freedom of electrons, and anyons are merely represented as excited states in this Hilbert space. While an Abelian anyon is represented by one excited state up to a phase factor, a non-Abelian anyon is represented by an equivalent class of adiabatically related electronic excited states in a multi-dimensional subspace of the Hilbert space\cite{Iguchi1997, Guruswamy1999a, Barkeshli2013, Hu2013, Li2018, Li2019a}\footnote{This is why non-Abelian anyons can support a scheme of quantum computation---topological quantum computation\cite{Fan2022}.}. Therefore, a physical non-Abelian anyon may have a nontrivial internal space spanned by local degrees of freedom\cite{Hu2018, zhao2022, zhao2024}, which is, in the context of topological field theory, regarded as the gauge space of anyons. Nevertheless, TQFT treats anyons as elementary and indecomposable simple objects of a modular tensor category that have no finer structures and is therefore not able to sense and probe such internal spaces of anyons. In contrast, Hamiltonian models such as the string-net model represent each anyonic excitation as concrete excited states. Anyons are manifested by orthogonal subspaces (one- or multi-dimensional) in the Hilbert space consisting of the fundamental local degrees of freedom of the model. We show that in the string-net model, it is possible to unitarily transform the fundamental degrees of freedom while preserving topological invariance and the Hamiltonian. When this unitary transformation preserves and/or transforms within the subspaces representing the anyons, it is qualified as a gauge transformation of the model, revealing an associated gauge invariance of the \((2+1)\)-dimensional topological order. When this transformation can permute anyon types, it is a global symmetry transformation of the model, as well as the topological order described by the model.

We construct the symmetry transformations of \((2+1)\)-dimensional topological orders in two steps (see Figure \ref{fig:sketch}). The first step establishes explicit duality maps between different string-net models describing the same topological order and classify all such equivalent string-net models\footnote{Such duality maps may also be realized by constant-depth quantum circuits\cite{lootens2022}.}. To be specific, while the fundamental degrees of freedom in a usual gauge theory take value in a group, those in the Turaev-Viro TQFT or a string-net model are simple objects in a unitary fusion category (UFC) $\Fus$. The string-net model with input UFC $\Fus$ describes a \((2+1)\)-dimensional topological order whose topological properties are encapsulated by unitary modular tensor category (UMTC) $\mathcal{Z}(\Fus)$ that is the \emph{Drinfeld morita center} of the input UFC $\Fus$. Two string-net models describe the same topological order if and only if their input UFCs are \emph{categorically Morita equivalent}\cite{etingof2016}---namely, they have isomorphic Drinfeld morita centers. We classify all Morita equivalent input UFCs and construct duality maps relating these equivalent string-net models describing the same topological order. Our duality generalizes the concept of electromagnetic duality\cite{Buerschaper2013, Wang2020, Hu2020} in topological orders.

In the second step, when two equivalent string-net models have isomorphic input UFCs, we can compose the duality in the first step with an isomorphism to form a symmetry transformation on the Hilbert space of the same model. Such a symmetry transformation can be a global symmetry or a gauge invariance transformation depending on whether it permutes anyon species. Such a symmetry, when it is invertible, is described by a group, while when it is noninvertible, is described by a fusion \(2\)-category.

We first apply our framework to reproduce the familiar \(\Z_2\) \(em\) exchange global symmetry in the \(\Z_2\) toric-code topological order in the string-net model. Though elementary, this example represents the essential progresses of our duality maps and symmetry constructions. We then move to the main part of this article: the Fibonacci gauge invariance of the doubled Fibonacci topological order in the string-net model, where the richer structure necessitates more intricate constructions and exposes novel phenomena.

\section{The \texorpdfstring{$em$}{Lg} exchange global symmetry of the \texorpdfstring{$\Z_2$}{Lg} toric code string-net model}

\subsection{The String-Net Model}

For our study, we take the form of the string-net model adapted from that in Ref. \cite{Hu2018} because the Hilbert space encompasses the full anyon spectra of the corresponding topological orders. The model is defined on a \(2\)-dimensional lattice, such as that in Fig. \ref{fig:lattice}. Each vertex is trivalent. Each plaquette hosts a tail, attached to any of its edges\footnote{The original string-net model in Ref. \cite{Levin2004}, which bears no such tails, cannot fully describe charge excitations. These added tails record the charges of anyons, thus enlarging the Hilbert space to encompass the complete anyon spectrum.}. Each edge or tail carries a label---fundamental degree of freedom---taking value in the simple objects of the input fusion category \(\Fus\) of the model. The Hilbert space of the model is spanned by all possible assignments of the labels, constrained by that the labels on the three edges (tails) meeting at any vertex must satisfy the fusion rules of \(\Fus\).

To illustrate our approach, we begin with a simple but canonical example---the \(\Z_2\) toric-code string-net model\cite{Kitaev2003a}. Following our procedure, we explicitly construct the \(\Z_2\) global symmetry that exchanges the charge and flux excitations of the toric code model.

\begin{figure}\centering
\Lattice
\caption{Part of the string-net model lattice. A tail (wavy line) is attached to an arbitrary edge of every plaquette.}
\label{fig:lattice}
\end{figure}
	
The input fusion category of the toric code string-net model is the \(\Z_2\) fusion category, which uses the two group elements \(\pm 1 \in \Z_2\) as its simple objects, whose quantum dimensions are both \(d_{+1} = d_{-1} = 1\). The fusion rules of this category capture the group multiplicity rules of \(\Z_2\) group:
\eq{
\delta_{ijk} = \frac{ijk + 1}{2},
}
The Hilbert space is exapnded by the configurations that each edge or tail on the lattice carries a group element \(\pm 1 \in \Z_2\), subject to the constraints that the degrees of freedom \(i, j, k\) on any three edges or tails meeting at a vertex must satisfy the fusion rule \(\delta_{ijk} = 1\). The Hamiltonian of the toric code string-net model is a sum of commutative projectors \(A_P\) and \(B_P\),
\eqn{H_{\rm TC} := - \sum_{\text{Plaquettes } P}(A_P + B_P),}
where \(A_P\) acts on tails in plaquettes \(P\), and \(B_P\) acts on edges surrounding plaquettes \(P\):
\eqn{
A_P \quad \ToricCodeA \quad := \quad \delta_{p,\ +1}\quad \ToricCodeA \quad,
}
\eqn{
B_P \quad \ToricCodeA \quad := \quad \frac{1}{2} \quad \ToricCodeA \quad + \quad \frac{1}{2} \quad \ToricCodeB \quad.
}
The ground states are common eigenstates of all \(A_P\) and \(B_P\) operators with \(+1\) eigenvalues, while an excited state \(\ket{\psi}\) is another common eigenstate satisfying \(A_P\ket{\psi} = 0\) (or \(B_P\ket{\psi} = 0\)) for one or more plaquettes \(P\), in each of which there resides a chargeon \(e\) (or a fluxon \(m\)). Unlike the original version of the string-net model, where chargeons reside on vertices, both chargeons and fluxons are situated in the plaquettes of our model. If \(A_P\ket{\psi} = B_P\ket{\psi} = 0\) in plaquette \(P\), there is a dyon \(\epsilon\) in plaquette \(P\). We also refer to the ground state as the trivial excited state, in which there are trivial anyons \(1\) in all plaquettes.

\subsection{The Duality Maps}

Given a fusion category \(\Fus\), there exist \emph{Frobenius algebras} in \(\Fus\). It is a mathematical theorem\cite{etingof2016} that the bimodules---a special class of representations of a given Frobenius algebra \(\A\) in \(\Fus\)---form another fusion category \(\Bimod_\Fus(\A)\) that is \emph{categorically Morita equivalent} to \(\Fus\). Conversely, every fusion category that is categorically Morita equivalent to \(\Fus\) is naturally isomorphic to the bimodule category \(\Bimod_\Fus(\A)\) over certain Frobenius algebra \(\A\) in \(\Fus\). These mathematical facts tell us:

\begin{itemize}
\item The string-net model with \(\Bimod_\Fus(\A)\) as its input UFC is equivalent to the string-net model with \(\Fus\) as its input UFC in that they describe the same topological order, where \(\A\) is a Frobenius algebra in \(\Fus\).

\item We classify all equivalent string-net models describing the same \((2+1)\)-dimensional topological order by all Frobenius algebras in one certain input UFC \(\Fus\) in the equivalent class.
\end{itemize}

We now explicitly establish this equivalence by a duality map \(\D\) between these two different models in this section. Precise definitions of Frobenius algebras and their bimodules are deferred to Appendix \ref{sec:frob}. Informally speaking, a \emph{Frobenius algebra} is an associative algebra taking simple objects of \(\Fus\) as its basis. In the \(\Z_2\) fusion category, there are exactly two Frobenius algebras. The first is the one-dimensional trivial algebra
\[\A_0 := 
\Big\{\ \alpha[+1]\ \Big|\ \alpha\in\mathbb{C}\ \Big\},\]
generated solely by the trivial simple object \(+1\) of \(\Z_2\) fusion category. The algebra multiplication rule is simply \([+1]\times [+1] = [+1]\). The second is the two-dimensional algebra 
\[\A_{\Z_2} := \Big\{\ \alpha[+] + \beta[-]\ \Big|\ \alpha, \beta\in\mathbb{C}\ \Big\}\]
with the multiplication rules \([+]\times [+] = [-]\times[-] = [+], [+]\times[-] = [-]\times [+] = [-]\), where \([\pm]\) refers to group elements \(\pm1\in\Z_2\).

A bimodule \(M\) over Frobenius algebra \(\A\) in UFC \(\Fus\) is another linear space \(V_M\) spanned by simple objects of \(\Fus\), equipped with a function \(P_M: \A^2\times V_M^2\times L_\Fus \to \mathbb{C}\), representing a pair of algebra elements \((a, b) \in \A^2\) as rank-3 tensors on the representation spaces \(V_M\), where \(L_\Fus\) is the set of all simple objects of fusion category \(\Fus\). These tensors indicate that two algebra elements \(a\) and \(b\) in \(\A\) act sequentially on \(x \in V_M\) from both sides, transforming it to \(y \in V_M\) with the coefficient \(\sum_{u} [P_M]^{ab}_{xuy}\). This bimodule action shall commute with the algebra multiplications of Frobenius algebra \(\A\) (see Appendix \ref{sec:bimod} for precise definition). The intermediate object \(u\) varies over \(L_\Fus\) to make the action satisfying the fusion rule \(\delta_{axu} = \delta_{ubz} = 1\).

Now we restrict attention to bimodules over the non-trivial Frobenius algebra \(\A_{\Z_2}\). This algebra has two inequivalent simple (i.e., irreducible) bimodules, denoted by \(M_{\pm}\). We write their underlying vector spaces and action coefficients as
\begin{align*}
V_+ = V_- = \Big\{\ \alpha[+] + \beta[-]\ \Big|\ \alpha, \beta\in\mathbb{C}\Big\},\qquad [P_+]^{ab}_{xuy} = \delta_{axu} &\delta_{uby},\\
[P_-]^{++}_{+++} = [P_-]^{++}_{---} = 1, \qquad [P_-]^{--}_{+-+} = [P_-]^{--}_{-+-} = -1,&\\
[P_-]^{+-}_{++-} = [P_-]^{-+}_{-++} = \mathbb{i}, \qquad [P_-]^{+-}_{--+} = [P_-]^{-+}_{+--} = -\mathbb{i}.&
\end{align*}
The bimodules \(M_{\pm}\) share the same underlying representation space but differ in
their action coefficients \(P_{\pm}\). They realize, respectively, the trivial and the sign representations of the group \(\mathbb{Z}_{2}\). Therefore, the bimodule category \(\Bimod_{\Z_2}(\A_{\Z_2})\), which takes simple bimodules \(M_\pm\) as simple objects, is naturally isomorphic to the representation category \({\tt Rep}(\Z_2)\) of group \(\Z_2\). Namely, the respected electromagnetic duality of \(\Z_2\) toric-code model\cite{Buerschaper2013, Wang2020, Hu2020} is a special case of our duality maps.

The string-net model with the input fusion category \(\Bimod_{\Z_2}(\A_{\Z_2})\) describes the same topological order---the toric-code topological order---as the original \(\Z_2\) string-net model. The fundamental degrees of freedom on edges and tails of the dual model are simple bimodules \(M_\pm\). We construct a duality map \(\D_{\text{TC}}\) that embeds the fundamental degrees of freedom \(M_\pm\) of the dual model into the original model, based on the definition of the bimodules:
\eqn{
\D_{\text{TC}}\ \ \Edge{M_i}\quad :=\quad \frac{1}{4}\sum_{a, b, x, u, y = \pm 1}\ [P_i]^{ab}_{xuy}\quad \ToricCodeD\ ,
}
where \(M_i = M_\pm\) are simple objects in UFC \(\Bimod_{\Z_2}(\A_{\Z_2})\). In this expression, the factor \(4\) in the denominator arises from \(d^2_{\A_{\Z_2}}\), where \(d_{\A_{\Z_2}} = d_{+1} + d_{-1} = 2\) is the total quantum dimension of the Frobenius algebra \(\A_{\Z_2}\). The black line refers to both edges and tails. The red lines are auxiliary tails that will be annihilated by topological moves (see Appendix \ref{sec:pachner}), resulting in a unitary transformation between the two Hilbert spaces of the dual \(\Bimod_{\Z_2}(\A_{\Z_2})\) string-net model and the original \(\Z_2\) string-net model, which do not modify the lattice shapes and can be understood plaquette by plaquette:
\eqn{
\FibonacciE\ ,
}
where \(I_k, E_k, M = M_\pm\), and ``\(\cdots\)'' denotes the expansion coefficients.

On a side note, for the trivial Frobenius algebra \(\A_{0}\) in any UFC \(\Fus\), simple bimodules over \(\A_{0}\) are in one-to-one correspondence with the simple objects of UFC \(\Fus\), and the bimodule actions are all trivial:
\[V_x = \{\ \alpha x\ |\ \alpha\in\mathbb{C}\ \},\qquad P^{11}_{xxx} = 1,\qquad \forall x \in L_\Fus,\]
where \(1\) is the trivial simple object in \(\Fus\). Hence \(\Bimod_{\Fus}(\A_{0})\) is naturally isomorphic to the original UFC \(\Fus\), and the corresponding duality map \(\D_0\) is simply the identity map of the original string-net model.

\subsection{The Symmetry Transformations}

We can further construct a symmetry transformation on the Hilbert space of the original \(\Z_2\) string-net model because \(\Z_2\) fusion category is isomorphic\footnote{In general, the isomorphism is not unique when the input UFC admits nontrivial automorphisms; different choices of isomorphism, when composed with the same duality map, yield distinct symmetry transformations.} to \(\Bimod_{\Z_2}(\A_{\Z_2})\) fusion category:
\eqn[eq:TCIso]{
\F_{\text{TC}}: \Z_2 \to \Bimod_{\Z_2}(\A_{\Z_2}),\qquad +1 \mapsto M_+,\qquad -1 \mapsto M_-.
}
Such isomorphism \(\F_{\text{TC}}\) induces an isomorphic map \(\varphi_{\text{TC}}\) between the original \(\Z_2\) string-net model and the dual \(\Bimod_{\Z_2}(\A_{\Z_2})\) string-net model, and thus a \emph{unitary transformation} \(\G_{\text{TC}}\) of the original \(\Z_2\) string-net model:
\eqn[eq:global]{\G_{\text{TC}} := \D_{\text{TC}}\circ\varphi_{\text{TC}}\ ,}
where
\eqn{\varphi_{\text{TC}}\quad\Edge{+1}\quad :=\quad \Edge{M_+},\qquad \varphi_{\text{TC}}\quad\Edge{-1}\quad := \quad \Edge{M_-}.}
Here, the line refers to both edges and tails. Consequently, the unitary transformation \(\G_{\text{TC}}\) transforms the local degrees of freedom \(\pm 1\in\Z_2\) on edges (tails) to
\eqn{\Edge{+1}\ \Longrightarrow\ \frac{1}{4}\sum_{a,b,x,u,y = \pm 1}[P_+]^{ab}_{xuy}\ \ToricCodeD\ ,\qquad \Edge{\tau}\ \Longrightarrow\ \frac{1}{4}\sum_{a,b,x, u, y = \pm 1}\ [P_-]^{ab}_{xuy}\ \ToricCodeD\ .}
The red lines will be annihilated by topological moves. 

The symmetry transformation \(\G_{\text{TC}}\) is a \(\Z_2\) global symmetry transformation of the toric code string-net model because:
\begin{enumerate}
\item \(\G_{\text{TC}}\) is a unitary \(\Z_2\) transformation:
\eqn{
\G^\dagger_{\text{TC}} = \G^{-1}_{\text{TC}} = \G_{\text{TC}}.
}
\item \(\G_{\text{TC}}\) preserves the model's Hamiltonian \(H_{\rm TC}\):
\eqn{
\G^\dagger_{\text{TC}} H_{\rm TC} \G_{\text{TC}} = H_{\rm TC}.
}
\item Transformation \(\G_{\text{TC}}\) preserves the ground-state Hilbert space of the model but exchanges \(A_P\) and \(B_P\) operators:
\eqn{
\G^\dagger_{\text{TC}} A_P \G_{\text{TC}} = B_P, \qquad\qquad \G^\dagger_{\text{TC}} B_P \G_{\text{TC}} = A_P,
}
and hence exchanges chargeons and fluxons because an \(A_P\) (\(B_p\)) measures the chargeon (fluxon) in plaquette \(P\)\footnote{In the original string-net model, chargeons reside on vertices and fluxons are located in plaquettes, necessitating lattice dualization after the symmetry transformation. In contract, our model places both chargeons and fluxons in plaquettes, eliminating the need to alter the lattice shape.}.
\end{enumerate}

\subsection{Criterion for Distinguishing Gauge Invariances from Global Symmetries}\label{subsec:criterion}

The symmetry transformation \eqref{eq:global} of the \(\Z_2\) toric code topological order is a global symmetry transformation that exchanges the anyon species of chargeons and fluxons. In contract, as to be seen in the subsequent sections, the symmetry transformation of the doubled Fibonacci topological order is a gauge invariance that preserves all anyon species but only transforms excited states within the internal Hilbert space of each anyon. To determine whether a symmetry transformation defined by a Frobenius algebra \(\A\) in a fusion category \(\Fus\) is a global symmetry or a gauge invariance, we use the criterion based on \emph{algebraic Morita equivalence} between Frobenius algebras\cite{etingof2016}. This concept of Morita equivalence between two Frobenius algebras differs from the concept of categorical Morita equivalence between fusion categories introduced before.

Algebraic Morita equivalence implies that the two Frobenius algebras have the isomorphic modules. According to the boundary-bulk correspondence, two string-net models with bimodule categories over two Morita-equivalent Frobenius algebras not only have the same anyon species but also exhibit the same relationships between each anyon species and the fundamental degrees of freedom of the models. As previously noted, the fusion category \(\Fus\) itself is the bimodule category over the trivial one-dimensional Frobenius algebra \(\A_0 = \{1\}\).

Therefore, a symmetry transformation is a gauge invariance if and only if its defining Frobenius algebra \(\A\) is algebraically Morita equivalent to the one-dimensional trivial Frobenius algebra \(\A_0\) spanned solely by the trivial simple object. This is the case for the nontrivial Frobenius algebra in the Fibonacci fusion category. Otherwise, it is a global symmetry, as in the case of $\Z_2$ toric code model.

An exceptional case arises when the defining Frobenius algebra \(\A_0\) of the duality map is trivial, for then \(\Bimod_{\Fus}(\A_0)\simeq\Fus\) and the duality map reduces to the identity of the original model. Nonetheless, the resulting symmetry transformation can still be nontrivial, since one may compose this identity duality map with any nontrivial automorphisms of \(\Fus\). If the automorphisms is an inner automorphisms, the symmetry transformation corresponds to a gauge redundancy, whereas those coming from outer automorphisms implement genuine global symmetries permuting the anyon charges.

\section{Fibonacci String-Net Model and Internal Spaces of Anyons}\label{sec:fibo}

Having completed the analysis of the \(em\) exchange symmetry in the \(\Z_2\) toric-code model, we now turn to a second case study---the doubled Fibonacci topological order realized by the Fibonacci string-net construction. This section offers a concise review of the Fibonacci string-net model, while the subsequent sections develop the duality maps and gauge invariance transformations.

The input fusion category of the Fibonacci string-net model, the Fibonacci fusion category, denoted by \(\Fibo\), contains two simple objects \(1\) and \(\tau\) that are the values of the basic degree of freedom on any edge/tail. These simple objects satisfy the fusion rules 
\eq{
\delta_{111} = \delta_{1\tau\tau} = \delta_{\tau 1\tau} = \delta_{\tau\tau 1} = \delta_{\tau\tau\tau} = 1,\qquad \delta_{11\tau} = \delta_{1\tau 1} = \delta_{\tau 11} = 0.
}
The Hilbert space \(\Hil_\Fibo\) of the model is spanned by all possible assignments of the simple objects \(1\) and \(\tau\) on all edges and tails of the lattice, constrained by the fusion rules \(\delta_{ijk} = 1\) for the three labels \(i\), \(j\), \(k\) on any three edges or tails meeting at any vertex of the lattice.

The Hamiltonian of the Fibonacci string-net model reads
\eqn[eq:FiboHamil]{
H_\Fibo := -\sum_{{\rm Plaquettes}\ P}Q_P,
}
where the commuting projectors \(Q_P\) are defined as
\eq{
&Q_P := \frac{1}{1 + \phi^2}(Q_P^1 + \phi Q_P^\tau),\\
Q_P^s\quad \ToricCodeA \quad :=& \quad\delta_{p,1}\quad \sum_{j_k\in\{1, \tau\}}\Bigg[\prod_{k = 0}^5G^{e_{k+1}i_kj_k}_{sj_{k+1}i_{k+1}}\sqrt{d_{i_k}d_{j_k}}\Bigg]\FibonacciA\ .
}
Here, \(s\in\{1,\tau\}\), \(d_1 = 1, d_\tau = \phi = (\sqrt{5} + 1)/2\), and the nonzero \(G\)-symbols are
\eq{
G^{111}_{111} = 1,\quad G^{111}_{\tau\tau\tau} = \frac{1}{\sqrt{\phi}},\quad G^{1\tau\tau}_{1\tau\tau} = G^{1\tau\tau}_{\tau\tau\tau} = \frac{1}{\phi},\quad G^{\tau\tau\tau}_{\tau\tau\tau} = -\frac{1}{\phi^2},\quad G^{abm}_{cdn} = G^{mab}_{ncd} = G^{cdm}_{abn}.
}
Unlike in the original string-net model or in the \(\Z_2\) toric code model, we unify the charge and flux measurement operators into a common \(Q_P\) operator. The ground states are common eigenstates of all \(Q_P\) operators with \(+1\) eigenvalues. An excited state \(\ket{\psi}\) is another common eigenstate that satisfies \(Q_P\ket{\psi} = 0\) for one or more plaquettes \(P\), in which we say there reside anyons. We also dub a ground state a trivial excited state, which has a trivial anyon in each plaquette. There are four anyon species of the doubled Fibonacci topological order:
\eqn[eq:FiboAnyons]{
1\bar 1,\qquad \tau\bar 1,\qquad 1\bar\tau,\qquad \tau\bar\tau,
}
where \(1\bar 1\) is the trivial anyon. The measurement operator \(\Pi_P^J\) measuring whether there is an anyon \(J\) in plaquette \(P\) is
\eqn[eq:FiboMeasure]{\Pi_P^J\quad\ToricCodeA\quad :=\quad &\sum_{s,t\in\{1, \tau\}}\Bigg[\prod_{k = 0}^5G^{e_{k+1}i_ki_{k+1}}_{sj_{k+1}j_k}\sqrt{d_{i_k}d_{j_k}}\Bigg]G^{pi_0i_6}_{j_6st}G^{pj_0j_6}_{i_0ts}\ \times\\
&\sqrt{\frac{d_sd_t}{d_p}}\ z^{J;t}_{pps}\ \FibonacciB\ ,}
where the nonzero elements of \(z^J\) tensors are
\eq{
&z_{111}^{1\bar 1; 1} = z_{11\tau}^{1\bar 1; \tau} = 1\ ;\\
z_{\tau\tau 1}^{\tau\bar 1; \tau} = 1,\qquad z_{\tau\tau\tau}^{\tau\bar 1; 1} = -\frac{\phi}{2} - &\frac{i}{2\phi}\sqrt{\phi^2+1},\qquad z_{\tau\tau\tau}^{\tau\bar 1; \tau} = -\frac{1}{2\phi} + \frac{i}{2}\sqrt{\phi^2 + 1}\ ;\\
z_{\tau\tau 1}^{1\bar\tau; \tau} = 1,\qquad z_{\tau\tau\tau}^{\tau\bar 1; 1} = -\frac{\phi}{2} + &\frac{i}{2\phi}\sqrt{\phi^2+1},\qquad z_{\tau\tau\tau}^{\tau\bar 1; \tau} = -\frac{1}{2\phi} - \frac{i}{2}\sqrt{\phi^2 + 1}\ ;\\
z_{111}^{\tau\bar\tau; 1} = 1,\qquad z_{11\tau}^{\tau\bar\tau; \tau} = -\frac{1}{\phi^2},&\qquad z_{\tau\tau 1}^{\tau\bar\tau; \tau} = 1,\qquad z_{\tau\tau\tau}^{\tau\bar\tau; 1} = 1,\qquad z_{\tau\tau\tau}^{\tau\bar\tau; \tau} = \frac{1}{\phi^2}\ .
}

It is the kairos to bring up the concept of anyon's \emph{internal space}. In contrast to Abelian anyons, which do not have nontrivial internal spaces, a non-Abelian anyon does have an internal space because multiple non-Abelian anyons occupy a well-defined multi-dimensional Hilbert space\cite{Iguchi1997, Guruswamy1999a, Barkeshli2013, Hu2013, Li2018, Li2019a}. Such internal spaces are generally hidden in the language of TQFT unless the anyons carry group representations. The string-net model is however able to manifest such internal spaces by representing a non-Abelian anyon on a certain multi-dimensional Hilbert subspace of excited states of the model. Consequently, in the string-net model, an anyon appearing in an excited state is not only labeled by its anyon species \(J\) in each plaquette but also by each anyon's internal charge \(p\)---the degree of freedom on the tail where the anyon resides. A non-abelian anyon carries more than one charge type, and the gauge invariance transformation to be constructed will be able to mix the internal charges \(p\) while preserving the anyon species \(J\) of a non-Abelian anyon. 

In the Fibonacci string-net model, there are five allowed pairs \((J, p)\) of anyon species \(J\) and its charges \(p\):
\eq{
(1\bar 1, 1),\qquad (1\bar\tau, \tau),\qquad (\tau\bar 1, \tau),\qquad (\tau\bar\tau, 1),\qquad (\tau\bar\tau, \tau).
}
An anyon \(\tau\bar\tau\) in a certain plaquette \(P\) has two different possible charge choices \(1\) and \(\tau\), apparently expanding a $2$-dimensional internal space. Anyons \(\tau\bar 1\) and \(1\bar\tau\) are both non-abelian anyons but they seem to both carry only one type of charge \(\tau\) in the string-net model. Nevertheless, as to be seen, to reveal the gauge structure of the doubled Fibonacci phase, it is inevitable to further enlarge the string-net model's Hilbert space in a natural way as follows, such that anyons \(\tau \bar 1\) and \(1\bar\tau\) will each carry two different internal charges.

\section{Enlarging the Hilbert Space and the Duality Map}\label{sec:enlarge}

The Fibonacci fusion category \(\Fibo\) has a nontrivial \(2\)-dimensional Frobenius algebra 
\eqn[eq:FiboFrob]{
\A := \Big\{\alpha 1 + \beta\tau\ \Big|\ \alpha, \beta\in\CC,\ 1^2 = 1,\ 1\tau = \tau 1 = \tau,\ \tau^2 = 1 + \phi^{-\frac{3}{4}}\tau\Big\},
}
where \(1,\tau\) are the two simple objects of \(\Fibo\), regarded as the two basis elements of the algebra. Frobenius algebra \(\A\) has two simple bimodules: The two-dimensional trivial bimodule \(M_1 = (P_1, V_1)\), whose representation space \(V_1\) is
\eq{
V_1 := \{\alpha 1 + \beta \tau | \alpha, \beta\in \CC\},
}
and the three-dimensional nontrivial bimodule \(M_\tau = (P_\tau, V_\tau)\) with the representation space
\eq{
V_\tau := \{\alpha 1 + \beta \tau_1 + \gamma\tau_2 | \alpha, \beta, \gamma\in \CC\}.
}
Here, \(\tau_1\) and \(\tau_2\) are both the simple objects \(\tau\), but regarded as two different basis elements in \(V_\tau\) because they are acted on differently by \(\A\)\footnote{This is analogous to the scenario where an irreducible representation of a group can appear more than once in a certain reducible representation of the group. When physics kicks in, the different occurrences of the same irreducible representation are distinguishable.}. We refer to the indices \(1\) and \(2\) of \(\tau\) in \(M_\tau\) the multiplicity label of \(\tau\). The precise components of the representation tensors \(P_i: \A^2 \times V_i \times \{1, \tau\} \times V_i \to \C\) can be found in Appendix \ref{sec:frobdata}, where \(i = 1, \tau\).

The Fibonacci string-net model with the input fusion category \(\Bimod_\Fibo(\A)\), denoted as the \emph{dual Fibonacci string-net model}, describes the same topological order---the doubled Fibonacci topological order---as the original Fibonacci string-net model with the input fusion category \(\Fibo\). The fundamental degrees of freedom on the edges and tails of the dual model are simple bimodules \(M_1\) and \(M_\tau\), which are simple objects in \(\Bimod_\Fibo(\A)\). We construct a duality map \(\D\) that embeds the fundamental degrees of freedom of the dual model as superpositions of those in the original model:
\eqn[eq:FiboDual]{
\D\ \ \Edge{M_i}\quad :=\quad \frac{1}{\phi^4}\sum_{a,b,u = 1, \tau}\ \sum_{x, y \in L_{M_i}} [P_i]^{ab}_{xuy}\quad \ToricCodeD\ ,
}
where \(M_i = M_1, M_\tau\), and \(L_{M_1} = \{1, \tau\}, L_{M_\tau} = \{1, \tau_1, \tau_2\}\). The factor \(\phi^4\) in the denominator arises from \(d^2_{\A}\), where \(d_{\A} = d_1 + d_\tau = \phi^2\) is the total quantum dimension of Frobenius algebra \(\A\). The black line refers to both edges and tails. The red lines are auxiliary tails that will be annihilated by topological moves, resulting in a unitary transformation between the Hilbert spaces of the Fibonacci model and its dual model:
\eqn[eq:FiboDualPlaq]{
\FibonacciC\ ,
}
where \(I_k, E_k, M \in \{M_1, M_\tau\}\), and ``\(\cdots\)'' denotes the expansion coefficients.

After the topological moves, the degree of freedom \(\tau\) on any edge will cease to have any multiplicity index, while that on any tail will still have a multiplicity index if it belongs to \(L_{M_\tau}\). Therefore, to make sense of this duality and make it unitary, we are urged to enlarge the Hilbert space \(\Hil_\Fibo\) of the original Fibonacci string-net model to \(\Hil^\ast\) by distinguishing \(\tau_1\) and \(\tau_2\) on each tail but not on the edges. This enlargement is also physically sound\footnote{As an analogy, recall that a massless photon has only two physical degrees of freedom---the transverse polarizations---at each momentum. In principle, just two independent functions would be enough to specify the electromagnetic field completely. Yet when we are considering the gauge transformations, we introduce the full four-potential $A_\mu(k)$, which contains unphysical longitudinal (and timelike) components. These extra modes are not observable and can always be removed by imposing an appropriate gauge-fixing.}: An anyon excitation resides in a plaquette, in which the tail carries the internal charge of the anyon that reflects the action of Frobenius algebra \(\A\); the precise action of \(\A\) can only be told when different occurrences of \(\tau\) in the representation (bimodule) of \(\A\) are distinguished by multiplicity indices, viz \(\tau_1\) and \(\tau_2\). In contrast, the degrees of freedom on edges are pertaining to ground states because any path along edges only has to be a closed loop. At any vertex along such a closed loop, fusion rules are met by definition of the model, and fusion rules regard \(\tau_1\) and \(\tau_2\) the same\footnote{As an analogy: It makes no sense to question the electric charge in a closed electric flux loop because the Gauss law (analogous to fusion rules) is met everywhere along the loop. Only when the loop is cut open to be a path, one can ask about the charges at the ends of the path where the Gauss law is broken, which corresponds to the tails in our string-net model.}.  

The two degrees of freedom \(M_{1}\) and \(M_{\tau}\) on any given tail in the dual model are orthogonal:
\eq{\Bigg\langle\quad \Tail{M_i}\ \Bigg|\quad \Tail{M_j}\ \Bigg\rangle = \delta_{ij},\qquad M_i, M_j\in\{M_1, M_\tau\},}
and they shall remain orthogonal in the original model after applying the duality map \(\D\). Due to this orthonormality condition, the actual enlargement is done by embedding \(\Hil_\Fibo\) in \(\Hil^\ast\) as 
\eqn[eq:FiboEmbed]{
\Tail{1}\quad \Longrightarrow\quad \Tail{1}\quad,\qquad \Tail{\tau} \quad\Longrightarrow\quad \Bigg(\frac{1}{2\phi} + \frac{\sqrt{\phi}}{2} \Bigg)\quad\Tail{\tau_1}\quad + \quad \Bigg( \frac{1}{2\phi} - \frac{\sqrt{\phi}}{2} \Bigg)\quad \Tail{\tau_2}
}
for each \emph{tail}, as shown in Fig. \ref{fig:gaugeA}. The physical Hilbert space \(\Hil_\Fibo\) of the string-net model is a subspace of this enlarged Hilbert space \(\Hil^\ast\).

\section{Gauge Invariance of the Doubled Fibonacci Topological Order}\label{sec:gauge}

We can further construct a symmetry transformation of the doubled Fibonacci phase based on the duality \(\D\) defined in Eqs. \eqref{eq:FiboDual} and \eqref{eq:FiboDualPlaq} because their input fusion categories are isomorphic:
\eqn[eq:FiboIso]{
\F_\A: \Fibo \to \Bimod_\Fibo(\A),\qquad 1 \mapsto M_1,\qquad \tau \mapsto M_\tau.
}
Such isomorphism \(\F_\A\) induces an isomorphic map \(\varphi_\A\) between the Fibonacci string-net model and its dual model, and thus a \emph{unitary transformation} \(\G\) of the Fibonacci string-net model in the enlarged Hilbert space \(\Hil^\ast\):
\eqn[eq:FiboGauge]{\G := \D\circ\varphi_\A\ ,}
where
\eqn[eq:FiboIsoHil]{\varphi_\A\quad\Edge{1}\quad :=\quad \Edge{M_1},\qquad \varphi_\A\quad\Edge{\tau}\quad := \quad \Edge{M_\tau}.}
Here, the line refers to both edges and tails. Consequently, the unitary transformation \(\G\) transforms the local degrees of freedom \(1\) and \(\tau\) on edges (tails) to
\eqn[eq:FiboGaugeDef]{\Edge{1}\ \Longrightarrow\ \frac{1}{\phi^4}\sum_{a,b,x,u, y = 1, \tau}[P_1]^{ab}_{xuy}\ \ToricCodeD\ ,\qquad \Edge{\tau}\ \Longrightarrow\ \frac{1}{\phi^4}\sum_{\substack{a,b,u = 1, \tau\\x, y = 1, \tau_1, \tau_2}}\ [P_\tau]^{ab}_{xuy}\ \ToricCodeD\ .}
The red lines will be annihilated by topological moves. 

The unitary transformation \(\G\) \eqref{eq:FiboGauge} does not preserve the physical Hilbert space \(\Hil_\Fibo\) of the string-net model but rather rotates it within the enlarged Hilbert space \(\Hil^\ast\). Nevertheless, unitary transformation \(\G\) \eqref{eq:FiboGauge} is not yet a gauge transformation of the pure topological gauge field. In the string-net model, a tail can be taken as boundary conditions specifying punctures in the space in TQFT perspective, incorporating both the pure gauge field and charges on the boundary and representing how matters (anyons) couple to the gauge field. The pure gauge field's dof should be still simple objects \(1\) and \(\tau\) of \(\Fibo\), just like the dofs on edges. Thus, the appropriate symmetry transformation of the doubled Fibonacci topological \emph{gauge field} should be a projection of $\G$ back into \(\Hil_\Fibo\):
\eqn[eq:FiboGaugeProj]{
\tilde{\G} := \Proj\G\Proj,
}
where \(\Proj\) projects \(\Hil^\ast\) to \(\Hil_\Fibo\). The transformation \(\tilde{\G}\) preserves the physical Hilbert space \(\Hil_\Fibo\). The projected symmetry transformation \(\tilde{\G}\) is not unitary and is noninvertible. Specifically, the composition of symmetry transformation \(\tilde{\G}\) is given by the projection of multiplying unitary transformation:
\eqn[eq:compose]{
\tilde\G^{(n)} := \Proj\G^n \Proj
}
for \(n > 1\). This composition differs from the traditional way of composing symmetry transformations by simply multiplying the matrix \(\tilde{\G}\), which results in \(\tilde{\G}^n\ne \tilde\G^{(n)}\) for \(n > 1\) and does not represent valid symmetry transformations.

We now discuss how the unitary transformation \(\G\) \eqref{eq:FiboGauge} and the symmetry transformation \(\tilde{\G}\) \eqref{eq:FiboGaugeProj} transform the spectrum of the Fibonacci string-net model. The unitary transformation \(\G\) preserves the anyon species but acts nontrivially on the (enlarged) internal space of each anyon---the local Hilbert space expanded by the basic degrees of freedom \(\{1, \tau_1, \tau_2\}\) on the tail where the anyon is located. Different anyon species experience distinct actions. Specifically, we can block-diagonalize the \(\G\) \eqref{eq:FiboGauge} within the enlarged Hilbert subspaces representing different anyons:
\eqn[eq:FiboGaugeSpec]{\G = \prod_{{\rm Plaquettes\ }P}\sum_{{\rm Anyons\ }J}\G_P^J\Pi_P^J,}
where \(J = 1\bar 1, 1\bar\tau, \tau\bar 1, \tau\bar\tau\) is the anyon species, \(\Pi_P^J\) is the measurement operator \eqref{eq:FiboMeasure} measuring whether there is an anyon \(J\) in plaquette \(P\), and \(\G_P^J\) is block-diagonal in \(\Hil^\ast\), acting nontrivially only on the Hilbert subspace spanned by excited states with the same anyon species in all plaquettes, identical charges in all plaquettes except plaquette \(P\), and varying charges of anyon \(J\) on the tail in plaquette \(P\). 

Similarly, the symmetry transformation \(\tilde{\G}\), being the projection of \(\G\), represents a gauge invariance transformation that preserves the anyon species. The projected transformation \(\tilde\G\) can also be block-diagonalized along with \(\G\):
\eqn[eq:FiboGaugeProjSpec]{
\tilde\G = \Proj\G\Proj = \prod_{{\rm Plaquettes\ }P}\sum_{{\rm Anyons\ }J}\tilde\G_P^J\Pi_P^J,\qquad\qquad \tilde\G_P^J := \Proj\G_P^J\Proj.
}

\begin{figure}\centering
\subfloat[]{\FiboGaugeS\label{fig:gaugeA}}\hspace{40pt}
\subfloat[]{\FiboGaugeA\label{fig:gaugeB}}\hspace{60pt}
\subfloat[]{\FiboGaugeB\label{fig:gaugeC}}
\caption{(a) The physical basic degrees of freedom \(1\) and \(\tau\) embedded in the \(\{1, \tau_1, \tau_2\}\) enlarged Hilbert space. The black lines refer to the physical degrees of freedom \(1\) and \(\tau\). (2) The action of the unitary symmetry transformations \(\G\) (the blue vectors) and the symmetry transformation \(\tilde\G\) (the orange vectors) for anyon species \(\tau\bar 1, 1\bar\tau\). (c) The action of the unitary symmetry transformations \(\G\) (the blue vectors) and the symmetry  transformation \(\tilde\G\) (the orange vectors) for anyon species \(\tau\bar\tau\).}
\label{fig:gauge}
\end{figure}

\begin{enumerate}
\item The trivial anyon \(1\bar 1\) has only one charge \(1\) that transforms trivially under the symmetry transformation:
\eqn[eq:FiboII]{
\G_P^{1\bar 1} = \tilde\G_P^{1\bar 1} = \idm,\qquad\qquad \G_P^{1\bar 1}\quad\Tail{1}\quad =\quad \tilde\G_P^{1\bar 1}\quad\Tail{1}\quad = \quad \Tail{1}\ .
}
The ground states of the doubled Fibonacci topological order are invariant under the gauge invariance transformation.
\item Anyon \(\tau\bar 1\) has only one charge \(\tau\) in the string-net model. Nevertheless, we have enlarged the degree of freedom \(\tau\) on a tail to a \(2\)-dimensional space spanned by degrees of freedom \(\tau_1\) and \(\tau_2\). We can now express the symmetry transformation \(\G\) \eqref{eq:FiboGauge} in this enlarged internal space for \(\tau\bar 1\) anyon:
\begin{align}
&\G_P^{\tau\bar 1} = \mat{1 & 0 & 0 \\ 
0 & \frac{\phi\sqrt{\phi^2 + 1} - 1}{2\phi\sqrt{\phi}} & \frac{\sqrt{\phi^2 + 1} + \phi^2}{2\phi^2} \\ 
0 & -\frac{\sqrt{\phi^2 + 1} + \phi^2}{2\phi^2} & \frac{\phi\sqrt{\phi^2 + 1} - 1}{2\phi\sqrt{\phi}}},\nonumber\\
&\tilde\G_P^{\tau\bar 1} = \mat{1 & 0 & 0 \\ 
0 & \frac{(\phi\sqrt{\phi^2 + 1} - 1)(\sqrt{\phi} + 1)}{4\phi^2} & \frac{1 - \phi\sqrt{\phi^2 + 1}}{4\phi^2\sqrt{\phi}}\\
0 & \frac{1 - \phi\sqrt{\phi^2 + 1}}{4\phi^2\sqrt{\phi}} & \frac{(\phi\sqrt{\phi^2 + 1} - 1)(\sqrt{\phi} - 1)}{4\phi^2}
},\label{eq:FiboTI}\\
&\G_P^{\tau\bar 1}\quad\Tail{\tau}\quad = \quad\Tail{\tau_{\tau\bar 1}}\quad = \quad\frac{\sqrt[4]{5} - 1}{2\sqrt{\phi}}\quad\Tail{\tau_1}\quad - \frac{\sqrt[4]{5} + 1}{2\sqrt{\phi}}\quad\Tail{\tau_2}\quad,\nonumber\\
&\tilde\G_P^{\tau\bar 1}\quad\Tail{\tau}\quad = \quad\Tail{\tilde\tau_{\tau\bar 1}}\quad  =\quad\frac{\phi\sqrt{\phi^2 + 1} - 1}{2\phi\sqrt{\phi}}\quad\Tail{\tau}\quad.\nonumber
\end{align}
Here, the physical degree of freedom \(\tau\) of the anyon \(\tau \bar{1}\) is a superposition \eqref{eq:FiboEmbed} of \(\tau_1\) and \(\tau_2\). The unitary transformation \(\G\) \eqref{eq:FiboGauge} rotates this physical charge \(\tau\) of the \(\tau \bar{1}\) anyon out of the physical Hilbert space \(\Hil_\Fibo\), necessitating the application of the projected transformation \(\tilde{\G}_P^{\tau \bar{1}}\) to preserve the physical Hilbert space, as shown in Fig. \ref{fig:gaugeB}. Note that because \(\det(\tilde{\G}_P^{\tau \bar{1}}) = 0\), the matrix \(\tilde{\G}_P^{\tau \bar{1}}\) does not have an inverse matrix \([\tilde{\G}_P^{\tau \bar{1}}]^{-1}\). For this reason, the symmetry transformation \(\tilde{\G}\) \eqref{eq:FiboGaugeProj} represents a noninvertible symmetry. 
\item Anyon \(1\bar\tau\) also has a single charge \(\tau\). The unitary transformation \(\G\) \eqref{eq:FiboGauge} rotates the physical charge \(\tau\) \eqref{eq:FiboEmbed} of the \(1\bar\tau\) anyon within the \(\{\tau_1, \tau_2\}\) space but in a different way compared to the \(\tau\bar 1\) anyon in Eqs. \eqref{eq:FiboTI}:
\begin{align}
&\G_P^{1\bar\tau} = \mat{1 & 0 & 0 \\ 
0 & -\frac{\phi\sqrt{\phi^2 + 1} + 1}{2\phi\sqrt{\phi}} & \frac{\phi\sqrt{\phi} - \sqrt[4]{5}}{2\phi\sqrt{\phi}} \\ 
0 & \frac{\sqrt[4]{5} - \phi\sqrt{\phi}}{2\phi\sqrt{\phi}} & -\frac{\phi\sqrt{\phi^2 + 1} + 1}{2\phi\sqrt{\phi}}},\nonumber\\
&\tilde\G_P^{\tau\bar 1} = \mat{1 & 0 & 0 \\ 
0 & -\frac{(\phi\sqrt{\phi^2 + 1} + 1)(\sqrt{\phi} + 1)}{4\phi^2} & \frac{1 + \phi\sqrt{\phi^2 + 1}}{4\phi^2\sqrt{\phi}}\\
0 & \frac{1 + \phi\sqrt{\phi^2 + 1}}{4\phi^2\sqrt{\phi}} & -\frac{(\phi\sqrt{\phi^2 + 1} + 1)(\sqrt{\phi} - 1)}{4\phi^2}
} ,\label{eq:FiboIT}\\
&\G_P^{1\bar\tau}\quad\Tail{\tau}\quad = \quad\Tail{\tau_{1\bar\tau}}\quad = \quad - \frac{\sqrt[4]{5} + 1}{2\sqrt{\phi}}\quad\Tail{\tau_1}\quad + \quad\frac{\sqrt[4]{5} - 1}{2\sqrt{\phi}}\quad\Tail{\tau_2}\quad,\nonumber\\
&\tilde\G_P^{1\bar\tau}\quad\Tail{\tau}\quad = \quad\Tail{\tilde\tau_{1\bar\tau}}\quad =\quad -\frac{\phi\sqrt{\phi^2 + 1} + 1}{2\phi\sqrt{\phi}}\quad\Tail{\tau}\ .\nonumber
\end{align}
The transformations are depicted in Fig. \ref{fig:gaugeB}. The matrix \(\tilde{\G}_P^{1\bar\tau}\) is noninvertible.
\item Anyon \(\tau\bar\tau\) has two gauge charge \(1\) and \(\tau\) in the string-net model. In the enlarged \(3\)-dimensional space spanned by basic degrees of freedom \(1\), \(\tau_1\), and \(\tau_2\), the unitary transformation \(\G\) \eqref{eq:FiboGauge} rotates these two physical charges \(1\) and \(\tau\) \eqref{eq:FiboEmbed} (see Fig. \ref{fig:gaugeC}.):
\begin{align}
&\G_P^{\tau\bar\tau} = \mat{\frac{1}{\phi^2} & \frac{\phi^2\sqrt{5} - \sqrt{\phi}}{2\phi^4}\sqrt[4]{5} & -\frac{\phi^2\sqrt{5} + \sqrt{\phi}}{2\phi^4}\sqrt[4]{5}\\ 
\frac{\sqrt[4]{5} + \phi\sqrt{\phi^2 + 1}}{2\phi^2} & \frac{1 - \phi^2\sqrt{5\phi}}{2\phi^4} & \frac{\phi\sqrt{5} - 2\sqrt{\phi}}{2\phi^3} \\ 
\frac{\sqrt[4]{5} - \phi\sqrt{\phi^2 + 1}}{2\phi^2} & -\frac{\phi\sqrt{5} + 2\sqrt{\phi}}{2\phi^3} & -\frac{1 + \phi^2\sqrt{5\phi}}{2\phi^4}\\ 
},\nonumber\\ 
&\tilde\G_P^{\tau\bar\tau} = \mat{\frac{1}{\phi^2} & \frac{\sqrt[4]{5} + \phi\sqrt{\phi^2 + 1}}{2\phi^2\sqrt{\phi}} & \frac{\sqrt[4]{5} - \phi\sqrt{\phi^2 + 1}}{2\phi^2\sqrt{\phi}}\\
\frac{\sqrt[4]{5} + \phi\sqrt{\phi^2 + 1}}{2\phi^2} & - \frac{1 + \sqrt{\phi}}{2\phi^3} & \frac{1}{2\phi^3\sqrt{\phi}} \\
\frac{\sqrt[4]{5} - \phi\sqrt{\phi^2 + 1}}{2\phi^2} & \frac{1}{2\phi^3\sqrt{\phi}} & \frac{1 - \sqrt{\phi}}{2\phi^3}
},\nonumber\\
&\G_P^{\tau\bar\tau}\quad\Tail{1}\quad = \tilde\G_P^{\tau\bar\tau}\quad\Tail{1}\quad = \quad\Tail{1_{\tau\bar\tau}}\quad = \quad \frac{1}{\phi^2}\quad\Tail{1}\quad + \quad \frac{\sqrt[4]{5}}{\phi}\quad\Tail{\tau}\ ,\\
&\G_P^{\tau\bar\tau}\quad\Tail{\tau}\quad = \quad\Tail{\tau_{\tau\bar\tau}}\quad = \quad \frac{\sqrt{\phi^2 + 1}}{\phi^2}\quad\Tail{1} \quad - \quad \frac{\sqrt{\phi}}{\phi^2}\quad\Tail{\tau_1} \quad - \quad \frac{\sqrt{\phi}}{\phi^2}\quad\Tail{\tau_2}\quad ,\nonumber\\
&\tilde\G_P^{\tau\bar\tau}\quad\Tail{\tau}\quad = \quad\Tail{\tilde\tau_{\tau\bar\tau}}\quad = \quad \frac{\sqrt{\phi^2 + 1}}{\phi^2}\quad\Tail{1} \quad - \quad \frac{\sqrt{\phi}}{\phi^3}\quad\Tail{\tau} \quad .\nonumber
\end{align}
When \(\G_P^{\tau\bar{\tau}}\) acts on an excited state with a \(\tau\bar{\tau}\) anyon in plaquette \(P\), the charges on the tail in plaquette \(P\) transform into a superposition of charges \(1\), \(\tau_1\), and \(\tau_2\). This transformation does not mean that only the degrees of freedom on the tail in \(P\) change while leaving the degrees of freedom on all other edges and tails invariant. Instead, the excited states are transformed into a superposition of \emph{excited states} where the anyon \(\tau\bar{\tau}\) in plaquette \(P\) has charges \(1\) and \(\tau\) (or \(1\), \(\tau_1\), and \(\tau_2\)).

The matrix \(\tilde{\G}_P^{\tau\bar{\tau}}\) is noninvertible. 
\end{enumerate}
 
We have to note that while the explicit representation matrices for the symmetry transformations are defined in the string-net model, the symmetry itself is an intrinsic property of the doubled Fibonacci topological order and is independent of the specific model realization. We will discuss the structure of this symmetry in the next section and show that it is a categorical gauge invariance described by a fusion \(2\)-category.

\section{Fibonacci Categorical gauge invariance}\label{sec:cat}

We now show that the symmetry of the doubled Fibonacci topological order is a \emph{categorical gauge invariance} characterized by a fusion \(2\)-category---the Fibonacci fusion \(2\)-category. 

Recall that in a usual gauge theory with gauge group \(G\), a gauge-field value \(g \in G\) is transformed to \(g' = hgh^{-1} \in G\) by a gauge transformation characterized by \(h \in G\). The gauge group \(G\) is both the space of the gauge field and is space of the gauge transformations. The question in our case is: What is the structure of the symmetry of the doubled Fibonacci topological order that is analogous to the gauge group \(G\) together with its symmetry transformation in a usual gauge theory?

Looking back to how \(\G\) transforms the basic dof as in Eq. \eqref{eq:FiboGaugeDef}, the two sides of \eqref{eq:FiboGaugeDef} seem different in lattice structures, although the red lines on the RHS are auxiliary. But the two sides do have the same lattice structure even before annihilating the red lines. In \(\Fibo\), \(1\) and \(\tau\) are simple bimodules over the trivial Frobenius algebra \(\A_0 = \C[1]\), so \(\Fibo = \Bimod_{\Fibo}(\A_0)\). Thus, an edge/tail labeled by simple object \(i\in\{1, \tau\}\) in the original model must also carry two red lines: 
\eqn[eq:expandtrans]{\FibonacciD{i} \Rightarrow \frac{1}{\phi^4}\sum_{a,b,u = 1, \tau}\sum_{x, y \in L_{M_i}} [P_\tau]^{ab}_{xuy} \ToricCodeD\ ;}
however, these two red lines are omitted in the original model because they are labeled by trivial object \(1\).

It turns out that
\eq{
\Bimod_\Fibo(\A) \subset \Fibo
\quad\text{and}\quad \Fibo \cong \Bimod_\Fibo(\A).
}
The bimodules \(M_1, M_\tau\) are composite objects in \(\Fibo\). In words, our transformation \(\G\) \eqref{eq:FiboGauge} transforms the string-net model with input fusion category \(\Fibo=\Bimod_{\Fibo}(\A_0)\) to to be that with input subcategory \(\Bimod_{\Fibo}(\A)\subset \Fibo\). The input data \(\Fibo\) is invariant. This is indeed a gauge invariance transformation because it preserves the Hamiltonian \(\Hil^\ast\) and transforms the internal spaces of anyons only. 

Inspired by the above discussions, we find that a fusion \(2\)-category---the \textbf{Fibonacci fusion 2-category} exists to describe this gauge invariance coherently. It is defined\footnote{For a general definition and properties of fusion \(2\)-categories, refer to Ref. \cite{douglas2018}.} by the following three ingredients: 
\begin{enumerate}
\item The objects in the Fibonacci fusion \(2\)-category are bimodule categories over Frobenius algebras in \(\Fibo\). These bimodule categories are subcategories of and isomorphic to \(\Fibo\). Two such objects are \(\Bimod_\Fibo(\A_0)=\Fibo\) and \(\Bimod_\Fibo(\A)\subset \Fibo\).
\item The \(1\)-morphisms are isomorphism functors between objects. Now that all bimodule categories in the Fibonacci fusion \(2\)-category are isomorphic, we only need to define the functors \(\F_\A'\) from \(\Fibo = \Bimod_\Fibo(\A_0)\) to the other bimodule category \(\Bimod_\Fibo(\A')\). That is, each \(1\)-morphism is labeled by a Frobenius algebra in \(\Fibo\). Two special examples of \(1\)-morphisms are the identity functor 
\eq{
{\rm id}(1) = 1,\qquad {\rm id}(\tau) = \tau,
}
and \(\F_\A\) defined in Eq. \eqref{eq:FiboIso}:
\eq{
\F_\A(1) = M_1,\qquad \F_\A(\tau) = M_\tau.
}
\item The \(2\)-morphisms are natural transformations of \(1\)-morphisms. We are particularly interested in two types of \(2\)-morphisms: the composition of \(1\)-morphisms and the direct sum of \(1\)-morphisms. Since all bimodule categories are isomorphic, the composition of \(1\)-morphisms is well-defined and maps simple objects \(1\) and \(\tau \in \Fibo\) to bimodules over Frobenius algebras in the bimodule categories. These compositions are also \(1\)-morphisms because the bimodules in the bimodule categories are composite objects in the parent fusion category \(\Fibo\) and are bimodules over some Frobenius algebras in \(\Fibo\). The explicit structures of these compositions are complex and will not be discussed here.

Given two \(1\)-morphisms \(\mathcal{F}_{\A_1}\) and \(\mathcal{F}_{\A_2}\), the direct sum is
\eq{
(\mathcal{F}_{\A_1}\oplus\mathcal{F}_{\A_2})(1) = \mathcal{F}_{\A_1}(1)\oplus\mathcal{F}_{\A_2}(1),\qquad (\mathcal{F}_{\A_1}\oplus\mathcal{F}_{\A_2})(\tau) = \mathcal{F}_{\A_1}(\tau)\oplus\mathcal{F}_{\A_2}(\tau).
}
The direct sums \(\mathcal{F}_{\A_1}(1)\oplus\mathcal{F}_{\A_2}(1)\) and \(\mathcal{F}_{\A_1}(\tau)\oplus\mathcal{F}_{\A_2}(\tau)\) are the two simple bimodules over disconnected Frobenius algebra \(\A_1\oplus\A_2\).
\end{enumerate}

In the framework of the Fibonacci fusion \(2\)-category, the transformation \(\G\) \eqref{eq:FiboGauge} is a representation of the \(1\)-morphism \(\phi\) over the enlarged Hilbert space \(\Hil^\ast\):
\eq{
\rho(\id) = \idm,\qquad \rho(\F_\A) = \G,
}
which is compatible with the \(2\)-morphisms:
\eq{
\rho(\mathcal{F}_1\circ\mathcal{F}_2) = \rho(\mathcal{F}_2)\rho(\mathcal{F}_1),\qquad \rho(\mathcal{F}_1\oplus\mathcal{F}_2) = \rho(\mathcal{F}_2) + \rho(\mathcal{F}_1).
}
Now, to answer the question raised at the beginning of this section, the Fibonacci fusion \(2\)-category is the mathematical structure representing the \emph{phase space} of the doubled Fibonacci order, comprising both the topological gauge field’s dofs and the conjugate momenta---symmetry transformations. The dof space is UFC \(\Fibo\), and gauge transformations are \(1\)-morphisms\footnote{In a usual (in particular a lattice) gauge theory, the gauge field's dof space is the gauge group, which per se is the representation space of the gauge transformations---the adjoint representation of the gauge group. Consider gauge group $U(1)$ as a simple example, the gauge transformation $\df/\df\theta$ is the conjugate momentum of $\exp{\ii\theta}\in U(1)$. }. The action of \(1\)-morphisms on the representation space \(\Fibo\) forms the ``adjoint representation'' of UFC \(\Fibo\), as all resultant bimodule categories are subcategories of \(\Fibo\).

We name the gauge symmetry of the doubled Fibonacci order and the Fibonacci Turaev-Viro TQFT as the \emph{Fibonacci 2-categorical gauge symmetry}. Each bimodule category in the Fibonacci fusion \(2\)-category is a choice of gauge fixing, and the chosen bimodule category's two simple objects are anyons' gauge charges. Since we chose \(\Fibo\) as our gauge fixing, after the symmetry transformation, we must project the transformed states back into \(\Hil_\Fibo\), such that anyons' charges are measured in the basis comprising the original dof \(1\) and \(\tau\).

Besides, the global symmetry transformation \(\G\) maps the degrees of freedom \(\pm1\)---the simple objects in the \(\Z_2\) fusion category---to the degrees of freedom \(M_\pm\), which are the simple objects in the bimodule category \(\Bimod_{\Z_2}(\A_{\Z_2})\). Analogous to our discussion about the Fibonacci categorical gauge invariance, this $em$ exchange global symmetry of the toric code topological order is also fundamentally a \emph{categorical symmetry} described by a fusion \(2\)-category---the \(\Z_2\) fusion \(2\)-category defined based on the \(\Z_2\) fusion category. It reduces to the traditional \(\Z_2\) group description because the transformation is invertible.

\section{Local Symmetry Transformations}

\begin{figure}\centering
\subfloat[]{\includegraphics[width = 6cm]{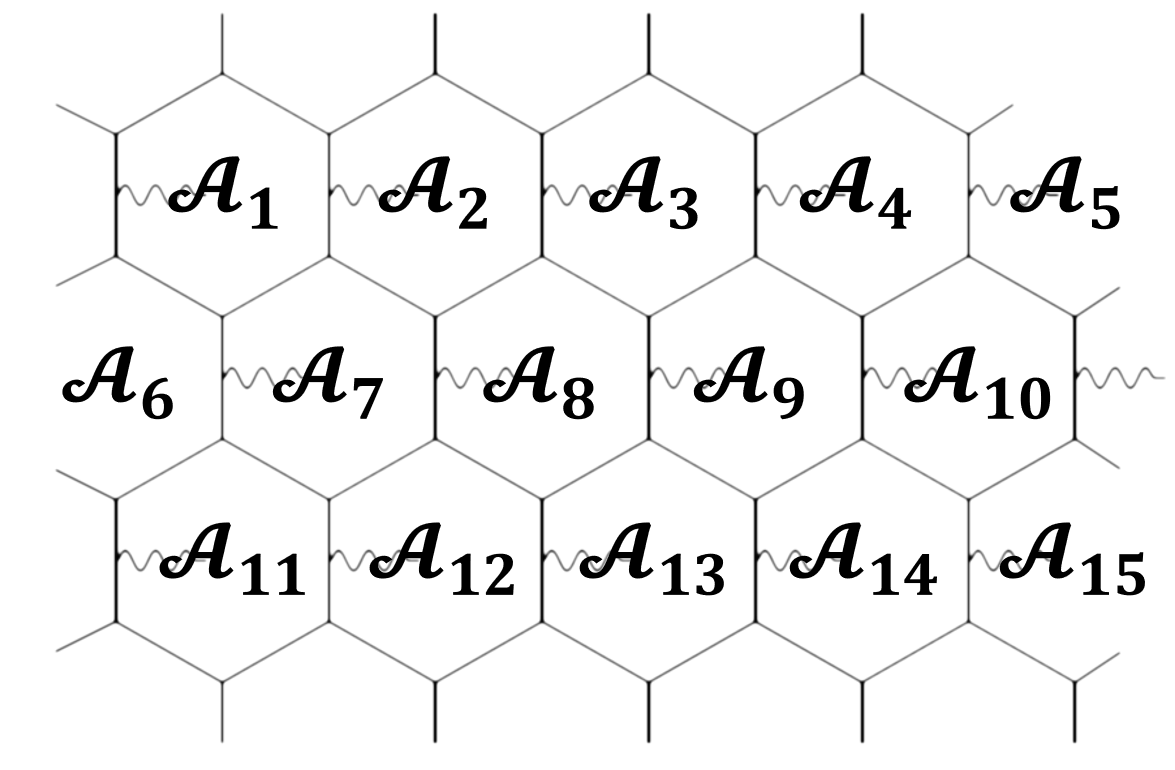}\label{fig:localA}}\hspace{50pt}
\subfloat[]{\includegraphics[width = 5cm]{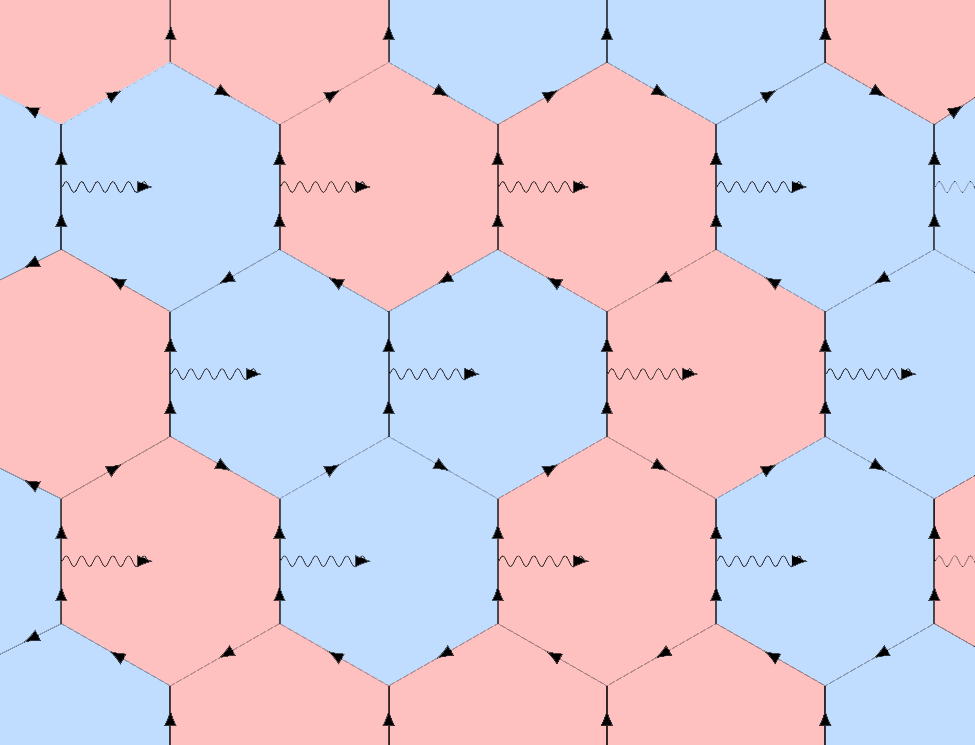}\label{fig:localB}}
\caption{(a) In the local dual model, each plaquette $P$ is equipped with a Frobenius algebra $\A_P$. The tail in $P$ carries simple $\A_P$ bimodule objects and the edge between adjacent plaquettes $P, Q$ carries simple $\A_P\text{-}\A_Q$ bimodules. (b) Under a local global symmetry transformation, different global symmetry sectors (red and blue regions) are separated by symmetric gapped domain walls---simple $\A_{\text{red}}-\A_{\text{blue}}$ bimodules on the edges, where $\A_{\text{red}}-\A_{\text{blue}}$ bimodule category is not isomorphic to \(\Bimod_\Fus(\A_{\text{red}})\cong\Bimod_\Fus(\A_{\text{blue}})\).}
\label{fig:local}
\end{figure}

Up to this point we have considered only \emph{uniform} gauge transformations that apply the same transformation everywhere on the lattice simultaneously. Gauge transformations, however, are \emph{local}: the transformations may differ at different locations.

To extend our construction to local gauge transformations, we introduce the notion of \emph{local dual model} and \emph{local duality map} \(\D\) that is no longer defined by a single Frobenius algebra:

\begin{itemize}
\item A local dual model is defined by labeling each plaquette \(P\) of the lattice with a Frobenius algebra \(\A_{P}\) in UFC \(\Fus\); Different plaquettes may be labeled by different algebras, and duality transformations are to be defined locally dictated by plaquettes' algebras. See Figure \ref{fig:localA}. In this section, we only consider those Frobenius algebras \(\A_P\) in UFC \(\Fus\), such that \(\Bimod_\Fus(\A_P)\) is always isomorphic to UFC \(\Fus\).

\item In the local dual model, every tail inside plaquette \(P\) is labeled by a simple bimodule over Frobenius algebra \(\A_P\), while every edge separating two neighbouring plaquettes \(P\) and \(Q\) is labeled by a simple \(\A_{P}\text{-}\A_{Q}\) bimodule \(M\), i.e.\ a linear space \(V_{M}\) spanned by simple objects of \(\Fus\), equipped with a function \(P_{M}:\A_{P}\times\A_{Q}\times V_{M}^{2}\times L_{\Fus}\to\mathbb{C}\) encoding the left action of elements of \(A_{P}\) and the subsequent right action of elements of \(A_{Q}\) on \(V_{M}\).

\item The local duality map \(\D\) sends each basis degree of freedom \(M\)---an \(\A_P\text{-}\A_Q\) bimodules---on edges and tails (\(P = Q\) in this case) of the local dual model back into superposition states in the original model by
\eqn{\Edge{M}\ \Longrightarrow\ \frac{1}{d_{\A_P}d_{\A_Q}}\sum_{a\in L_{\A_P}}\sum_{b\in L_{\A_Q}}\sum_{x, y\in L_M}\sum_{u\in L_\Fus}[P_M]^{ab}_{xuy}\ \ToricCodeD\ ,}
where \(L_{\A}\) (\(L_M\)) is the set of simple objects in Frobenius algebra \(\A\) (Bimodule \(M\)).

\item Annihilating all auxiliary tails by topological moves. This is always valid because auxiliary tails are annihilated plaquettes by plaquettes, while the degrees of freedom on all auxiliary tails in a plaquette \(P\) belongs to the same Frobenius algebra \(\A_P\).
\end{itemize}

This construction provides criterion equivalent to that in Section \ref{subsec:criterion} for distinguishing global symmetries from gauge invariances. If, for any two Frobenius algebras \(A_P, A_Q\) appearing in certain plaquettes of the local dual model, the \(\A_{P}\text{-}\A_{Q}\) bimodules always form a fusion category, which is isomorphic to the original fusion category \(\Fus\), then the local duality map can be composed with an isomorphism to be a local gauge invariance; otherwise it may yield a global symmetry. In fact, the modules over Frobenius algebra \(\A\) in Section \ref{subsec:criterion} are just \(\A_0\text{-}A\) bimodules, where \(\A_0\) is the trivial Frobenius algebra in \(\Fus\).

In the case of gauge invariance, the local degrees of freedom on every edge or tail---the simple \(\A_P\text{-}\A_Q\) bimodules---form a UFC isomorphic to \(\Fus\). Accordingly, for every edge or tail \(e\) we can choose an isomorphism
\[
\varphi_{e}: \Edge{x_e}\quad\Longrightarrow\quad\Edge{M_e},
\]
where \(M_e\) is a simple \(\A_{P}\text{-}\A_{Q}\) bimodule if edge \(e\) separates plaquettes \(P\) and \(Q\) (or a simple \(\A_{P}\)-bimodule if \(e\) is a tail within plaquette \(P\)). The corresponding local gauge transformation acting on the internal spaces of the anyons factorizes as 
\eqn{\G = \prod_{{\rm Plaquettes\ }P}\sum_{{\rm Anyons\ }J}\G_{P;\A_P}^J\Pi_P^J,}
where \(\G^J_{P;\A_{P}}\) is the symmetry action on the internal space of anyon \(J\) in plaquette \(P\), determined only by the Frobenius algebra \(\A_{P}\) locally in plaquette \(P\).

A richer scenario arises when, \(\Bimod_\Fus(\A_P)\) for all \(\A_P\) are always isomorphic to the original fusion category \(\Fus\), yet the mixed \(\A_P\text{-}\A_Q\) bimodules on edges do \emph{not} form a fusion category. In this setting we can apply local isomorphisms only to the tail degrees of freedom, yielding a global symmetry action on anyons in each plaquette, while the
edge-carried \(\A_P\text{-}\A_Q\) bimodules persist as \emph{symmetric domain walls} that separate distinct global symmetry sectors of the underlying topological
order described by the \(\Fus\)-model (see Figure \ref{fig:localB}). As to be reported in our subsequent works, the resultant lattice model realises a symmetry-enriched topological
(SET) phase, and if we introduce these edge bimodules as new fundamental degrees of freedom of our lattice model, we effectively \emph{gauge} the original model: The global symmetry of the \(\Fus\)-string-net become local gauge invariances of the parent theory, and the phase ascends to a parent (gauged) topological order.

\begin{acknowledgments}
The authors thank Davide Gaiotto, Lukas Mueller, Ling-Yan Hung, Hongguang Liu, Yuting Hu, Chenjie Wang, Nianrui Fu, and Yinan Wang for inspiring and helpful discussions. YW is supported by NSFC Grant No. KRH1512711, the Shanghai Municipal Science and Technology Major Project (Grant No. 2019SHZDZX01), Science and Technology Commission of Shanghai Municipality (Grant No. 24LZ1400100), and the Innovation Program for Quantum Science and Technology (No. 2024ZD0300101). The authors are grateful for the hospitality of the Perimeter Institute during his visit, where the main part of this work is done. This research was supported in part by the Perimeter Institute for Theoretical Physics. Research at Perimeter Institute is supported by the Government of Canada through the Department of Innovation, Science and Economic Development and by the Province of Ontario through the Ministry of Research, Innovation and Science. 
\end{acknowledgments}

\appendix

\section{Review of the Extended String-net Model}\label{sec:review}

In this section, we briefly review the string-net model defined in Ref. \cite{Hu2018}. The string-net model is an exactly solvable Hamiltonian model defined on a \(2\)-dimensional lattice. An example lattice is depicted in Fig. \ref{fig:lattice}. All vertices are trivalent. Within each plaquette of the lattice, a tail is attached to an arbitrary edge of the plaquette, pointing inward. We will later demonstrate that different choices of the edge to which the tail is attached are equivalent. In general cases, each edge and tail is oriented, and different choices of directions are equivalent. Nevertheless, for the case of the Fibonacci string-net model presented in the main body, different direction choices are the same, so we omit the directions of edges and tails in the main body.


The input data of the string-net model is a fusion category \(\Fus\), described by a finite set \(L_\Fus\), whose elements are called \emph{simple objects}, equipped with three functions \(N: L_\Fus^3 \to \NN\), \(d: L_\Fus \to \RR^+\), and \(G: L^6_\Fus \to \C\). The function \(N\) sets the \emph{fusion rules} of the simple objects, satisfying
\eq{
\sum_{e \in L_\Fus} N_{ab}^e N_{ec}^d = \sum_{f \in L_\Fus} N_{af}^d N_{bc}^f, \qquad\qquad N_{ab}^c = N_{c^\ast a}^{b^\ast}.
}
There exists a special simple object \(1 \in L_\Fus\), called the \emph{trivial object}, such that for any \(a, b \in L_\Fus\),
\eq{
N_{1a}^b = N_{1b}^a = \delta_{ab},
}
where \(\delta\) is the Kronecker symbol. For each \(a \in L_\Fus\), there exists a unique simple object \(a^\ast \in L_\Fus\), called the \emph{opposite object} of \(a\), such that
\eq{
N_{ab}^1 = N_{ba}^1 = \delta_{ba^\ast}.
}
We only consider the case where for any \(a, b, c \in L_\Fus\), \(N_{ab}^c = 0\) or \(1\). In this case, we define
\eq{
\delta_{abc} = N_{ab}^{c^\ast} \in \{0, 1\}.
}
In this work we assume commutative fusion rules, \(\delta_{abc} = \delta_{bac}\); however, our construction extends readily to noncommutative fusion categories (see the Appendices of Ref. \cite{zhao2025}).

The basic configuration of the string-net model is established by labeling each edge and tail with a simple object in \(L_\Fus\), subject to the constraint on all vertices that \(\delta_{ijk} = 1\) for the three incident edges or tails meeting at this vertex, all pointing toward the vertex and respectively counterclockwise labeled by \(i, j, k \in L_\Fus\). We can reverse the direction of any edge or tail and simultaneously conjugate its label as \(j \to j^\ast\), which keeps the configuration invariant. The Hilbert space \(\Hil\) of the model is spanned by all possible configurations of these labels on the edges and tails.

The function \(d\) returns the \emph{quantum dimensions} of the simple objects in \(L_\Fus\). It is the largest eigenvalues of the fusion matrix and forms the \(1\)-dimensional representation of the fusion rule.
\eq{d_ad_b = \sum_{c\in L_\Fus}N_{ab}^cd_c.}
In particular, \(d_1 = 1\), and for any \(a\in L_\Fus, d_a = d_{a^\ast}\ge 1\). 

The function \(G\) defines the \(6j\)-\emph{symbols} of the fusion algebra. It satisfies
\eq{\sum_nd_nG^{pqn}_{v^*u^*a}G^{uvn}_{j^*i^*b}G^{ijn}_{q^*p^*c} = G^{abc}_{i^*pu^*}&G^{c^*b^*a^*}_{vq^*j},\qquad\sum_nd_nG^{ijp}_{kln}G^{j^*i^*q}_{l^*k^*n} = \frac{\delta_{pq^*}}{d_p}\delta_{ijp}\delta_{klq},\\
G^{ijm}_{kln} = G^{klm^*}_{ijn^*} = G^{jim}_{lkn^*}= G^{mij}_{nk^*l^*} &= \alpha_m\alpha_n\overline{G^{j^*i^*m^*}_{l^*k^*n^*}},\qquad \Big|G^{abc}_{1bc}\Big| = \frac{1}{\sqrt{d_bd_c}}\delta_{abc}.
}
where \(\alpha_m = G^{1mm^\ast}_{1m^\ast m}\in\{\pm 1\}\) is the Frobenius-Schur indicator of simple object \(m\).

The Hamiltonian of the string-net model reads
\eqn{H := - \sum_{{\rm Plaquettes}\ P}Q_P,\qquad\qquad Q_P := \frac{1}{D}\sum_{s\in L_\Fus}d_sQ_P^s,\qquad D := \sum_{a\in L_\Fus}d_a^2,}
where the plaquette operator \(Q_P^s\) acts on edges surrounding plaquette \(P\) and has the following matrix elements on a hexagonal plaquette\footnote{We only show the actions of \(Q_P\) operator on a hexagonal plaquette. The matrix elements of \(Q_P\) operators on other types of plaquettes are defined similarly.}:
\eq{Q_P^s\ \PlaquetteSrc\
:=\ &\delta_{p,1}\ \delta_{j_1,j_7}\ \sum_{j_k\in L_\Fus}\  \prod_{k = 1}^{6}\ \Bigg(\sqrt{d_{i_k}d_{j_k}}\ G^{e_ki_ki_{k+1}^\ast}_{sj_{k+1}^\ast j_k}\Bigg)\PlaquetteTar\ .}
Here, we omit the ``$\ket{\cdot}$'' labels surrounding all diagram for simplicity, unless they are specifically required.

It turns out that
\eq{
(Q_P^s)^\dagger = Q_P^{s^\ast},\qquad Q_P^rQ_P^s = \sum_{t\in L_\Fus} N_{rs}^tQ_P^t,\qquad Q_P^2 = Q_P,\qquad Q_{P_1}Q_{P_2} = Q_{P_2}Q_{P_1}.
}
The summands \(Q_P\) in Hamiltonian \(H\) are commuting projectors, so the Hamiltonian is exactly solvable. The ground-state subspace \(\Hil_0\) of the system is the projection
\eqn{
\Hil_0 = \Bigg[\prod_{{\rm Plaquettes\ } P}Q_P\Bigg]\Hil.
}
If the lattice has the sphere topology, the model has a unique ground state \(\ket\Phi\) up to scalars.

\subsection{Topological Features}\label{sec:pachner}

We briefly review the topological nature of the ground-state subspace of the string-net model defined in Ref. \cite{Hu2018}. Any two lattices with the same topology can be transformed into each other by so-called \emph{Pachner moves}. There are unitary linear maps between the Hilbert spaces of two string-net models with the same input fusion category on different lattices associated with these moves, denoted as \(\T\). The ground states are invariant under such linear transformations. There are three kinds of elementary Pachner moves, whose corresponding linear transformations are:
\eqn[eq:pachner]{
\T \quad \PachnerOne\ ,\\
\T \quad \PachnerTwo\ ,\\
\T \quad \PachnerThree\ .}
Here we use red ``\({\color{red}\times}\)'' to mark the plaquettes to contract. Any other Pachner moves and their corresponding linear transformations of Hilbert spaces are compositions of these three elementary moves. Given initial and final lattices, there are multiple ways to compose these elementary moves, but different ways result in the same transformation matrices on the ground-state Hilbert space.

We have also noted that different selections of the edge to which the tail is attached are equivalent. These variations lead to distinct lattice configurations and, consequently, different Hilbert spaces for the lattice model. The equivalence of states in such Hilbert spaces is established by the following linear transformation \(\T'\):
\eqn{
\T'\quad\PachnerFour\ .
}
The states where tails attach to other edges can be obtained recursively in this manner.

For convenience, in certain instances, we will temporarily incorporate auxiliary states with multiple tails within a single plaquette. These states, despite having multiple tails in one plaquette, are all equivalent to states within the Hilbert space:
\eqn{
\PachnerFive\ .
}

\subsection{Excited States}\label{sec:spec}

An \emph{excited state} \(\ket\varphi\) of the string-net model is an eigenstate of the Hamiltonian such that \(Q_P\ket\varphi = 0\) at some plaquettes \(P\). In such a state, there are \emph{anyons} in these plaquettes \(P\). We also refer to the ground states as trivial excited states, in which there are only \emph{trivial anyons} in all plaquettes. We assume the sphere topology, in which the model has a unique ground state; nevertheless, the results in this section apply to other topologies.

We start with the simplest excited states with a pair of anyons in two \emph{adjacent} plaquettes with a common edge \(E\). This state can be generated by ribbon operator \(W_E^{J;pq}\):
\eqn{
W_E^{J; pq} \ExcitedA := \sum_{k \in L_\Fus} \sqrt{\frac{d_k}{d_j}} \ \overline{z_{pqj}^{J;k}} \ \ \ExcitedB \ ,
}
where \(j\) is the label on edge \(E\), and \(\bar{z}\) is the complex conjugate. Here, the four-indexed tensor \(z_{pqj}^{J; k}\) is called the \emph{half-braiding tensor}, defined by the following equation:
\eq{
\frac{\delta_{jt}N_{rs}^t}{d_t} z_{pqt}^{J;w} = \sum_{u,l,v\in L_\Fus} z_{lqr}^{J;v} z_{pls}^{J;u} \cdot d_u d_v G^{r^*s^*t}_{p^*wu^*} G^{srj^*}_{qw^*v} G^{s^*ul^*}_{rv^*w}.
}
The label \(J\), called the \emph{anyon species}, labels different minimal solutions of the \(z\) tensor that cannot be the sum of any other nonzero solutions. The ribbon operator \(W_E^{J;pq}\) creates in the two adjacent plaquettes a pair of anyons \(J^\ast\) and \(J\) with charges \(p^\ast\) and \(q\). An anyon species \(J\) may have multiple possible charges \(p\), causing multiple excited states of the string-net model to represent the same anyon. Categorically, anyon species \(J\) are labeled by simple objects in the \emph{center} of the input fusion category \(\Fus\), a modular tensor category whose categorical data record all topological properties of the topological order that the string-net model describes:
\eq{
J \in L_{\mathcal{Z}(\Fus)}\ .
}

States with two quasiparticles in two non-adjacent plaquettes are generated by ribbon operators along longer paths. These ribbon operators result from concatenating shorter ribbon operators. For example, to create two quasiparticles \(J^\ast\) and \(J\) with charges \(p^\ast_0\) and \(p_n\) in two non-adjacent plaquettes \(P_0\) and \(P_n\), we can choose a sequence of plaquettes \((P_0, P_1, \cdots, P_n)\), where \(P_i\) and \(P_{i+1}\) are adjacent plaquettes with their common edge \(E_i\). The ribbon operator \(W_{P_0P_n}^{J; p_0p_n}\) is
\eq{
W_{P_0P_n}^{J; p_0p_n} := \left[\sum_{p_1 p_2 \cdots p_{n-1} \in L_\Fus} \prod_{k = 1}^{n-1} \left(d_{p_k} B_{P_k} W_{E_k}^{J; p_k p_{k+1}}\right)\right] W_{E_0}^{J; p_0 p_1}.
}
Different choices of plaquette paths \((P_0, P_1, \cdots, P_n)\) give the same operator \(W_{P_0P_n}^{J; p_0 p_n}\) if these sequences can deform continuously from one to another. Following the same method, we can also define the creation operator of three or more anyons.

At the end of this section, we define the measurement operator \(\Pi_P^J\) measuring whether there is an anyon \(J\) excited in plaquette \(P\):
\eqn{\Pi_P^J\ \PlaquetteSrc\qquad
:=\qquad  \sum_{s, t\in L_\Fus}\ \frac{d_sd_t}{d_p}z_{pps}^{J; t}\quad \PlaquetteMsr\quad.}
The set of measurement operators are orthonormal and complete:
\eq{\Pi_P^J\Pi_P^K = \delta_{JK}\Pi_P^J,\qquad \sum_{J\in L_{\mathcal{Z}(\Fus)}}\Pi_P^J = \idm.}

\section{Frobenius Algebras and Bimodules}\label{sec:theory}

It is a mathematical theorem \cite{etingof2016} that two fusion categories \(\Fus\) and \(\Fus'\) have isomorphic centers if and only if they are \emph{categorically Morita equivalent}. That is, two string-net models with categorically Morita equivalent input fusion categories describe the same topological order. Category theory also tells that if a fusion category \(\Fus'\) is categorically Morita equivalent to \(\Fus\), there must be a \emph{Frobenius algebra} \(\A\) in \(\Fus\), such that \(\Fus'\) is isomorphic to the \emph{bimodule category} over \(\A\) in \(\Fus\):
\eqn{
\Fus' \cong \Bimod_\Fus(\A).
}
Therefore, different string-net models describing the same topological order are classified by all Frobenius algebras \(\A\) in a particular input fusion category \(\Fus\). Such equivalent models have bimodule categories \(\Bimod_\Fus(\A)\) as their input fusion categories. We can establish the duality maps between these equivalent models. In this section, we briefly review the definition of Frobenius algebras in a given fusion category and their bimodules and leave the duality maps for the next section.

\subsection{Frobenius Algebra}\label{sec:frob}

A Frobenius algebra \(\A\) in a fusion category \(\Fus\) is characterized by a pair of functions \((n, f)\).  Function \(n: L_\Fus \to \NN\) returns the \emph{multiplicity} \(n_a\) of \(a\in L_\Fus\) appearing in the Frobenius algebra \(\A\), satisfying \(n_1 = 1\) and \(n_a = n_{a^\ast}\). The simple objects of \(\A\) are labeled by pairs \(a_\alpha\), where \(a \in L_\Fus\) satisfies \(n_a > 0\), and \(\alpha = 1, 2, \ldots, n_a\) is the \emph{multiplicity index}. We denote the set of all simple objects in \(\A\) as \(L_\A\). 

The algebraic multiplication of \(\A\) is given by a function \(f: L^3_\A \to \C\), satisfying:
\eqn[eq:frob]{\sum_{t_\tau \in L_\A} f_{r_\rho s_\sigma t_\tau} f_{a_\alpha b_\beta t_\tau^\ast} G^{rst}_{abc} \sqrt{d_c d_t} &= \sum_{\gamma = 1}^{n_c} f_{a_\alpha c_\gamma s_\sigma} f_{r_\rho c_\gamma^\ast b_\beta}\ ,\\ \\
\sum_{a_\alpha b_\beta \in L_\A} f_{a_\alpha b_\beta c_\gamma} f_{b_\beta^\ast a_\alpha^\ast c_\gamma^\ast} \sqrt{d_a d_b} = d_\A \sqrt{d_c},\qquad f_{a_\alpha b_\beta c_\gamma} &= f_{b_\beta c_\gamma a_\alpha}, \qquad f_{0 a_\alpha b_\beta} = \delta_{ab^\ast} \delta_{\alpha\beta},}
where
\eqn{
d_\A := \sum_{a \in L_\Fus} n_a d_a
}
is the \emph{quantum dimension} of the Frobenius algebra \(\A\). This definition aligns with the one in the main body, where a Frobenius algebra \(\A\) is expressed as a vector space spanned by simple objects, and the algebraic multiplicity rule is given by function \(f\):
\eq{
\A = \C[L_\A],\qquad a_\alpha b_\beta = \sum_{c_\gamma \in L_\A} f_{a_\alpha b_\beta c_\gamma^\ast} \, c_\gamma \in \C[L_\A].
}

For convenience, in a lattice model, we use red edges or tails to indicate that this edge or tail is labeled by a simple object in Frobenius algebra \(\A\), and a red dot on a vertex to represent a coefficient \(f\) multiplied to this state.
\eqn{
\FrobeniusA\ .
}
We also use dashed red edges or tails to represent that we are summing over all states with labels on this edge in \(L_\A\). The definition \eqref{eq:frob} of Frobenius algebra \(\A\) can then be illustrated graphically by the Pachner moves.
\eq{\FrobeniusB\ ,}
\eq{\FrobeniusC\ .}

\subsection{Bimodules over a Frobenius Algebra}\label{sec:bimod}

A bimodule \(M\) over a Frobenius algebra \(\A\) in a fusion category \(\Fus\) is characterized by a pair of functions \((n^M, P_M)\). The function \(n^M: L_\Fus \to \NN\) returns the \emph{multiplicity} \(n^M_a\) of \(a \in L_\Fus\) appearing in bimodule \(M\), satisfying \(n^M_a = n^M_{a^\ast}\). The simple objects of \(M\) are labeled by pairs \(a_i\), where \(a \in L_\Fus\) satisfies \(n_a^M > 0\), and \(i = 1, 2, \ldots, n^M_a\) labels the multiplicity index. We denote the set of all simple objects in bimodule \(M\) as \(L_M\). 

The action of Frobenius algebra \(\A\) on bimodule \(M\) is characterized by function \(P_M: L_\A^2 \times L_M \times L_\Fus \times L_M \to \C\), satisfying the following defining equations:
\eqn[eq:bimod]{&\sum_{uv\in L_\Fus}\ \sum_{y_\upsilon\in L_M}\ [P_M]^{a_\alpha r_\rho}_{x_\chi u y_\upsilon}\ [P_M]^{b_\beta s_\sigma}_{y_\upsilon v z_\zeta}\ G^{v^\ast by}_{urw}\ G^{w^\ast bu}_{axc}\ G^{sz^\ast v}_{wrt^\ast}\ \sqrt{d_ud_vd_wd_yd_cd_t}\\ 
=\ &\sum_{\gamma = 1}^{n_c}\ \sum_{\tau = 1}^{n_t}\ P^{c_\gamma t_\tau}_{x_\chi w z_\zeta}f_{a_\alpha c_\gamma^\ast b_\beta}\ f_{r_\rho s_\sigma t_\tau},\\
&\qquad\qquad [P_M]^{00}_{x_\chi y z_\zeta} = \delta_{xy}\delta_{yz}\delta_{\chi\upsilon}\delta_{\upsilon\zeta},\qquad [P_M]^{a_\alpha b_\beta}_{x_\chi y z_\zeta} = [P_M]^{b_\beta a_\alpha}_{z_\zeta^\ast y^\ast x_\chi^\ast}.
}
This definition aligns with the one in the main body, where a bimodule \(M\) is expressed as a vector space spanned by simple objects \(M = \C[L_M]\). A pair of Frobenius algebra elements \((a_\alpha, b_\beta)\in \C[L_\A]^2\) is represented as a three-index tensor \(P_M\) on the bimodule space \(\C[L_M]\).

For convenience, in a lattice model, we use a blue line to indicate that this line is labeled by a simple object in bimodule \(M\) and a wavy blue line to represent summing over all states with labels on this edge in \(L_\Fus\) with coefficient \(P_M\):
\eqn[eq:bimoddef]{
\BimoduleA\ .
}
The definition \eqref{eq:bimod} of bimodule \(M\) can then be depicted graphically by Pachner moves:
\eq{
\BimoduleB\ .
}

\subsection{The Bimodule Fusion Category over a Frobenius Algebra}\label{sec:bimodcat}

The set of all bimodules over a given Frobenius algebra \(\A\) in a fusion category \(\Fus\) forms a fusion category, denoted as \(\Bimod_\Fus(\A)\). In this section, we briefly introduce the categorical data of the fusion category \(\Bimod_\Fus(\A)\).

\begin{enumerate}
\item A bimodule \(M\) in \(\Bimod_\Fus(\A)\) is \emph{simple} if it cannot be written as a direct sum of two other bimodules. That is, we cannot find two bimodules \(M_1\) and \(M_2\) such that:
\eq{
n^M_a = n^{M_1}_a + n^{M_2}_a,\quad [P_M]^{a_\alpha b_\beta}_{x_\chi y z_\zeta} = \begin{cases}[P_{M_1}]^{a_\alpha b_\beta}_{x_\chi y z_\zeta},\qquad\qquad\qquad (\chi \le n^{M_1}_x, \zeta \le n^{M_1}_z),\\ \\
[P_{M_2}]^{a_\alpha b_\beta}_{x_{(\chi - n^{M_1}_x)} y z_{(\zeta - n^{M_2}_z)}},\quad (\chi > n^{M_1}_x, \zeta > n^{M_1}_z),\\ \\ 0.\qquad\qquad\qquad\qquad\qquad (\rm otherwise).
\end{cases}
}

\item The quantum dimension of a bimodule \(M\) in \(\Bimod_\Fus(\A)\) is
\eqn{
d_M = \frac{1}{d_\A}\sum_{a\in L_\Fus}n^M_ad_a.
}

\item For any three bimodules \(M_1\), \(M_2\), and \(M_3\), we can represent their fusion rules in terms of their simple objects. Define the matrix \([\Delta_{M_1 M_2 M_3}]\) that represents how the basis elements in the bimodule spaces are connected when the three bimodules fuse:
\eqn{\ [\Delta_{M_1M_2M_3}]_{r_\rho s_\sigma t_\tau}^{x_\chi y_\upsilon z_\zeta} := &\frac{1}{d_\A^3}\sum_{a_\alpha b_\beta c_\gamma\in L_\A}\sum_{p\in L_\Fus}\sum_{u_\rho\in L_{M_1}}\sum_{v_\sigma\in L_{M_2}}\sum_{w_\lambda\in L_{M_3}}\ [P_1]^{b_\beta c_\gamma^\ast}_{x_\chi u r_\rho}\ [P_2]^{c_\gamma a_\alpha^\ast}_{y_\upsilon v s_\sigma}\ \times\\
&\ [P_3]^{a_\alpha b_\beta^\ast}_{z_\zeta w t_\tau}\ G^{bxu^\ast}_{c^\ast r^\ast p}\ G^{swp}_{br^\ast t^\ast}\ G^{pvz}_{aw^\ast s^\ast}\ G^{xyz}_{vpc}\ \sqrt{d_ud_vd_wd_ad_bd_cd_rd_sd_t}\ d_p\ .}
This definition can be depicted graphically:
\eq{
\BimoduleC\ .
}
The fusion rule of three bimodules \(M_1, M_2, M_3\) is
\eqn{\delta_{M_1M_2M_3} = {\rm Tr}[\Delta_{M_1M_2M_3}] = \sum_{r_\rho, s_\sigma, t_\tau\in L_M}[\Delta_{M_1M_2M_3}]_{r_\rho s_\sigma t_\tau}^{r_\rho s_\sigma t_\tau}.}
Here, the three indices \(r_\rho, s_\sigma, t_\tau\) of the matrix in the superscripts or subscripts should be understood as a pair, labeling the fusion vertex. In this paper, we focus on the case in which the fusion coefficients \(N_{M_1 M_2}^{M_3}\) in the bimodule category \(\Bimod_\Fus(\A)\) can only be \(0\) or \(1\), ensuring that \(\delta_{M_1 M_2 M_3}\) is well-defined.
\item The Frobenius algebra \(\A\) itself is the trivial bimodule \(M_0\) over \(\A\):
\eqn{
L_{M_0} = L_\A,\qquad [P_{M_0}]^{a_\alpha b_\beta}_{x_\chi y z_\zeta} = \sum_{\upsilon = 1}^{n_y}f_{a_\alpha x_\chi y^\ast_\upsilon} f_{y_\upsilon b_\beta z_\zeta^\ast}.
}
Given a bimodule \(M\), its opposite bimodule \(M^\ast\) is
\eqn{
L_{M^\ast} = L_M,\qquad [P_{M^\ast}]^{a_\alpha b_\beta}_{x_\chi yz_\zeta} = ([P_M]^{a_\alpha b_\beta}_{x_\chi yz_\zeta})^\ast.
}
\item The bimodule conditions \eqref{eq:bimod} induce that matrix \(\Delta_{M_1M_2M_3}\) is a projector:
\eq{
\Delta_{M_1M_2M_3}^2 = \Delta_{M_1M_2M_3}.
}
If \(\delta_{M_1M_2M_3}\ne 0\), we can find the normalized eigenvectors \(\V_{M_1M_2M_3}^{r_\rho s_\sigma t_\tau}\in\C\), such that
\eqn{
\sum_{r_\rho\in L_{M_1}}\sum_{s_\sigma\in L_{M_2}}\sum_{t_\tau\in L_{M_3}}[\Delta_{M_1M_2M_3}]^{x_\chi y_\upsilon z_\zeta}_{r_\rho s_\sigma t_\tau}\ \V_{M_1M_2M_3}^{r_\rho s_\sigma t_\tau} = \V_{M_1M_2M_3}^{x_\chi y_\upsilon z_\zeta},\\
\sum_{x_\chi\in L_{M_1}}\sum_{y_\upsilon\in L_{M_2}}\sum_{z_\zeta\in L_{M_3}}|\V_{M_1M_2M_3}^{x_\chi y_\upsilon z_\zeta}|^2\ \sqrt{d_xd_yd_z} = d_\A^2\sqrt{d_{M_1}d_{M_2}d_{M_3}}\ .}

For convenience, in a lattice model, we use blue lines labeled by a bimodule \(M\) to represent summing over all states with labels in \(L_M\) on this line. A vertex state with a blue dot at the certex is a superposition with coefficients $\mathcal{V}$:
\eqn[eq:vertexdef]{
\BimoduleD\ .
}
Note that two basis states in the RHS of Eq. \eqref{eq:vertexdef} for the same simple object labels $x, y, z$ but different multiplicity indices $\chi, \upsilon, \zeta$ are regarded orthogonal states. Such a state is invariant under \(\mathcal{D}_{M_1M_2M_3}\) matrix. The \(6j\)-symbol of \(\Bimod_\Fus(\A)\) category is
\eqn{
\BimoduleE\ .
}

\end{enumerate}

\section{General Constructions of Dualities and Symmetry Transformations in the Extended String-Net Model}\label{sec:trans}

Given a fusion category \(\Fus\) and a Frobenius algebra \(\A \in \Fus\), two string-net models with \(\Fus\) and \(\Bimod_\Fus(\A)\) as the input data describe the same topological order. Categorically, \(\Bimod_\Fus(\A)\) is defined by an injective functor
\eqn{
\D: \Bimod_\Fus(\A) \to \Fus, \qquad M \mapsto \bigoplus_{a \in L_\Fus} n^M_a a,
}
and for any morphisms \(\phi_{M_1 M_2}^{M_3} \in \Bimod_\Fus(\A): M_1 \otimes M_2 \to M_3\) and \(\varphi_{xy}^z \in \Fus: x \otimes y \to z\),
\eq{
\D(\phi_{M_1 M_2}^{M_3}) = \sum_{z_\zeta \in L_{M_3}} \Bigg[ \sum_{x_\chi \in L_{M_1}} \sum_{y_\upsilon \in L_{M_2}} \mathcal{V}_{M_1 M_2 M_3^\ast}^{x_\chi y_\upsilon z_\zeta^\ast} \ \varphi_{x_\chi y_\upsilon}^{z_\zeta} \Bigg],
}
where \(x_\chi\), \(y_\upsilon\), and \(z_\zeta\) are respectively the \(\chi\)-th \(x\) object,  \(\upsilon\)-th \(y\), and \(\zeta\)-th \(z\) in the direct sum \(\D(M)\). Note that two morphisms $\varphi_{x_\chi y_\upsilon}^{z_\zeta}$ for the same simple objects $x, y, z$ but different multiplicity indices $\chi, \upsilon, \zeta$ are regarded orthogonal morphisms.

Such a functor \(\D\) induces a duality between the two models with \(\F\) and \(\Bimod_\Fus(\A)\) as the input data:
\eqn{
\Edge{M} \Longrightarrow\ \sum_{a_\alpha, b_\beta\in L_\A}\sum_{x_\chi, z_\zeta\in L_M}\BimoduleG\ = \sum_{a_\alpha, b_\beta\in L_\A}\sum_{x_\chi, z_\zeta\in L_M}\sum_{y\in L_\Fus}[P_M]^{a_\alpha b_\beta}_{x_\chi yz_\zeta}\BimoduleX.
}
This duality induces a unitary morphism between the Hilbert spaces \(\Hil_{\Bimod_\Fus(\A)}\) and \(\Hil_{\Fus}\) of these two models. Such a unitary linear transformation can be understood plaquette by plaquette:
\begin{align}
\BimoduleH\ .\label{eq:pladuality}
\end{align}
Note that the black edges and tails labeled by \(M_i, N_i \in \Bimod_\Fus(\A)\) represent basis states in the dual model, where \(\Bimod_\Fus(\A)\) is the input fusion category and \(M_i, N_i\) are simple objects. In contract, the blue edges and tails labeled by \(M_i, N_i \in \Bimod_\Fus(\A)\) represent superposition states in the original model with \(\Fus\) as the input fusion category, where the superpositions are defined in Eqs. \eqref{eq:bimoddef} and \eqref{eq:vertexdef}. In the RHS of Eq. \eqref{eq:pladuality}, each edge and tail of the plaquette is labeled by simple modules' components $x_i, y_i\in L_{M_i}, p_\alpha, q_\beta\in L_M, e_i\in L_{N_i}$, which may carry multiplicity indices. Nevertheless, after pachner moves contracting all extra plaquettes labeled by ``{\color{red} $\times$}'', the multiplicity indices of degrees of freedom on edges are reduced. Only $q_\beta$ on the endpoint of the tail still carry multiplicity indices.

After the above transformation \eqref{eq:pladuality} is applied to all plaquettes, the resulting basis state \(\ket{\psi}\) satisfies
\eqn{
\langle \psi | \psi \rangle = d_\A^{g-2},
}
where \(g\) is the genus of the surface on which the lattice is embedded. By applying the duality map and normalizing the resulting basis states, we obtain a unitary morphism between the two string-net models.

After the topological moves in Eq. \eqref{eq:pladuality}, the degree of freedom on any edge will cease to have any multiplicity index of simple objects in bimodules, while that on any tail will still have a multiplicity index. Therefore, to make sense of this duality and make it unitary, we are urged to enlarge the Hilbert space of the original Fibonacci string-net model on each tail but not on the edges, such that two simple objects \(a_\alpha, a_\beta \in L_M\) with different multiplicity indices \(\alpha\ne\beta\) are distinguishable on tails.

\subsection{Enlarging the Hilbert Space}\label{sec:generalenlarge}

In the enlarged Hilbert space, each tail carries a degree of freedom labeled by a pair \(a_\alpha\), where 
\eqn{a\in L_\mathscr{F},\qquad \alpha = 1, 2, \cdots, N^\mathcal{A}_a,\qquad N^\mathcal{A}_a = \max_{M\in L_{{\rm Bimod}_\mathscr{F}}(\mathcal{A})}\{n^M_a\},
}
where \(L_{\Bimod_\F(\A)}\) is the set of all simple bimodules over Frobenius algebra \(\A\). But the basic degrees of freedom on edges remain to take value varying the simple objects of the input fusion category \(\Fus\). The Hilbert space on the tail is spanned by all enlarged degrees of freedom on tails and original degrees of freedom on edges, subject to the fusion rules on all vertices. 

For any bimodule \(M\), its basis element \(x^M_\chi\in L_M\) corresponds to a superposition state \(\ket{x^M_\chi}\) in the local Hilbert space of a tail:
\eqn[eq:localstatemulti]{
\ket{x^M_\chi} := \sum_{i = 1}^{N^\mathcal{A}_x} A^{x,M}_{\chi,i}\ket{x_i}.
}
The coefficients $A_{\chi, i}^{x, M}$ are determined by solving the orthonormality conditions on the local states in Eq. \eqref{eq:localstatemulti}:
\eqn{\BimoduleF\ .}

\subsection{Duality}\label{sec:dual}

The duality map \eqref{eq:pladuality} can be simplified. We can represent the unitary duality map vertex by vertex:
\eqn{
\D := \frac{1}{d_\A^{N_P - 1 + \frac{g}{2}}}\prod_{{\rm Edge\ }e}E_e\prod_{{\rm Vertex\ }v}\D_v,
}
where \(N_P\) is the number of plaquettes in the lattice, and \(\D_v\) acts on vertex \(v\):
\eqn{
\TransB\ .
}
Note that each edge connects two vertices that are acted upon by two \(\D_v\) operators independently. Nevertheless, an edge \(e\) can only carry one label. We use \(E_e\) to ensure this uniqueness:
\eqn{
\TransF\quad .
}
The \(E_e\) moves erase the multiplicity indices of labels on edges. But the multiplicity labels on tails are retained.

The Hilbert space is not preserved under the duality map:
\eq{
\Hil_{\Fus} \ne \D \Hil_{\Bimod_\Fus(\A)},
}
where \(\Hil_{\Fus}\) and \(\Hil_{\Bimod_\Fus(\A)}\) are the Hilbert spaces of the string-net model with input fusion category \(\Fus\) and \(\Bimod_\Fus(\A)\), respectively, considered as subspaces of the enlarged Hilbert space. Nevertheless, since the two models describe the same topological order, the ground-state subspace \(\Hil_0\) is preserved under the duality map:
\eqn{
\Hil_{0, \Fus} = \Hil_{0, \Bimod_\Fus(\A)},
}
where \(\Hil_{0, \Fus}\) and \(\Hil_{0, \Bimod_\Fus(\A)}\) are the ground-state subspaces of the string-net model with input fusion category \(\Fus\) and \(\Bimod_\Fus(\A)\), respectively.

\subsection{Symmetry Transformation}\label{sec:symm}

In certain cases, \(\Fus\) and \(\Bimod_\Fus(\A)\) are isomorphic fusion category. That is, there exists an isomorphic functor \(\F_\A\) that maps simple objects of \(\Fus\) to simple objects in \(\Bimod_\Fus(\A)\): 
\eqn{
\varphi_\A(a) = M_a\in\Bimod_\Fus(\A).
}
Such isomorphic functor induces a linear isomorphism \(\varphi_\A\) between the Hilbert space of these two models that maps basic degrees of freedom on edges and tails to basic degrees of freedom:
\eqn{
\varphi_\A: \Hil_{\Fus}\to \Hil_{\Bimod_\Fus(\A)},\qquad\qquad\qquad \TransE\ .
}
The composition
\eqn{\G := \D\circ i: \Hil_\Fus\to \Hil_\Fus}
is just a unitary transformation of the same model with \(\Fus\) as the input fusion category, and the symmetry transformation is the composition of the unitary transformation and projection back into the original degrees of freedom. The set of all symmetry transformations in the Hilbert space of the string-net model with \(\Fus\) as the input fusion category forms the symmetry of the topological order.

In particular, consider the trivial Frobenius algebra \(\A_0\):
\eqn{
L_{\A_0} = \{0\},\qquad f_{000} = 1,
}
whose simple bimodules are labeled by simple objects in \(L_\Fus\): 
\eqn{L_{M_a} = \{a\},\qquad P^{00}_{aaa} = 1.}
The gauge transformation induced by Frobenius algebra \(\A_0\) is the identity transformation of the string-net model.

\subsection{Braiding of Bimodules}\label{sec:braid}
The input fusion category \(\Fus\) is a fusion category that lacks a braiding structure for exchanging two simple objects \(a\) and \(b\). Nevertheless, the braiding of the trivial object \(0\) with any other simple object \(a\) always exists as the trivial braiding, which can be represented graphically as:
\eq{
\BraidA \quad = \quad \BraidC\quad = \quad \BraidB
}
The last equality holds because the fusion of \(0\) with any simple object \(a\) is also trivial.

Similarly, in the bimodule category \(\Bimod_\Fus(\A)\), the trivial bimodule \(\A\) braids trivially with any other bimodule \(M \in \Bimod_\Fus(\A)\), based on the definition of bimodules:
\eqn[eq:trivialbraid]{
\BraidD \quad = \quad \BraidF \quad= \quad \frac{1}{d_\A^2} \ \BraidE\quad .
}

As a practical example, consider the situation where we contract a plaquette with a tail labeled by the trivial bimodule \(\A\) in the original model. Since the Pachner moves \eqref{eq:pachner} can only contract plaquettes without nontrivial tails inside them, one must first ``pull'' the trivial tail out of the plaquette and then annihilate the plaquettes, as shown in Fig. \ref{fig:braid}.

\begin{figure}\centering
\BraidG
\caption{Contract a plaquette with a tail labeled by the trivial bimodule \(\A\) in the original model.}
\label{fig:braid}
\end{figure}

\section{Frobenius Algebra of Fibonacci Fusion Categories and Its Simple Bimodules}\label{sec:frobdata}

In this section, we list the categorical data of the Fibonacci fusion category. The Fibonacci fusion category has two simple objects, denoted as $1$ and $\tau$. The nonzero fusion rules are $\delta_{111} = \delta_{1\tau\tau} = \delta_{\tau\tau\tau} = 1$, and the quantum dimensions are 
$$d_1 = 1,\qquad d_\tau = \phi = \frac{\sqrt{5} + 1}{2}.$$
The nonzero $6j$ symbols are
$$G^{111}_{111} = 1,\qquad G^{111}_{\tau\tau\tau} = \frac{1}{\sqrt{\phi}},\qquad G^{1\tau\tau}_{1\tau\tau} = G^{1\tau\tau}_{\tau\tau\tau} = \frac{1}{\phi},\qquad G^{\tau\tau\tau}_{\tau\tau\tau} = -\frac{1}{\phi^2}.$$

The Fibonacci fusion category has a nontrivial Frobenius algebra $\mathcal{A}$, such that
$$L_\mathcal{A} = \{1, \tau\},\qquad f_{111} = f_{1\tau\tau} = f_{\tau 1\tau} = f_{\tau\tau 1} = 1,\qquad f_{\tau\tau\tau} = -\frac{1}{\phi^\frac{3}{4}}.$$
There are two simple bimodules over $\mathcal{A}$, denoted as $M_0$ and $M_1$, such that
$$L_{M_0} = \{1, \tau\},$$
$$[P_{M_0}]^{11}_{111} = [P_{M_0}]^{11}_{\tau\tau\tau} = 1,\qquad [P_{M_0}]^{1\tau}_{11\tau} = [P_{M_0}]^{1\tau}_{\tau\tau 1} = 1,\qquad [P_{M_0}]^{1\tau}_{\tau\tau\tau} = \frac{1}{\phi^\frac{3}{4}},$$
$$[P_{M_0}]^{\tau 1}_{\tau 11} = [P_{M_0}]^{\tau 1}_{1\tau\tau} = 1,\qquad [P_{M_0}]^{\tau 1}_{\tau\tau\tau} = \frac{1}{\phi^\frac{3}{4}},$$
$$[P_{M_0}]^{\tau\tau}_{1\tau1} = [P_{M_0}]^{\tau\tau}_{\tau 1\tau} = 1,\qquad [P_{M_0}]^{\tau\tau}_{1\tau\tau} = [P_{M_0}]^{\tau\tau}_{\tau\tau 1} = \frac{1}{\phi^\frac{3}{4}},\qquad [P_{M_0}]^{\tau\tau}_{\tau\tau\tau} = \frac{1}{\phi^\frac{3}{2}}\ .$$
$$L_{M_1} = \{1, \tau_1, \tau_2\},$$

$$[P_{M_1}]^{11}_{111} = [P_{M_1}]^{11}_{\tau_0\tau\tau_0} = [P_{M_1}]^{11}_{\tau_1\tau\tau_1} = 1\ ,$$

$$[P_{M_1}]^{1\tau}_{11\tau_0} = [P_{M_1}]^{1\tau}_{\tau_1\tau 1} = [P_{M_1}]^{\tau 1}_{\tau_011} = [P_{M_1}]^{\tau 1}_{1\tau\tau_1} = \frac{1}{2\phi} + \frac{\sqrt{\phi}}{2}\ii\ ,$$

$$[P_{M_1}]^{1\tau}_{11\tau_1} = [P_{M_1}]^{1\tau}_{\tau_0\tau 1} = [P_{M_1}]^{\tau 1}_{\tau_111} = [P_{M_1}]^{\tau 1}_{1\tau\tau_0} = \frac{1}{2\phi} - \frac{\sqrt{\phi}}{2}\ii\ ,$$

$$[P_{M_1}]^{1\tau}_{\tau_0\tau\tau_0} = [P_{M_1}]^{1\tau}_{\tau_1\tau\tau_1} = [P_{M_1}]^{\tau 1}_{\tau_0\tau\tau_0} = [P_{M_1}]^{\tau 1}_{\tau_1\tau\tau_1} = -\frac{\sqrt[4]{\phi}}{2\phi^2}\ ,$$

$$[P_{M_1}]^{1\tau}_{\tau_0\tau\tau_1} = [P_{M_1}]^{\tau 1}_{\tau_1\tau\tau_0} = -\frac{\sqrt[4]{\phi}}{2} - \frac{\phi^\frac{3}{4}}{2}\ii\ ,\qquad [P_{M_1}]^{1\tau}_{\tau_1\tau\tau_0} = [P_{M_1}]^{\tau 1}_{\tau_0\tau\tau_1} = -\frac{\sqrt[4]{\phi}}{2} + \frac{\phi^\frac{3}{4}}{2}\ii\ ,$$

$$[P_{M_1}]^{\tau\tau}_{1\tau1} = -\frac{1}{\phi}\ ,\qquad [P_{M_1}]^{\tau\tau}_{1\tau\tau_0} =[P_{M_1}]^{\tau\tau}_{1\tau\tau_1} =[P_{M_1}]^{\tau\tau}_{\tau_0\tau 1} =[P_{M_1}]^{\tau\tau}_{\tau_1\tau 1} = -\frac{\sqrt[4]\phi}{\phi}\ ,$$

$$[P_{M_1}]^{\tau\tau}_{\tau_01\tau_0} = -\frac{1}{2\phi} + \frac{\ii}{2\sqrt{\phi}}\ ,\qquad [P_{M_1}]^{\tau\tau}_{\tau_11\tau_1} = -\frac{1}{2\phi} - \frac{\ii}{2\sqrt{\phi}}\ ,\qquad [P_{M_1}]^{\tau\tau}_{\tau_01\tau_1} = [P_{M_1}]^{\tau\tau}_{\tau_11\tau_0} = \frac{1}{2}\ ,$$

$$[P_{M_1}]^{\tau\tau}_{\tau_0\tau\tau_1} = [P_{M_1}]^{\tau\tau}_{\tau_1\tau\tau_0} = \frac{\sqrt{\phi}}{2\phi^2}\ ,\qquad [P_{M_1}]^{\tau\tau}_{\tau_0\tau\tau_0} = -\frac{\sqrt{\phi}}{2\phi^3} - \frac{\phi}{2}\ii\ ,\qquad [P_{M_1}]^{\tau\tau}_{\tau_1\tau\tau_1} = -\frac{\sqrt{\phi}}{2\phi^3} + \frac{\phi}{2}\ii\ .$$

The nonzero vertex coefficients are
$$\mathcal{V}_{M_0M_0M_0}^{111} = \mathcal{V}_{M_0M_0M_0}^{1\tau\tau} = \mathcal{V}_{M_0M_0M_0}^{\tau 1\tau} = \mathcal{V}_{M_0M_0M_0}^{\tau\tau 1} = 1,\qquad \mathcal{V}_{M_0M_0M_0}^{\tau \tau\tau} = \frac{1}{\phi^\frac{3}{4}},$$

$$\mathcal{V}_{M_0M_1M_1}^{111} = \mathcal{V}_{M_0M_1M_1}^{1\tau_1\tau_1} = \mathcal{V}_{M_0M_1M_1}^{1\tau_2\tau_2} = 1,\qquad \mathcal{V}_{M_0M_1M_1}^{\tau \tau_1\tau_1} = \mathcal{V}_{M_0M_1M_1}^{\tau\tau_2\tau_2} = -\frac{1}{2\phi^{\frac{7}{4}}},$$
$$\mathcal{V}_{M_0M_1M_1}^{\tau 1\tau_1} = \mathcal{V}_{M_0M_1M_1}^{\tau\tau_21} = \frac{1}{2\phi} - \frac{\sqrt{\phi}}{2}i,\qquad \mathcal{V}_{M_0M_1M_1}^{\tau 1\tau_2} = \mathcal{V}_{M_0M_1M_1}^{\tau\tau_11} = \frac{1}{2\phi} + \frac{\sqrt{\phi}}{2}i,\qquad $$
$$\mathcal{V}_{M_0M_1M_1}^{\tau \tau_1\tau_2} = -\frac{\sqrt[4]{\phi}}{2} + \frac{\phi^\frac{3}{4}}2{}i,\qquad \mathcal{V}_{M_0M_1M_1}^{\tau\tau_2\tau_1} = -\frac{\sqrt[4]{\phi}}{2} - \frac{\phi^\frac{3}{4}}{2}i,$$

$$\mathcal{V}_{M_1M_1M_1}^{111} = \frac{1}{\phi^\frac{3}{4}},\qquad \mathcal{V}_{M_1M_1M_1}^{\tau_1\tau_1\tau_1} = \mathcal{V}_{M_1M_1M_1}^{\tau_2\tau_2\tau_2} = -\frac{1}{2}\sqrt{\frac{5}{\phi}},$$
$$\mathcal{V}_{M_1M_1M_1}^{1\tau_1\tau_1} = \mathcal{V}_{M_1M_1M_1}^{\tau_11\tau_1} = \mathcal{V}_{M_1M_1M_1}^{\tau_1\tau_11} = \mathcal{V}_{M_1M_1M_1}^{1\tau_2\tau_2} = \mathcal{V}_{M_1M_1M_1}^{\tau_21\tau_2} = \mathcal{V}_{M_1M_1M_1}^{\tau_2\tau_21} = -\frac{1}{2\phi^\frac{7}{4}},$$
$$\mathcal{V}_{M_1M_1M_1}^{\tau_1\tau_1\tau_2} = \mathcal{V}_{M_1M_1M_1}^{\tau_1\tau_2\tau_1} = \mathcal{V}_{M_1M_1M_1}^{\tau_2\tau_1\tau_1} = \mathcal{V}_{M_1M_1M_1}^{\tau_1\tau_2\tau_2} = \mathcal{V}_{M_1M_1M_1}^{\tau_2\tau_1\tau_2} = \mathcal{V}_{M_1M_1M_1}^{\tau_2\tau_2\tau_1} = \frac{1}{2sqrt{\phi}},$$
$$\mathcal{V}_{M_1M_1M_1}^{1\tau_1\tau_2} = \mathcal{V}_{M_1M_1M_1}^{\tau_11\tau_2} = \mathcal{V}_{M_1M_1M_1}^{\tau_1\tau_21} = -\frac{\sqrt[4]{\phi}}{2} - \frac{\phi^\frac{3}{4}}{2}i,$$
$$\mathcal{V}_{M_1M_1M_1}^{1\tau_2\tau_1} = \mathcal{V}_{M_1M_1M_1}^{\tau_2\tau_11} = \mathcal{V}_{M_1M_1M_1}^{\tau_11\tau_2} = -\frac{\sqrt[4]{\phi}}{2} + \frac{\phi^\frac{3}{4}}{2}i.$$

The fusion category ${\rm Bimod}_{\rm Fibo}(\mathcal{A})$ is isomorphic to the Fibonacci fusion category by replacing $1$ as $M_0$, and $\tau$ as $M_1$:
$$\delta_{M_0M_0M_0} = \delta_{M_0M_1M_1} = \delta_{M_1M_1M_1} = 1,\qquad d_{M_0} = 1,\qquad d_{M_1} = \phi,$$
$$G^{M_0M_0M_0}_{M_0M_0M_0} = 1,\quad G^{M_0M_0M_0}_{M_1M_1M_1} = \frac{1}{\sqrt{\phi}},\quad G^{M_0M_1M_1}_{M_0M_1M_1} = G^{M_0M_1M_1}_{M_1M_1M_1} = \frac{1}{\phi},\quad G^{M_1M_1M_1}_{M_1M_1M_1} = -\frac{1}{\phi^2}.$$

Bimodule $M_1$ has multiplicity $n^{M_1}_\tau = 2$, so we have to enlarge the Hilbert spaces on tails from $2$-dimensional local Hilbert spaces spanned by $1$ and $\tau$ to $3$-dimensional local Hilbert spaces spanned by $1$, $\tau_1$ and $\tau_2$. To ensure the orthonormality, the basic state $\ket{\tau}$ labeled by the simple object $\tau$ in the trivial bimodule $M_0$ is a superposition state of two basic state $\ket{\tau_1}, \ket{\tau_2}$ labeled by the simple object $\tau$ in the nontrivial bimodule $M_1$:
\eq{
\Tail{1}\quad \Longrightarrow\quad \Tail{1}\quad,\qquad\qquad \Tail{\tau} \quad\Longrightarrow\quad \Bigg(\frac{1}{2\phi} + \frac{\sqrt{\phi}}{2} \Bigg)\quad\Tail{\tau_1}\quad + \quad \Bigg( \frac{1}{2\phi} - \frac{\sqrt{\phi}}{2} \Bigg)\quad \Tail{\tau_2}
}
This $\tau$ charge on a tail is unique up to exchanging $\tau_1$ and $\tau_2$ labels.

\bibliographystyle{apsrev4-1}
\bibliography{StringNet}
\end{document}

%% file: input.tex
\newcommand{\Lattice}{
\begin{tikzpicture}[baseline={([yshift=-.8ex]current bounding box.center)},scale=1]
\draw [decorate,decoration={snake,segment length=.12cm, amplitude=.04cm}] (1.7,6.8) -- (2.3,6.8);
\draw [] (1.7,6.4) -- (1.7,6.8);
\draw [] (1.7,6.8) -- (1.7,7.2);
\draw [] (1.7,7.2) -- (2.4,7.6);
\draw [] (2.4,7.6) -- (3.1,7.2);
\draw [] (2.4,6.) -- (1.7,6.4);
\draw [] (3.1,6.4) -- (2.4,6.);
\draw [] (2.4,5.2) -- (2.4,5.6);
\draw [] (2.4,5.6) -- (2.4,6.);
\draw [decorate,decoration={snake,segment length=.12cm, amplitude=.04cm}] (2.4,5.6) -- (3.,5.6);
\draw [decorate,decoration={snake,segment length=.12cm, amplitude=.04cm}] (3.1,6.8) -- (3.7,6.8);
\draw [] (3.1,6.4) -- (3.1,6.8);
\draw [] (3.1,6.8) -- (3.1,7.2);
\draw [] (3.1,7.2) -- (3.8,7.6);
\draw [] (3.8,7.6) -- (4.5,7.2);
\draw [] (3.8,6.) -- (3.1,6.4);
\draw [] (4.5,6.4) -- (3.8,6.);
\draw [] (3.8,5.2) -- (3.8,5.6);
\draw [] (3.8,5.6) -- (3.8,6.);
\draw [decorate,decoration={snake,segment length=.12cm, amplitude=.04cm}] (3.8,5.6) -- (4.4,5.6);
\draw [decorate,decoration={snake,segment length=.12cm, amplitude=.04cm}] (4.5,6.8) -- (5.1,6.8);
\draw [] (4.5,6.4) -- (4.5,6.8);
\draw [] (4.5,6.8) -- (4.5,7.2);
\draw [] (4.5,7.2) -- (5.2,7.6);
\draw [] (5.2,7.6) -- (5.9,7.2);
\draw [] (5.2,6.) -- (4.5,6.4);
\draw [] (5.9,6.4) -- (5.2,6.);
\draw [] (5.2,5.2) -- (5.2,5.6);
\draw [] (5.2,5.6) -- (5.2,6.);
\draw [decorate,decoration={snake,segment length=.12cm, amplitude=.04cm}] (5.2,5.6) -- (5.8,5.6);
\draw [decorate,decoration={snake,segment length=.12cm, amplitude=.04cm}] (5.9,6.8) -- (6.5,6.8);
\draw [] (5.9,6.4) -- (5.9,6.8);
\draw [] (5.9,6.8) -- (5.9,7.2);
\draw [] (5.9,7.2) -- (6.6,7.6);
\draw [] (6.6,7.6) -- (7.3,7.2);
\draw [] (6.6,6.) -- (5.9,6.4);
\draw [] (7.3,6.4) -- (6.6,6.);
\draw [] (6.6,5.2) -- (6.6,5.6);
\draw [] (6.6,5.6) -- (6.6,6.);
\draw [decorate,decoration={snake,segment length=.12cm, amplitude=.04cm}] (6.6,5.6) -- (7.2,5.6);
\draw [decorate,decoration={snake,segment length=.12cm, amplitude=.04cm}] (1.7,4.4) -- (2.3,4.4);
\draw [] (1.7,4.) -- (1.7,4.4);
\draw [] (1.7,4.4) -- (1.7,4.8);
\draw [] (1.7,4.8) -- (2.4,5.2);
\draw [] (2.4,5.2) -- (3.1,4.8);
\draw [] (2.4,3.6) -- (1.7,4.);
\draw [] (3.1,4.) -- (2.4,3.6);
\draw [] (2.4,3.2) -- (2.4,3.6);
\draw [decorate,decoration={snake,segment length=.12cm, amplitude=.04cm}] (3.1,4.4) -- (3.7,4.4);
\draw [] (3.1,4.) -- (3.1,4.4);
\draw [] (3.1,4.4) -- (3.1,4.8);
\draw [] (3.1,4.8) -- (3.8,5.2);
\draw [] (3.8,5.2) -- (4.5,4.8);
\draw [] (3.8,3.6) -- (3.1,4.);
\draw [] (4.5,4.) -- (3.8,3.6);
\draw [] (3.8,3.2) -- (3.8,3.6);
\draw [decorate,decoration={snake,segment length=.12cm, amplitude=.04cm}] (4.5,4.4) -- (5.1,4.4);
\draw [] (4.5,4.) -- (4.5,4.4);
\draw [] (4.5,4.4) -- (4.5,4.8);
\draw [] (4.5,4.8) -- (5.2,5.2);
\draw [] (5.2,5.2) -- (5.9,4.8);
\draw [] (5.2,3.6) -- (4.5,4.);
\draw [] (5.9,4.) -- (5.2,3.6);
\draw [] (5.2,3.2) -- (5.2,3.6);
\draw [decorate,decoration={snake,segment length=.12cm, amplitude=.04cm}] (5.9,4.4) -- (6.5,4.4);
\draw [] (5.9,4.) -- (5.9,4.4);
\draw [] (5.9,4.4) -- (5.9,4.8);
\draw [] (5.9,4.8) -- (6.6,5.2);
\draw [] (6.6,5.2) -- (7.3,4.8);
\draw [] (6.6,3.6) -- (5.9,4.);
\draw [] (7.3,4.) -- (6.6,3.6);
\draw [] (6.6,3.2) -- (6.6,3.6);
\draw [decorate,decoration={snake,segment length=.12cm, amplitude=.04cm}] (7.3,6.8) -- (7.9,6.8);
\draw [] (7.3,6.4) -- (7.3,6.8);
\draw [] (7.3,6.8) -- (7.3,7.2);
\draw [] (7.3,7.2) -- (7.6,7.4);
\draw [] (8.,6.) -- (7.3,6.4);
\draw [] (8.3,6.2) -- (8.,6.);
\draw [] (8.,5.2) -- (8.,5.6);
\draw [] (8.,5.6) -- (8.,6.);
\draw [decorate,decoration={snake,segment length=.12cm, amplitude=.04cm}] (8.,5.6) -- (8.6,5.6);
\draw [decorate,decoration={snake,segment length=.12cm, amplitude=.04cm}] (7.3,4.4) -- (7.9,4.4);
\draw [] (7.3,4.) -- (7.3,4.4);
\draw [] (7.3,4.4) -- (7.3,4.8);
\draw [] (7.3,4.8) -- (8.,5.2);
\draw [] (8.,5.2) -- (8.3,5.);
\draw [] (8.,3.6) -- (7.3,4.);
\draw [] (8.,3.2) -- (8.,3.6);
\draw [] (6.6,7.6) -- (6.6,8.);
\draw [] (5.2,7.6) -- (5.2,8.);
\draw [] (3.8,7.6) -- (3.8,8.);
\draw [] (2.4,7.6) -- (2.4,8.);
\draw [] (8.3,3.8) -- (8.,3.6);
\draw [] (1.7,4.) -- (1.3,3.8);
\draw [] (1.7,4.8) -- (1.3,5.);
\draw [] (1.7,6.4) -- (1.3,6.2);
\draw [] (1.7,7.2) -- (1.3,7.4);
\end{tikzpicture}
}

\newcommand{\FibonacciA}{
\begin{tikzpicture}[baseline={([yshift=-.8ex]current bounding box.center)},scale=.45]

\draw [] (3.2,5.6) -- (3.2,6.8);
\draw [] (3.2,7.2) -- (4.6,8.);
\draw [] (4.6,8.) -- (6.,7.2);
\draw [] (6.,5.6) -- (4.6,4.8);
\draw [] (4.6,4.8) -- (3.2,5.6);
\draw [] (6.,7.2) -- (6.,5.6);
\draw [decorate,decoration={snake,segment length=.12cm, amplitude=.04cm}] (3.2,6.4) -- (4.4,6.4);
\draw [] (2.6,7.6) -- (3.2,7.2);
\draw [] (4.6,8.8) -- (4.6,8.);
\draw [] (6.6,7.6) -- (6.,7.2);
\draw [] (6.6,5.2) -- (6.,5.6);
\draw [] (4.6,4.) -- (4.6,4.8);
\draw [] (2.6,5.2) -- (3.2,5.6);
\draw [] (3.2,6.6) -- (3.2,7.2);
\node [centered, rotate=0.] at (2.3,7.8) {\scriptsize $e_1$};
\node [centered, rotate=0.] at (5.,8.8) {\scriptsize $e_2$};
\node [centered, rotate=0.] at (6.9,7.8) {\scriptsize $e_3$};
\node [centered, rotate=0.] at (6.9,5.) {\scriptsize $e_4$};
\node [centered, rotate=0.] at (5.,4.) {\scriptsize $e_5$};
\node [centered, rotate=0.] at (2.3,5.) {\scriptsize $e_6$};
\node [centered, rotate=0.] at (2.85,6.) {\scriptsize $j_0$};
\node [centered, rotate=0.] at (2.85,6.9) {\scriptsize $j_0$};
\node [centered, rotate=0.] at (3.6,7.95) {\scriptsize $j_1$};
\node [centered, rotate=0.] at (5.6,7.95) {\scriptsize $j_2$};
\node [centered, rotate=0.] at (6.4,6.4) {\scriptsize $j_3$};
\node [centered, rotate=0.] at (5.6,4.9) {\scriptsize $j_4$};
\node [centered, rotate=0.] at (3.7,4.9) {\scriptsize $j_5$};
\node [centered, rotate=0.] at (4.7,6.4) {\scriptsize $1$};
\end{tikzpicture}
}

\newcommand{\FibonacciB}{
\begin{tikzpicture}[baseline={([yshift=-.8ex]current bounding box.center)},scale=.45]

\draw [] (3.2,5.6) -- (3.2,6.8);
\draw [] (3.2,7.2) -- (4.6,8.);
\draw [] (4.6,8.) -- (6.,7.2);
\draw [] (6.,5.6) -- (4.6,4.8);
\draw [] (4.6,4.8) -- (3.2,5.6);
\draw [] (6.,7.2) -- (6.,5.6);
\draw [decorate,decoration={snake,segment length=.12cm, amplitude=.04cm}] (3.2,6.4) -- (4.4,6.4);
\draw [] (2.6,7.6) -- (3.2,7.2);
\draw [] (4.6,8.8) -- (4.6,8.);
\draw [] (6.6,7.6) -- (6.,7.2);
\draw [] (6.6,5.2) -- (6.,5.6);
\draw [] (4.6,4.) -- (4.6,4.8);
\draw [] (2.6,5.2) -- (3.2,5.6);
\draw [] (3.2,6.6) -- (3.2,7.2);
\node [centered, rotate=0.] at (2.3,7.8) {\scriptsize $e_1$};
\node [centered, rotate=0.] at (5.,8.8) {\scriptsize $e_2$};
\node [centered, rotate=0.] at (6.9,7.8) {\scriptsize $e_3$};
\node [centered, rotate=0.] at (6.9,5.) {\scriptsize $e_4$};
\node [centered, rotate=0.] at (5.,4.) {\scriptsize $e_5$};
\node [centered, rotate=0.] at (2.3,5.) {\scriptsize $e_6$};
\node [centered, rotate=0.] at (2.85,6.) {\scriptsize $j_6$};
\node [centered, rotate=0.] at (2.85,6.9) {\scriptsize $j_0$};
\node [centered, rotate=0.] at (3.6,7.95) {\scriptsize $j_1$};
\node [centered, rotate=0.] at (5.6,7.95) {\scriptsize $j_2$};
\node [centered, rotate=0.] at (6.4,6.4) {\scriptsize $j_3$};
\node [centered, rotate=0.] at (5.6,4.9) {\scriptsize $j_4$};
\node [centered, rotate=0.] at (3.7,4.9) {\scriptsize $j_5$};
\node [centered, rotate=0.] at (4.7,6.4) {\scriptsize $p$};
\end{tikzpicture}
}

\newcommand{\FibonacciC}{\begin{tikzpicture}[baseline={([yshift=-.8ex]current bounding box.center)},scale=.5]
\draw [] (3.2,5.6) -- (3.2,6.5);
\draw [] (3.2,6.5) -- (3.2,7.2);
\draw [] (3.2,7.2) -- (4.6,8.);
\draw [] (4.6,8.) -- (6.,7.2);
\draw [] (6.,7.2) -- (6.,5.6);
\draw [] (6.,5.6) -- (4.6,4.8);
\draw [] (4.6,4.8) -- (3.2,5.6);
\draw [decorate,decoration={snake,segment length=.12cm, amplitude=.04cm}] (3.2,6.4) -- (4.4,6.4);
\node [centered, rotate=0.] at (2.7,6.) {\scriptsize $I_6$};
\node [centered, rotate=0.] at (2.7,6.9) {\scriptsize $I_0$};
\node [centered, rotate=0.] at (3.6,7.9) {\scriptsize $I_1$};
\node [centered, rotate=0.] at (5.6,7.9) {\scriptsize $I_2$};
\node [centered, rotate=0.] at (6.5,6.4) {\scriptsize $I_3$};
\node [centered, rotate=0.] at (5.6,4.9) {\scriptsize $I_4$};
\node [centered, rotate=0.] at (3.7,4.9) {\scriptsize $I_5$};
\node [centered, rotate=0.] at (4.7,6.4) {\scriptsize $M$};
\draw [] (2.6,7.6) -- (3.2,7.2);
\draw [] (4.6,8.8) -- (4.6,8.);
\draw [] (6.6,7.6) -- (6.,7.2);
\draw [] (6.6,5.2) -- (6.,5.6);
\draw [] (4.6,4.) -- (4.6,4.8);
\draw [] (2.6,5.2) -- (3.2,5.6);
\node [centered, rotate=0.] at (2.2,7.8) {\scriptsize $E_1$};
\node [centered, rotate=0.] at (4.7,9.) {\scriptsize $E_2$};
\node [centered, rotate=0.] at (7.,7.8) {\scriptsize $E_3$};
\node [centered, rotate=0.] at (7.,5.) {\scriptsize $E_4$};
\node [centered, rotate=0.] at (4.7,3.8) {\scriptsize $E_5$};
\node [centered, rotate=0.] at (2.2,5.) {\scriptsize $E_6$};
\end{tikzpicture}\qquad \Longrightarrow \qquad\sum_{i_k, e_k = 1, \tau}\ \ \sum_{p\in L_M}\cdots\begin{tikzpicture}[baseline={([yshift=-.8ex]current bounding box.center)},scale=.5]
\draw [] (3.2,5.6) -- (3.2,6.5);
\draw [] (3.2,6.5) -- (3.2,7.2);
\draw [] (3.2,7.2) -- (4.6,8.);
\draw [] (4.6,8.) -- (6.,7.2);
\draw [] (6.,7.2) -- (6.,5.6);
\draw [] (6.,5.6) -- (4.6,4.8);
\draw [] (4.6,4.8) -- (3.2,5.6);
\draw [decorate,decoration={snake,segment length=.12cm, amplitude=.04cm}] (3.2,6.4) -- (4.4,6.4);
\node [centered, rotate=0.] at (2.8,6.) {\scriptsize $i_6$};
\node [centered, rotate=0.] at (2.8,6.9) {\scriptsize $i_0$};
\node [centered, rotate=0.] at (3.6,7.9) {\scriptsize $i_1$};
\node [centered, rotate=0.] at (5.6,7.9) {\scriptsize $i_2$};
\node [centered, rotate=0.] at (6.4,6.4) {\scriptsize $i_3$};
\node [centered, rotate=0.] at (5.6,4.9) {\scriptsize $i_4$};
\node [centered, rotate=0.] at (3.7,4.9) {\scriptsize $i_5$};
\node [centered, rotate=0.] at (4.7,6.4) {\scriptsize $p$};
\draw [] (2.6,7.6) -- (3.2,7.2);
\draw [] (4.6,8.8) -- (4.6,8.);
\draw [] (6.6,7.6) -- (6.,7.2);
\draw [] (6.6,5.2) -- (6.,5.6);
\draw [] (4.6,4.) -- (4.6,4.8);
\draw [] (2.6,5.2) -- (3.2,5.6);
\node [centered, rotate=0.] at (2.2,7.8) {\scriptsize $e_1$};
\node [centered, rotate=0.] at (4.7,9.) {\scriptsize $e_2$};
\node [centered, rotate=0.] at (7.,7.8) {\scriptsize $e_3$};
\node [centered, rotate=0.] at (7.,5.) {\scriptsize $e_4$};
\node [centered, rotate=0.] at (4.7,3.8) {\scriptsize $e_5$};
\node [centered, rotate=0.] at (2.2,5.) {\scriptsize $e_6$};
\end{tikzpicture}}

\newcommand{\FibonacciD}[1]{
\begin{tikzpicture}[baseline={([yshift=-.8ex]current bounding box.center)},scale=1.]
\draw [] (0, 0) -- (0, 1.5);
\draw [red] (0, .4) -- (-.5, .4);
\draw [red] (0, 1.1) -- (.5, 1.1);

\node [centered, rotate=0.] at (.15,.15) {\scriptsize $#1$};
\node [centered, rotate=0.] at (.15,.75) {\scriptsize $#1$};
\node [centered, rotate=0.] at (.15,1.35) {\scriptsize $#1$};
\node [centered, rotate=0.] at (-.6,.4) {\scriptsize $1$};
\node [centered, rotate=0.] at (.6,1.1) {\scriptsize $1$};
\end{tikzpicture}
}

\newcommand{\FibonacciE}{\begin{tikzpicture}[baseline={([yshift=-.8ex]current bounding box.center)},scale=.5]
    \draw [] (3.2,5.6) -- (3.2,6.5);
    \draw [] (3.2,6.5) -- (3.2,7.2);
    \draw [] (3.2,7.2) -- (4.6,8.);
    \draw [] (4.6,8.) -- (6.,7.2);
    \draw [] (6.,7.2) -- (6.,5.6);
    \draw [] (6.,5.6) -- (4.6,4.8);
    \draw [] (4.6,4.8) -- (3.2,5.6);
    \draw [decorate,decoration={snake,segment length=.12cm, amplitude=.04cm}] (3.2,6.4) -- (4.4,6.4);
    \node [centered, rotate=0.] at (2.7,6.) {\scriptsize $I_6$};
    \node [centered, rotate=0.] at (2.7,6.9) {\scriptsize $I_0$};
    \node [centered, rotate=0.] at (3.6,7.9) {\scriptsize $I_1$};
    \node [centered, rotate=0.] at (5.6,7.9) {\scriptsize $I_2$};
    \node [centered, rotate=0.] at (6.5,6.4) {\scriptsize $I_3$};
    \node [centered, rotate=0.] at (5.6,4.9) {\scriptsize $I_4$};
    \node [centered, rotate=0.] at (3.7,4.9) {\scriptsize $I_5$};
    \node [centered, rotate=0.] at (4.7,6.4) {\scriptsize $M$};
    \draw [] (2.6,7.6) -- (3.2,7.2);
    \draw [] (4.6,8.8) -- (4.6,8.);
    \draw [] (6.6,7.6) -- (6.,7.2);
    \draw [] (6.6,5.2) -- (6.,5.6);
    \draw [] (4.6,4.) -- (4.6,4.8);
    \draw [] (2.6,5.2) -- (3.2,5.6);
    \node [centered, rotate=0.] at (2.2,7.8) {\scriptsize $E_1$};
    \node [centered, rotate=0.] at (4.7,9.) {\scriptsize $E_2$};
    \node [centered, rotate=0.] at (7.,7.8) {\scriptsize $E_3$};
    \node [centered, rotate=0.] at (7.,5.) {\scriptsize $E_4$};
    \node [centered, rotate=0.] at (4.7,3.8) {\scriptsize $E_5$};
    \node [centered, rotate=0.] at (2.2,5.) {\scriptsize $E_6$};
    \end{tikzpicture}\qquad \Longrightarrow \qquad\sum_{i_k, e_k, p = \pm 1}\ \cdots\begin{tikzpicture}[baseline={([yshift=-.8ex]current bounding box.center)},scale=.5]
    \draw [] (3.2,5.6) -- (3.2,6.5);
    \draw [] (3.2,6.5) -- (3.2,7.2);
    \draw [] (3.2,7.2) -- (4.6,8.);
    \draw [] (4.6,8.) -- (6.,7.2);
    \draw [] (6.,7.2) -- (6.,5.6);
    \draw [] (6.,5.6) -- (4.6,4.8);
    \draw [] (4.6,4.8) -- (3.2,5.6);
    \draw [decorate,decoration={snake,segment length=.12cm, amplitude=.04cm}] (3.2,6.4) -- (4.4,6.4);
    \node [centered, rotate=0.] at (2.8,6.) {\scriptsize $i_6$};
    \node [centered, rotate=0.] at (2.8,6.9) {\scriptsize $i_0$};
    \node [centered, rotate=0.] at (3.6,7.9) {\scriptsize $i_1$};
    \node [centered, rotate=0.] at (5.6,7.9) {\scriptsize $i_2$};
    \node [centered, rotate=0.] at (6.4,6.4) {\scriptsize $i_3$};
    \node [centered, rotate=0.] at (5.6,4.9) {\scriptsize $i_4$};
    \node [centered, rotate=0.] at (3.7,4.9) {\scriptsize $i_5$};
    \node [centered, rotate=0.] at (4.7,6.4) {\scriptsize $p$};
    \draw [] (2.6,7.6) -- (3.2,7.2);
    \draw [] (4.6,8.8) -- (4.6,8.);
    \draw [] (6.6,7.6) -- (6.,7.2);
    \draw [] (6.6,5.2) -- (6.,5.6);
    \draw [] (4.6,4.) -- (4.6,4.8);
    \draw [] (2.6,5.2) -- (3.2,5.6);
    \node [centered, rotate=0.] at (2.2,7.8) {\scriptsize $e_1$};
    \node [centered, rotate=0.] at (4.7,9.) {\scriptsize $e_2$};
    \node [centered, rotate=0.] at (7.,7.8) {\scriptsize $e_3$};
    \node [centered, rotate=0.] at (7.,5.) {\scriptsize $e_4$};
    \node [centered, rotate=0.] at (4.7,3.8) {\scriptsize $e_5$};
    \node [centered, rotate=0.] at (2.2,5.) {\scriptsize $e_6$};
    \end{tikzpicture}}

\newcommand{\PlaquetteSrc}{
\begin{tikzpicture}[baseline={([yshift=-.8ex]current bounding box.center)},scale=.5]
\draw [->-] (3.2,5.6) -- (3.2,6.8);
\draw [->-] (3.2,7.2) -- (4.6,8.);
\draw [->-] (4.6,8.) -- (6.,7.2);
\draw [->-] (6.,5.6) -- (4.6,4.8);
\draw [->-] (4.6,4.8) -- (3.2,5.6);
\draw [->-] (6.,7.2) -- (6.,5.6);
\draw [decorate,decoration={snake,segment length=.12cm, amplitude=.04cm}, ->] (3.2,6.4) -- (4.4,6.4);
\draw [->-] (2.6,7.6) -- (3.2,7.2);
\draw [->-] (4.6,8.8) -- (4.6,8.);
\draw [->-] (6.6,7.6) -- (6.,7.2);
\draw [->-] (6.6,5.2) -- (6.,5.6);
\draw [->-] (4.6,4.) -- (4.6,4.8);
\draw [->-] (2.6,5.2) -- (3.2,5.6);
\draw [->-] (3.2,6.6) -- (3.2,7.2);
\node [centered, rotate=0.] at (2.4,7.8) {\scriptsize $e_1$};
\node [centered, rotate=0.] at (4.95,8.8) {\scriptsize $e_2$};
\node [centered, rotate=0.] at (6.9,7.8) {\scriptsize $e_3$};
\node [centered, rotate=0.] at (6.9,5.) {\scriptsize $e_4$};
\node [centered, rotate=0.] at (5.,4.) {\scriptsize $e_5$};
\node [centered, rotate=0.] at (2.3,5.) {\scriptsize $e_6$};
\node [centered, rotate=0.] at (2.85,6.) {\scriptsize $i_7$};
\node [centered, rotate=0.] at (2.85,6.9) {\scriptsize $i_1$};
\node [centered, rotate=0.] at (3.6,7.9) {\scriptsize $i_2$};
\node [centered, rotate=0.] at (5.6,7.9) {\scriptsize $i_3$};
\node [centered, rotate=0.] at (6.4,6.4) {\scriptsize $i_4$};
\node [centered, rotate=0.] at (5.6,4.9) {\scriptsize $i_5$};
\node [centered, rotate=0.] at (3.7,4.9) {\scriptsize $i_6$};
\node [centered, rotate=0.] at (4.7,6.4) {\scriptsize $p$};
\end{tikzpicture}
}

\newcommand{\PlaquetteTar}{
\begin{tikzpicture}[baseline={([yshift=-.8ex]current bounding box.center)},scale=.5]
\draw [->-] (3.2,5.6) -- (3.2,6.8);
\draw [->-] (3.2,7.2) -- (4.6,8.);
\draw [->-] (4.6,8.) -- (6.,7.2);
\draw [->-] (6.,5.6) -- (4.6,4.8);
\draw [->-] (4.6,4.8) -- (3.2,5.6);
\draw [->-] (6.,7.2) -- (6.,5.6);
\draw [decorate,decoration={snake,segment length=.12cm, amplitude=.04cm}] (3.2,6.4) -- (4.4,6.4);
\draw [->-] (2.6,7.6) -- (3.2,7.2);
\draw [->-] (4.6,8.8) -- (4.6,8.);
\draw [->-] (6.6,7.6) -- (6.,7.2);
\draw [->-] (6.6,5.2) -- (6.,5.6);
\draw [->-] (4.6,4.) -- (4.6,4.8);
\draw [->-] (2.6,5.2) -- (3.2,5.6);
\draw [->-] (3.2,6.6) -- (3.2,7.2);
\node [centered, rotate=0.] at (2.4,7.8) {\scriptsize $e_1$};
\node [centered, rotate=0.] at (4.95,8.8) {\scriptsize $e_2$};
\node [centered, rotate=0.] at (6.9,7.8) {\scriptsize $e_3$};
\node [centered, rotate=0.] at (6.9,5.) {\scriptsize $e_4$};
\node [centered, rotate=0.] at (5.,4.) {\scriptsize $e_5$};
\node [centered, rotate=0.] at (2.3,5.) {\scriptsize $e_6$};
\node [centered, rotate=0.] at (2.85,6.) {\scriptsize $j_7$};
\node [centered, rotate=0.] at (2.85,6.9) {\scriptsize $j_1$};
\node [centered, rotate=0.] at (3.6,7.9) {\scriptsize $j_2$};
\node [centered, rotate=0.] at (5.6,7.9) {\scriptsize $j_3$};
\node [centered, rotate=0.] at (6.4,6.4) {\scriptsize $j_4$};
\node [centered, rotate=0.] at (5.6,4.9) {\scriptsize $j_5$};
\node [centered, rotate=0.] at (3.7,4.9) {\scriptsize $j_6$};
\node [centered, rotate=0.] at (4.7,6.4) {\scriptsize $1$};
\end{tikzpicture}
}

\newcommand{\PachnerOne}{\begin{tikzpicture}[baseline={([yshift=-.8ex]current bounding box.center)},scale=.6]
\draw [->-] (0.,0.2) -- (.5,1.);
\draw [->-] (0.,1.8) -- (.5,1.);
\draw [->-] (2.,1.) -- (.5,1.);
\draw [->-] (2.5,1.8) -- (2.,1.);
\draw [->-] (2.5,0.2) -- (2.,1.);
\node at (1.25, 1.3) {\scriptsize $m$};
\node at (-.15, 2.) {\scriptsize $a$};
\node at (-.15, 0.) {\scriptsize $b$};
\node at (2.65, 2.) {\scriptsize $d$};
\node at (2.65, 0.) {\scriptsize $c$};
\end{tikzpicture}\qquad = \qquad \sum_{n\in L_{\Fus}}\ \sqrt{d_md_n}\ G^{abm}_{cdn}\quad \begin{tikzpicture}[baseline={([yshift=-.8ex]current bounding box.center)},scale=.6]
\draw [->-] (0.,0.) -- (.8,.5);
\draw [->-] (1.6,0.) -- (.8,.5);
\draw [->-] (.8,.5) -- (.8,2.);
\draw [->-] (0.,2.5) -- (.8,2.);
\draw [->-] (1.6,2.5) -- (.8,2.);
\node at (1.05, 1.3) {\scriptsize $n$};
\node at (-.15, 2.6) {\scriptsize $a$};
\node at (-.15, -.1) {\scriptsize $b$};
\node at (1.75, 2.6) {\scriptsize $d$};
\node at (1.75, -0.1) {\scriptsize $c$};
\end{tikzpicture}}

\newcommand{\PachnerTwo}{\begin{tikzpicture}[baseline={([yshift=-.8ex]current bounding box.center)},scale=.6]
\draw [->-] (1, 0) -- (1, 1);
\draw [->-] (1, 1) arc (270:90:.5);
\draw [->-] (1, 1) arc (-90:90:.5);
\draw [->-] (1, 2) -- (1, 3);
\node at (1.25, .3) {\scriptsize $a$};
\node at (1., 1.5) {\scriptsize \color{red}$\times$};
\node at (.2, 1.5) {\scriptsize $x$};
\node at (1.8, 1.5) {\scriptsize $y$};
\node at (1.25, 2.7) {\scriptsize $b$};
\end{tikzpicture}\qquad = \qquad \sqrt{\frac{d_xd_y}{d_a}}\ \delta_{ab}\ \delta_{xya^\ast}\quad \begin{tikzpicture}[baseline={([yshift=-.8ex]current bounding box.center)},scale=.5]
\draw [->-] (1, 0) -- (1, 3);
\node at (1.25, 1.5) {\scriptsize $a$};
\end{tikzpicture}}

\newcommand{\PachnerThree}{\begin{tikzpicture}[baseline={([yshift=-.8ex]current bounding box.center)},scale=.5]
\draw [->-] (1, 0) -- (1, 3);
\node at (1.25, 1.5) {\scriptsize $a$};
\end{tikzpicture}\quad = \quad \frac{1}{D}\sum_{xy\in L_{\Fus}} \sqrt{\frac{d_xd_y}{d_a}}\ \delta_{xya^\ast} \quad \begin{tikzpicture}[baseline={([yshift=-.8ex]current bounding box.center)},scale=.6]
\draw [->-] (1, 0) -- (1, 1);
\draw [->-] (1, 1) arc (270:90:.5);
\draw [->-] (1, 1) arc (-90:90:.5);
\draw [->-] (1, 2) -- (1, 3);
\node at (1.25, .3) {\scriptsize $a$};
\node at (.2, 1.5) {\scriptsize $x$};
\node at (1.8, 1.5) {\scriptsize $y$};
\node at (1.25, 2.7) {\scriptsize $a$};
\end{tikzpicture}}

\newcommand{\PachnerFour}{
\begin{tikzpicture}[baseline={([yshift=-.8ex]current bounding box.center)},scale=.6]
\draw [->-] (3.2,5.6) -- (3.2,6.8);
\draw [->-] (3.2,7.2) -- (4.6,8.);
\draw [->-] (4.6,8.) -- (6.,7.2);
\draw [->-] (6.,5.6) -- (4.6,4.8);
\draw [->-] (4.6,4.8) -- (3.2,5.6);
\draw [->-] (6.,7.2) -- (6.,5.6);
\draw [decorate,decoration={snake,segment length=.12cm, amplitude=.04cm}, ->] (3.2,6.6) -- (4.4,6.6);
\draw [->-] (2.6,7.6) -- (3.2,7.2);
\draw [->-] (4.6,8.8) -- (4.6,8.);
\draw [->-] (6.6,7.6) -- (6.,7.2);
\draw [->-] (6.6,5.2) -- (6.,5.6);
\draw [->-] (4.6,4.) -- (4.6,4.8);
\draw [->-] (2.6,5.2) -- (3.2,5.6);
\draw [->-] (3.2,6.6) -- (3.2,7.2);
\node [centered, rotate=0.] at (2.5,7.8) {\scriptsize $e_1$};
\node [centered, rotate=0.] at (2.9,6.) {\scriptsize $i_0$};
\node [centered, rotate=0.] at (2.9,6.9) {\scriptsize $i_1$};
\node [centered, rotate=0.] at (3.6,7.8) {\scriptsize $i_2$};
\node [centered, rotate=0.] at (4.7,6.6) {\scriptsize $p$};
\end{tikzpicture}\ =\quad\sum_{j\in L_\Fus}\sqrt{d_{i_1}d_j}\ G^{i_2^\ast e_1i_1}_{i_0 p^\ast j}\ 
\begin{tikzpicture}[baseline={([yshift=-.8ex]current bounding box.center)},scale=.6]
\draw [->-] (3.2,5.6) -- (3.2,7.2);
\draw [->-] (3.2,7.2) -- (3.55,7.4);
\draw [->-] (3.55,7.4) -- (4.6,8.);
\draw [->-] (4.6,8.) -- (6.,7.2);
\draw [->-] (6.,5.6) -- (4.6,4.8);
\draw [->-] (4.6,4.8) -- (3.2,5.6);
\draw [->-] (6.,7.2) -- (6.,5.6);
\draw [->-] (2.6,7.6) -- (3.2,7.2);
\draw [->-] (4.6,8.8) -- (4.6,8.);
\draw [->-] (6.6,7.6) -- (6.,7.2);
\draw [->-] (6.6,5.2) -- (6.,5.6);
\draw [->-] (4.6,4.) -- (4.6,4.8);
\draw [->-] (2.6,5.2) -- (3.2,5.6);
\draw [decorate,decoration={snake,segment length=.12cm, amplitude=.04cm}, ->] (3.59,7.42) -- (4.1,6.5);
\node [centered, rotate=0.] at (2.5,5.) {\scriptsize $e_n$};
\node [centered, rotate=0.] at (2.5,7.8) {\scriptsize $e_1$};
\node [centered, rotate=0.] at (2.9,6.5) {\scriptsize $i_0$};
\node [centered, rotate=0.] at (3.3,7.6) {\scriptsize $j$};
\node [centered, rotate=0.] at (3.9,8.) {\scriptsize $i_2$};
\node [centered, rotate=0.] at (4.3,6.3) {\scriptsize $p$};
\node [centered, rotate=0.] at (3.7,5.1) {\scriptsize $i_n$};
\end{tikzpicture}
}

\newcommand{\PachnerFive}{
\begin{tikzpicture}[baseline={([yshift=-.8ex]current bounding box.center)},scale=.6]
\draw [->-] (3.2,5.6) -- (3.2,6.1);
\draw [->-] (3.2,6.1) -- (3.2,6.7);
\draw [->-] (3.2,6.7) -- (3.2,7.2);
\draw [->-] (3.2,7.2) -- (4.6,8.);
\draw [->-] (4.6,8.) -- (6.,7.2);
\draw [->-] (6.,5.6) -- (4.6,4.8);
\draw [->-] (4.6,4.8) -- (3.2,5.6);
\draw [->-] (6.,7.2) -- (6.,5.6);
\draw [decorate,decoration={snake,segment length=.12cm, amplitude=.04cm}, ->] (3.2,6.1) -- (4.4,6.1);
\draw [decorate,decoration={snake,segment length=.12cm, amplitude=.04cm}, ->] (3.2,6.7) -- (4.4,6.7);
\draw [->-] (2.6,7.6) -- (3.2,7.2);
\draw [->-] (4.6,8.8) -- (4.6,8.);
\draw [->-] (6.6,7.6) -- (6.,7.2);
\draw [->-] (6.6,5.2) -- (6.,5.6);
\draw [->-] (4.6,4.) -- (4.6,4.8);
\draw [->-] (2.6,5.2) -- (3.2,5.6);
\node [centered, rotate=0.] at (2.9,5.8) {\scriptsize $i_0$};
\node [centered, rotate=0.] at (2.9,7.1) {\scriptsize $i_1$};
\node [centered, rotate=0.] at (4.7,6.7) {\scriptsize $r$};
\node [centered, rotate=0.] at (4.7,6.1) {\scriptsize $s$};
\node [centered, rotate=0.] at (2.9,6.4) {\scriptsize $j$};
\end{tikzpicture}\quad = \quad \sum_{u\in L_\Fus}\sqrt{d_jd_p}\ G^{r^\ast i_1^\ast j}_{i_0s^\ast p}\quad 
\begin{tikzpicture}[baseline={([yshift=-.8ex]current bounding box.center)},scale=.6]
\draw [->-] (3.2,5.6) -- (3.2,6.8);
\draw [->-] (3.2,7.2) -- (4.6,8.);
\draw [->-] (4.6,8.) -- (6.,7.2);
\draw [->-] (6.,5.6) -- (4.6,4.8);
\draw [->-] (4.6,4.8) -- (3.2,5.6);
\draw [->-] (6.,7.2) -- (6.,5.6);
\draw [decorate,decoration={snake,segment length=.12cm, amplitude=.04cm}, ->] (3.2,6.6) -- (4.4,6.6);
\draw [->-] (2.6,7.6) -- (3.2,7.2);
\draw [->-] (4.6,8.8) -- (4.6,8.);
\draw [->-] (6.6,7.6) -- (6.,7.2);
\draw [->-] (6.6,5.2) -- (6.,5.6);
\draw [->-] (4.6,4.) -- (4.6,4.8);
\draw [->-] (2.6,5.2) -- (3.2,5.6);
\draw [->-] (3.2,6.6) -- (3.2,7.2);
\node [centered, rotate=0.] at (2.9,6.) {\scriptsize $i_0$};
\node [centered, rotate=0.] at (2.9,6.9) {\scriptsize $i_1$};
\node [centered, rotate=0.] at (4.7,6.6) {\scriptsize $p$};
\end{tikzpicture}
}

\newcommand{\PlaquetteMsr}{
\begin{tikzpicture}[baseline={([yshift=-.8ex]current bounding box.center)},scale=.55]
\draw [->-] (3.2,5.6) -- (3.2,6.8);
\draw [->-] (3.2,7.2) -- (4.6,8.);
\draw [->-] (4.6,8.) -- (6.,7.2);
\draw [->-] (6.,5.6) -- (4.6,4.8);
\draw [->-] (4.6,4.8) -- (3.2,5.6);
\draw [->-] (6.,7.2) -- (6.,5.6);
\draw [->-] (2.6,7.6) -- (3.2,7.2);
\draw [->-] (4.6,8.8) -- (4.6,8.);
\draw [->-] (6.6,7.6) -- (6.,7.2);
\draw [->-] (6.6,5.2) -- (6.,5.6);
\draw [->-] (4.6,4.) -- (4.6,4.8);
\draw [->-] (2.6,5.2) -- (3.2,5.6);
\draw [->-] (3.2,6.6) -- (3.2,7.2);
\node [centered, rotate=0.] at (2.4,7.8) {\scriptsize $e_1$};
\node [centered, rotate=0.] at (4.95,8.8) {\scriptsize $e_2$};
\node [centered, rotate=0.] at (6.9,7.8) {\scriptsize $e_3$};
\node [centered, rotate=0.] at (6.9,5.) {\scriptsize $e_4$};
\node [centered, rotate=0.] at (5.,4.) {\scriptsize $e_5$};
\node [centered, rotate=0.] at (2.3,5.) {\scriptsize $e_6$};
\node [centered, rotate=0.] at (2.85,6.) {\scriptsize $i_7$};
\node [centered, rotate=0.] at (2.85,6.9) {\scriptsize $i_1$};
\node [centered, rotate=0.] at (3.6,7.9) {\scriptsize $i_2$};
\node [centered, rotate=0.] at (5.6,7.9) {\scriptsize $i_3$};
\node [centered, rotate=0.] at (6.4,6.4) {\scriptsize $i_4$};
\node [centered, rotate=0.] at (5.6,4.9) {\scriptsize $i_5$};
\node [centered, rotate=0.] at (3.7,4.9) {\scriptsize $i_6$};

\draw [decorate,decoration={snake,segment length=.12cm, amplitude=.04cm}, ->] (3.2,6.4) -- (5.,6.4);
\draw [->] (3.8, 6.4) arc (180:0:.8);
\draw [] (4.4, 6.4) arc (-180:0:0.5);
\node [centered, rotate=0.] at (4.8,6.7) {\scriptsize $p$};
\node [centered, rotate=0.] at (4.1,6.1) {\scriptsize $t$};
\node [centered, rotate=0.] at (3.5,6.1) {\scriptsize $p$};
\node [centered, rotate=0.] at (5.65,6.4) {\scriptsize $s$};
\node [centered, rotate=0.] at (4.6,5.4) {\color{red} $\times$};
\end{tikzpicture}
}

\newcommand{\FiboGaugeS}{
\begin{tikzpicture}[baseline={([yshift=-.8ex]current bounding box.center)},scale=1.7]
\fill [lightgray!70!] (0, 0) -- (-.8, -.3) -- (-.8, .7) -- (0, 1);
\draw [densely dashed, magenta, ->] (0, 0) -- (1, 0);
\draw [densely dashed, magenta, ->] (0, 0) -- (-.5, -.5);
\node [centered, rotate=0., magenta] at (-.3,-.5) {\scriptsize $\tau_1$};
\node [centered, rotate=0., magenta] at (.9,.15) {\scriptsize $\tau_2$};

\draw [ultra thick, ->] (0, 0) -- (0, 1);
\draw [ultra thick, ->] (0, 0) -- (-.8, -.3);

\node [centered, rotate=0.] at (.15,1.) {\scriptsize $1$};
\node [centered, rotate=0.] at (-.75,-.4) {\scriptsize $\tau$};
\end{tikzpicture}
}

\newcommand{\FiboGaugeA}{
\begin{tikzpicture}[baseline={([yshift=-.8ex]current bounding box.center)},scale=1.2]
\draw [densely dashed, magenta, ->] (0, 0) -- (1, 0);
\draw [densely dashed, magenta, ->] (0, 0) -- (0, 1);
\node [centered, rotate=0., magenta] at (1.2,0.) {\scriptsize $\tau_1$};
\node [centered, rotate=0., magenta] at (.15,.95) {\scriptsize $\tau_2$};

\draw [ultra thick, ->] (0, 0) -- (0.945, -0.327); 
\draw [ultra thick, blue, ->] (0, 0) -- (0.1947, -0.98086); 
\draw [ultra thick, blue, ->] (0, 0) -- (-0.98086, 0.1947); 

\node [centered, rotate=0.] at (1.05,-.4) {\scriptsize $\tau$};
\node [centered, rotate=0., blue] at (.25,-1.1) {\scriptsize $\tau_{\tau\bar 1}$};
\node [centered, rotate=0., blue] at (-1., 0.) {\scriptsize $\tau_{1\bar\tau}$};

\draw [very thick, dashed, blue, ->] (0.9, -.4) arc (-30:-68:1.3); 
\draw [very thick, dashed, blue, ->] (.98, -.25) arc (-20: 170:1.); 
\node [centered, rotate=0.] at (.9,-.8) {\scriptsize $\G_P^{\tau\bar 1}$};
\node [centered, rotate=0.] at (.8,1.1) {\scriptsize $\G_P^{1\bar\tau}$};

\draw [thick, dashed, orange, ->] (0.2, -.93) -- (0.5, -0.23); 
\draw [ultra thick, orange, ->] (0, 0) -- (0.6, -0.2);
\node [centered, rotate=0.] at (.5,-.5) {\scriptsize $\Proj$};
\node [centered, rotate=0., orange] at (.6,0.) {\scriptsize $\tilde\tau_{\tau\bar 1}$};

\draw [thick, dashed, orange] (-0.98086, 0.1947) --(-0.93, 0.31); 
\draw [ultra thick, orange, ->] (0, 0) -- (-0.93, 0.31);
\node [centered, rotate=0., orange] at (-.55,0.45) {\scriptsize $\tilde\tau_{1\bar\tau}$};

\end{tikzpicture}
}

\newcommand{\FiboGaugeB}{
\begin{tikzpicture}[baseline={([yshift=-.8ex]current bounding box.center)},scale=1.5]
\fill [lightgray!70!] (-0.5, .12) -- (.85, -.2) -- (.85, .8) -- (-0.5, 1.12);

\draw [ultra thick, ->] (0, 0) -- (0, 1);
\node [centered, rotate=0.] at (-.1,1.) {\scriptsize $1$};

\draw [densely dashed, magenta, ->] (0, 0) -- (1, 0);
\draw [densely dashed, magenta, ->] (0, 0) -- (.5, .5);
\node [centered, rotate=0., magenta] at (1.1,0.) {\scriptsize $\tau_1$};
\node [centered, rotate=0., magenta] at (.3,.5) {\scriptsize $\tau_2$};

\draw [ultra thick, ->] (0, 0) -- (0.85, -.2);
\node [centered, rotate=0.] at (.9,-.3) {\scriptsize $\tau$};
\draw [densely dashed, magenta] (0, 0.4) -- (.65, 0.24) -- (.65, -0.17);
\draw [ultra thick, blue, ->] (0, 0) -- (0.65, .23);
\node [centered, rotate=0., blue] at (.85,.3) {\scriptsize $1_{\tau\bar\tau}$};

\draw [densely dashed, magenta] (0, 0.) -- (-.5, 0.) -- (-.75, -.25) -- (-.25, -.25) -- (0, 0);
\draw [densely dashed, magenta] (0, 0.)  -- (-.75, -.25) -- (-.75, .4);
\draw [ultra thick, blue, ->] (0, 0) -- (-.75, .4);
\node [centered, rotate=0., blue] at (-.9,.5) {\scriptsize $\tau_{\tau\bar\tau}$};

\draw [very thick, dashed, blue, ->] (0.1, .95) arc (60:22:1.3); 
\draw [very thick, dashed, blue, ->] (0.8, -.3) arc (-30:-193:.85); 

\node [centered, rotate=0.] at (.6,.8) {\scriptsize $\G_P^{\tau\bar\tau}$};
\node [centered, rotate=0.] at (-.7,-.7) {\scriptsize $\G_P^{\tau\bar\tau}$};

\draw [densely dashed, magenta] (0, 0.6)  -- (-.32, .68) -- (-.28, .08) -- (0, 0);
\draw [ultra thick, orange, ->] (0, 0) -- (-.32, .68);
\draw [thick, dashed, orange, ->] (-.7, .45) -- (-.3, .65);
\node [centered, rotate=0., orange] at (-.3,.8) {\scriptsize $\tilde\tau_{\tau\bar\tau}$};
\node [centered, rotate=0.] at (-.6,.6) {\scriptsize $\Proj$};
\draw [ultra thick, blue, ->] (0, 0) -- (-.75, .4);

\end{tikzpicture}
}

\newcommand{\ExcitedA}{
\begin{tikzpicture}[baseline={([yshift=-.8ex]current bounding box.center)},scale=.7]
\draw [->-] (3.2,5.6) -- (3.2,7.2);
\node [centered, rotate=0.] at (2.9,6.4) {\scriptsize $j$};
\end{tikzpicture}
}

\newcommand{\ExcitedB}{
\begin{tikzpicture}[baseline={([yshift=-.8ex]current bounding box.center)},scale=.8]
\draw [->-] (3.2,5.6) -- (3.2,6.1);
\draw [->-] (3.2,6.1) -- (3.2,6.7);
\draw [->-] (3.2,6.7) -- (3.2,7.2);
\draw [decorate,decoration={snake,segment length=.12cm, amplitude=.04cm}, ->] (3.2,6.7) -- (4.,6.7);
\draw [decorate,decoration={snake,segment length=.12cm, amplitude=.04cm}, ->] (3.2,6.1) -- (2.4,6.1);
\node [centered, rotate=0.] at (3.,5.8) {\scriptsize $j$};
\node [centered, rotate=0.] at (3.,6.4) {\scriptsize $k$};
\node [centered, rotate=0.] at (3.4,7.) {\scriptsize $j$};
\node [centered, rotate=0.] at (4.2,6.7) {\scriptsize $q$};
\node [centered, rotate=0.] at (2.2,6.1) {\scriptsize $p^\ast$};
\end{tikzpicture}
}

\newcommand{\FrobeniusA}{\begin{tikzpicture}[baseline={([yshift=-.8ex]current bounding box.center)},scale=.8]
\draw [red, ->-] (.3, 0.5) -- (1, 1);
\draw [red, ->-] (1.7, 0.5) -- (1, 1);
\draw [red, ->-] (1., 2.) -- (1, 1);
\fill [red] (1., 1.) circle (.15);
\node at (1.3, 1.7) {\scriptsize $a_\alpha$};
\node at (.2, .8) {\scriptsize $b_\beta$};
\node at (1.75, .8) {\scriptsize $c_\gamma$};
\end{tikzpicture}\qquad := \qquad f_{a_\alpha b_\beta c_\gamma}\quad \begin{tikzpicture}[baseline={([yshift=-.8ex]current bounding box.center)},scale=.8]
\draw [->-] (.3, 0.5) -- (1, 1);
\draw [->-] (1.7, 0.5) -- (1, 1);
\draw [->-] (1., 2.) -- (1, 1);
\node at (1.25, 1.6) {\scriptsize $a_\alpha$};
\node at (.3, .8) {\scriptsize $b_\beta$};
\node at (1.6, .8) {\scriptsize $c_\gamma$};
\end{tikzpicture}}

\newcommand{\FrobeniusB}{\T\ \sum_{\gamma = 0}^{n_c}\begin{tikzpicture}[baseline={([yshift=-.8ex]current bounding box.center)},scale=.7]
\draw [red, ->-] (0.,0.2) -- (.5,1.);
\draw [red, ->-] (0.,1.8) -- (.5,1.);
\draw [red, ->-] (2.,1.) -- (.5,1.);
\draw [red, ->-] (2.5,1.8) -- (2.,1.);
\draw [red, ->-] (2.5,0.2) -- (2.,1.);
\fill [red] (.5, 1.) circle (.15);
\fill [red] (2., 1.) circle (.15);
\node at (1.25, 1.3) {\scriptsize $c_\gamma$};
\node at (-.1, 2.) {\scriptsize $s_\sigma$};
\node at (-.1, 0.) {\scriptsize $a_\alpha$};
\node at (2.6, 2.) {\scriptsize $r_\rho$};
\node at (2.8, 0.) {\scriptsize $b_\beta$};
\end{tikzpicture}\qquad =\qquad \sum_{t_\tau\in L_{\mathcal{A}}}\quad \begin{tikzpicture}[baseline={([yshift=-.8ex]current bounding box.center)},scale=.7]
\draw [red, ->-] (0.,0.) -- (.8,.5);
\draw [red, ->-] (1.6,0.) -- (.8,.5);
\draw [red, ->-] (.8,.5) -- (.8,2.);
\draw [red, ->-] (0.,2.5) -- (.8,2.);
\draw [red, ->-] (1.6,2.5) -- (.8,2.);
\fill [red] (.8, .5) circle (.15);
\fill [red] (.8, 2.) circle (.15);
\node at (1.2, 1.3) {\scriptsize $t_\tau$};
\node at (-.2, 2.7) {\scriptsize $s_\sigma$};
\node at (-.2, -.2) {\scriptsize $a_\alpha$};
\node at (1.9, 2.7) {\scriptsize $r_\rho$};
\node at (1.9, -0.2) {\scriptsize $b_\beta$};
\end{tikzpicture}\qquad =: \qquad \begin{tikzpicture}[baseline={([yshift=-.8ex]current bounding box.center)},scale=.7]
\draw [red, ->-] (0.,0.) -- (.8,.5);
\draw [red, ->-] (1.6,0.) -- (.8,.5);
\draw [red, thick, densely dashed] (.8,.5) -- (.8,2.);
\draw [red, ->-] (0.,2.5) -- (.8,2.);
\draw [red, ->-] (1.6,2.5) -- (.8,2.);
\fill [red] (.8, .5) circle (.15);
\fill [red] (.8, 2.) circle (.15);
\node at (-.2, 2.7) {\scriptsize $s_\sigma$};
\node at (-.2, -.2) {\scriptsize $a_\alpha$};
\node at (1.9, 2.7) {\scriptsize $r_\rho$};
\node at (1.9, -0.2) {\scriptsize $b_\beta$};
\end{tikzpicture}}

\newcommand{\FrobeniusC}{\T\quad \begin{tikzpicture}[baseline={([yshift=-.8ex]current bounding box.center)},scale=.7]
\draw [red, ->-] (1, 0) -- (1, 1);
\draw [red, thick, densely dashed] (1, 1) arc (-90:270:.5);
\draw [red, ->-] (1, 2) -- (1, 3);
\fill [red] (1., 1.) circle (.1);
\fill [red] (1., 2.) circle (.1);
\node at (1.35, .3) {\scriptsize $a_\alpha$};
\node at (1.35, 2.7) {\scriptsize $b_\beta$};
\node [red] at (1., 1.5) {$\times$};
\end{tikzpicture}\qquad = \qquad d_{\mathcal{A}}\ \delta_{ab}\ \delta_{\alpha\beta}\quad \begin{tikzpicture}[baseline={([yshift=-.8ex]current bounding box.center)},scale=.7]
\draw [red, ->-] (1, 0) -- (1, 3);
\node at (1.35, 1.5) {\scriptsize $a_\alpha$};
\end{tikzpicture}}

\newcommand{\BimoduleA}{\begin{tikzpicture}[baseline={([yshift=-.8ex]current bounding box.center)},scale=1.]
\draw [blue, ->-] (1, 0) -- (1, .6);
\draw [blue, snake it, ultra thick] (1, .6) -- (1, 1.3);
\draw [blue, ->-] (1, 1.3) -- (1, 2.);
\draw [red, -<-] (1, .6) -- (.3, .6);
\draw [red, -<-] (1, 1.3) -- (1.7, 1.3);
\node at (0.1, .6) {\scriptsize $a_\alpha$};
\node at (1.9, 1.3) {\scriptsize $b_\beta$};
\node at (1.3, .2) {\scriptsize $x_\chi$};
\node at (1.25,.9) {\scriptsize $M$};
\node at (1.25,1.8) {\scriptsize $z_\zeta$};
\end{tikzpicture}\qquad := \qquad\sum_{y\in L_\Fus}\ [P_M]^{a_\alpha b_\beta}_{x_\chi y z_\zeta}\quad\begin{tikzpicture}[baseline={([yshift=-.8ex]current bounding box.center)},scale=1.]
\draw [->-] (1, 0) -- (1, .6);
\draw [->-] (1, .6) -- (1, 1.3);
\draw [->-] (1, 1.3) -- (1, 2.);
\draw [-<-] (1, .6) -- (.3, .6);
\draw [-<-] (1, 1.3) -- (1.7, 1.3);
\node at (0.1, .6) {\scriptsize $a_\alpha$};
\node at (1.9, 1.3) {\scriptsize $b_\beta$};
\node at (1.25, .2) {\scriptsize $x_\chi$};
\node at (1.15,.9) {\scriptsize $y$};
\node at (1.25,1.8) {\scriptsize $z_\zeta$};
\end{tikzpicture}
}

\newcommand{\BimoduleB}{\T\ \sum_{y_\upsilon\in L_M}
\begin{tikzpicture}[baseline={([yshift=-.8ex]current bounding box.center)},scale=.7]
\draw [blue, ->-] (1, 0) -- (1, 1.);
\draw [blue, snake it, ultra thick] (1, 1.) -- (1, 1.8);
\draw [blue, ->-] (1, 1.8) -- (1, 2.6);
\draw [blue, snake it, ultra thick] (1, 2.6) -- (1, 3.4);
\draw [blue, ->-] (1, 3.4) -- (1, 4.4);

\draw [red, -<-] (1, 1.) -- (0, 1.);
\draw [red, -<-] (1, 1.8) -- (2, 1.8);
\draw [red, -<-] (1, 2.6) -- (0, 2.6);
\draw [red, -<-] (1, 3.4) -- (2, 3.4);

\node at (1.3, .5) {\scriptsize $x_\chi$};
\node at (-.3, 1.) {\scriptsize $a_\alpha$};
\node at (1.3, 1.4) {\scriptsize $M$};
\node at (2.3, 1.8) {\scriptsize $r_\rho$};
\node at (1.3, 2.2) {\scriptsize $y_\upsilon$};
\node at (-.3, 2.6) {\scriptsize $b_\beta$};
\node at (1.3, 3.) {\scriptsize $M$};
\node at (2.3, 3.4) {\scriptsize $s_\sigma$};
\node at (1.3, 3.9) {\scriptsize $z_\zeta$};
\end{tikzpicture}\qquad = \qquad
\begin{tikzpicture}[baseline={([yshift=-.8ex]current bounding box.center)},scale=.7]
\draw [blue, ->-] (1, 0) -- (1, 1.5);
\draw [blue, snake it, ultra thick] (1, 1.5) -- (1, 2.9);
\draw [blue, ->-] (1, 2.9) -- (1, 4.4);
\draw [red, -<-] (-.5, 1.5) -- (-1.3, 2.5);
\draw [red, -<-] (-.5, 1.5) -- (-1.3, .5);
\draw [red, densely dashed, very thick] (1, 1.5) -- (-.5, 1.5);
\fill [red] (-.5, 1.5) circle (0.15);
\draw [red, densely dashed, very thick] (1, 2.9) -- (2.5, 2.9);
\draw [red, -<-] (2.5, 2.9) -- (3.3, 3.9);
\draw [red, -<-] (2.5, 2.9) -- (3.3, 1.9);
\fill [red] (2.5, 2.9) circle (0.15);
\node at (1.35, .7) {\scriptsize $x_\chi$};
\node at (1.3, 2.2) {\scriptsize $M$};
\node at (1.35, 3.7) {\scriptsize $z_\zeta$};
\node at (-1.5, .3) {\scriptsize $a_\alpha$};
\node at (-1.5, 2.7) {\scriptsize $b_\beta$};
\node at (3.5, 1.7) {\scriptsize $r_\rho$};
\node at (3.5, 4.1) {\scriptsize $s_\sigma$};
\end{tikzpicture}}

\newcommand{\BimoduleC}{\T\quad
\begin{tikzpicture}[baseline={([yshift=-.8ex]current bounding box.center)},scale=1.]
\draw [blue, ->-] (2.,3.2) -- (2.,2.4);
\draw [blue, ultra thick, snake it] (2.,2.4) -- (2.,1.6);
\draw [blue, ->-] (2.,1.6) -- (2.,1.);
\draw [blue, ->-] (.2,-.2) -- (.95,.3);
\draw [blue, ultra thick, snake it] (.95,.3) -- (1.55,.7);
\draw [blue, ->-] (1.55,.7) -- (2.,1.);
\draw [blue, ->-] (3.8,-.2) -- (3.05,.3);
\draw [blue, ultra thick, snake it] (3.05,0.3) -- (2.45,0.7);
\draw [blue, ->-] (2.45,0.7) -- (2.,1.);
\draw [red, thick, densely dashed] (2., 2.4) ..controls (3., 2.4) and (2.9, 1.15) .. (2.45, 0.7);
\draw [red, thick, densely dashed] (2., 1.6) ..controls (1., 1.6) and (.4, .8) .. (.95, 0.3);
\draw [red, thick, densely dashed] (1.55, .7) ..controls (2., .1) and (2.6, -.3) .. (3.05, 0.3);
\node at (2.25, 2.9) {\scriptsize $x_\chi$};
\node at (2.25, 1.9) {\scriptsize $M_1$};
\node at (2.25, 1.3) {\scriptsize $r_\rho$};
\node at (.4, 0.2) {\scriptsize $y_\upsilon$};
\node at (1.1, 0.7) {\scriptsize $M_2$};
\node at (1.7, 1.) {\scriptsize $s_\sigma$};
\node at (2.15, .7) {\scriptsize $t_\tau$};
\node at (2.9, .65) {\scriptsize $M_3$};
\node at (3.5, .2) {\scriptsize $z_\zeta$};
\node [red] at (2.4, .35) {$\times$};
\node [red] at (2.4, 1.6) {$\times$};
\node [red] at (1.3, 1.1) {$\times$};
\end{tikzpicture}\qquad = \qquad d_{\mathcal{A}}^3\ [\Delta_{M_1M_2M_3}]^{x_\chi y_\upsilon z_\zeta}_{r_\rho s_\sigma t _\tau}\begin{tikzpicture}[baseline={([yshift=-.8ex]current bounding box.center)},scale=1.]
\draw [->-] (2.,3.2) -- (2.,1.);
\draw [->-] (.2,-.2) -- (2.,1.);
\draw [->-] (3.8,-.2) -- (2.,1.);
\node at (2.25, 2.) {\scriptsize $x_\chi$};
\node at (.8, .5) {\scriptsize $y_\upsilon$};
\node at (3.2, .5) {\scriptsize $z_\zeta$};
\end{tikzpicture}}

\newcommand{\BimoduleD}{\begin{tikzpicture}[baseline={([yshift=-.8ex]current bounding box.center)},scale=1.]
\draw [blue, thick, ->-] (.3, 0.5) -- (1, 1);
\draw [blue, thick, ->-] (1.7, 0.5) -- (1, 1);
\draw [blue, thick, ->-] (1., 2.) -- (1, 1);
\fill [blue] (1., 1.) circle (.15);
\node at (1.3, 1.7) {\scriptsize $M_1$};
\node at (.2, .8) {\scriptsize $M_2$};
\node at (1.75, .8) {\scriptsize $M_3$};
\end{tikzpicture}\qquad := \sum_{x_\chi\in L_{M_1}}\sum_{y_\upsilon\in L_{M_2}}\sum_{z_\zeta\in L_{M_3}}\mathcal{V}_{M_1M_2M_3}^{x_\chi y_\upsilon z_\zeta}\quad \begin{tikzpicture}[baseline={([yshift=-.8ex]current bounding box.center)},scale=.8]
\draw [->-] (.3, 0.5) -- (1, 1);
\draw [->-] (1.7, 0.5) -- (1, 1);
\draw [->-] (1., 2.) -- (1, 1);
\node at (1.3, 1.6) {\scriptsize $x_\chi$};
\node at (.3, .8) {\scriptsize $y_\upsilon$};
\node at (1.6, .8) {\scriptsize $z_\zeta$};
\end{tikzpicture}}

\newcommand{\BimoduleE}{G^{M_1M_2M}_{M_3M_4M'}\quad =\quad \frac{1}{d_\mathcal{A}^3\sqrt{d_{M_1}d_{M_2}d_{M_3}d_{M_4}d_{M}d_{M'}}}\ \ \mathcal{T}\ \ \begin{tikzpicture}[baseline={([yshift=-.8ex]current bounding box.center)},scale=1.]
\draw [blue, thick, -<-] (0, 1) -- (2, 1);
\draw [blue, thick, ->-] (1, 2) arc (90:180:1);
\draw [blue, thick, ->-] (1, 0) arc (270:180:1);
\draw [blue, thick, -<-] (2, 1) arc (0:90:1);
\draw [blue, thick, -<-] (2, 1) arc (0:-90:1);
\draw [blue, thick, -<-] (2, 1) arc (0:-90:1);
\draw [blue, thick] (1, 2) arc [start angle = 180, end angle = 90, x radius = .7, y radius = .5];
\draw [blue, thick] (1., 0) arc [start angle = 180, end angle = 270, x radius = .7, y radius = .5];
\draw [blue, thick, ->-] (1.7, 2.5) arc [start angle = 90, end angle = -90, x radius = 1.4, y radius = 1.5];
\node at (-.1, 1.7) {\scriptsize $M_1$};
\node at (-.1, .3) {\scriptsize $M_2$};
\node at (2.1, .3) {\scriptsize $M_3$};
\node at (2.1, 1.7) {\scriptsize $M_4$};
\node at (1., 1.3) {\scriptsize $M$};
\node at (3.4, 1.) {\scriptsize $M'$};
\node [red] at (1., 1.6) {$\times$};
\node [red] at (1., .6) {$\times$};
\node [red] at (2.6, 1.) {$\times$};
\fill [blue] (0, 1) circle (0.15);
\fill [blue] (2, 1) circle (0.15);
\fill [blue] (1, 2) circle (0.15);
\fill [blue] (1, 0) circle (0.15);
\end{tikzpicture}}

\newcommand{\BimoduleF}{\sum_{\alpha = 1}^{n^M_x}\sum_{\beta = 1}^{n^N_x}\begin{tikzpicture}[baseline={([yshift=-.8ex]current bounding box.center)},scale=1.]
\draw [thick, blue, ->- ] (0, 0) arc(-90:100:1);
\draw [thick, blue, ->- ] (0, 0) arc(270:80:1);
\fill [blue] (0, 0) circle (0.15);
\fill [blue] (0, 2) circle (0.15);
\draw [blue, ->-] (0, 0) -- (0, 1.);
\draw [blue, ->-] (0, 1.) -- (0, 2);
\node [] at (-1.4, 1.) {$M_1$};
\node [] at (1.4, 1.) {$M_2$};
\node [] at (.3, .5) {$x^M_\alpha$};
\node [] at (.3, 1.5) {$y^N_\beta$};
\node [red] at (.5, 1.) {$\times$};
\node [red] at (-.5, 1.) {$\times$};
\end{tikzpicture}\quad = \quad d_\mathcal{A}n^M_x\delta_{xy}\delta_{M_1M_2M_3}\delta_{MN}\sqrt{d_{M_1}d_{M_2}d_x}}

\newcommand{\BimoduleX}{
\begin{tikzpicture}[baseline={([yshift=-.8ex]current bounding box.center)},scale=.8]
\draw [->-] (1, 0) -- (1, .6);
\draw [->-] (1, .6) -- (1, 1.3);
\draw [->-] (1, 1.3) -- (1, 2.);
\draw [-<-] (1, .6) -- (.3, .6);
\draw [-<-] (1, 1.3) -- (1.7, 1.3);
\node at (0.1, .6) {\scriptsize $a_\alpha$};
\node at (1.9, 1.3) {\scriptsize $b_\beta$};
\node at (1.25, .2) {\scriptsize $x_\chi$};
\node at (1.15,.9) {\scriptsize $y$};
\node at (1.25,1.8) {\scriptsize $z_\zeta$};
\end{tikzpicture}
}

\newcommand{\TransB}{\mathcal{D}_v\quad
\begin{tikzpicture}[baseline={([yshift=-.8ex]current bounding box.center)},scale=.8]
\draw [->-] (0, 0) -- (1, .6);
\draw [->-] (2, 0) -- (1, .6);
\draw [-<-] (1, .6) -- (1, 1.8);
\node [] at (1.5, 1.4) {$M_1$};
\node [] at (0., .5) {$M_2$};
\node [] at (1.9, .5) {$M_3$};
\end{tikzpicture}\ \ := \ \ 
\begin{tikzpicture}[baseline={([yshift=-.8ex]current bounding box.center)},scale=.8]
\draw [->-, thick, blue] (0, 0) -- (1, .6);
\draw [->-, thick, blue] (2, 0) -- (1, .6);
\draw [-<-, thick, blue] (1, .6) -- (1, 1.8);
\fill [blue] (1, .6) circle (.15);
\node [] at (1.5, 1.4) {\scriptsize $M_1$};
\node [] at (0., .5) {\scriptsize $M_2$};
\node [] at (1.9, .5) {\scriptsize $M_3$};
\end{tikzpicture} \ \ = \ \ \sum_{x_\chi\in L_{M_1}} \sum_{y_\upsilon\in L_{M_2}} \sum_{z_\zeta\in L_{M_3}} \mathcal{V}_{M_1M_2M_3}^{x_\chi y_\upsilon z_\zeta}
\begin{tikzpicture}[baseline={([yshift=-.8ex]current bounding box.center)},scale=.8]
\draw [->-] (0, 0) -- (1, .6);
\draw [->-] (2, 0) -- (1, .6);
\draw [-<-] (1, .6) -- (1, 1.8);
\node [] at (1.4, 1.4) {\scriptsize $x_\chi$};
\node [] at (0.3, .5) {\scriptsize $y_\upsilon$};
\node [] at (1.7, .5) {\scriptsize $z_\zeta$};
\end{tikzpicture}
}

\newcommand{\TransE}{\varphi_\A\quad
\begin{tikzpicture}[baseline={([yshift=-.8ex]current bounding box.center)},scale=1.]
\draw (0, 0) -- (0, 2);
\node at (.2, 1.) {\scriptsize $a$};
\end{tikzpicture}\qquad := \qquad
\begin{tikzpicture}[baseline={([yshift=-.8ex]current bounding box.center)},scale=1.]
\draw [->-](1, 0) -- (1, 2.);
\node [] at (1.3, 1.) {\scriptsize $M_a$};
\end{tikzpicture}
}

\newcommand{\TransF}{E_e\quad
\begin{tikzpicture}[baseline={([yshift=-.8ex]current bounding box.center)},scale=1.]
\draw [blue, ->-] (0, 0) -- (.7, 0);
\draw [blue, ->-] (.7, 0) -- (1.4, 0);
\draw [blue, ->-] (0, 0) -- (-.3, -.5);
\draw [blue, ->-] (0, 0) -- (-.3, .5);
\draw [blue, ->-] (1.4, 0) -- (1.7, -.5);
\draw [blue, ->-] (1.4, 0) -- (1.7, .5);
\node [] at (-.3, 0.) {\scriptsize $v_1$};
\node [] at (1.7, 0.) {\scriptsize $v_2$};
\node [] at (.3, 0.3) {\scriptsize $x^{M_1}_\alpha$};
\node [] at (1.1, 0.3) {\scriptsize $x^{M_2}_\beta$};
\end{tikzpicture}\qquad := \qquad \sum_{i}(A^{x, M_1}_{\alpha, i})^\ast A^{x, M_2}_{\beta, i}\quad
\begin{tikzpicture}[baseline={([yshift=-.8ex]current bounding box.center)},scale=1.]
\draw [->-] (0, 0) -- (1., 0);
\draw [->-] (0, 0) -- (-.3, .5);
\draw [->-] (0, 0) -- (-.3, -.5);
\draw [->-] (1, 0) -- (1.3, .5);
\draw [->-] (1, 0) -- (1.3, -.5);
\node [] at (.5, 0.2) {\scriptsize $x$};
\end{tikzpicture}
}

\newcommand{\BimoduleG}{
\begin{tikzpicture}[baseline={([yshift=-.8ex]current bounding box.center)},scale=1.]
\draw [blue, ->-] (0, 0) -- (0, 0.5);
\draw [blue, snake it, ultra thick] (0, 0.5) -- (0, 1.);
\draw [blue, ->-] (0, 1) -- (0, 1.5);
\draw [red, -<-] (0, .5) -- (-.5, .5);
\draw [red, -<-] (0, 1.) -- (.5, 1.);

\node [centered, rotate=0.] at (.2,.15) {\scriptsize $x_\chi$};
\node [centered, rotate=0.] at (.2,.75) {\scriptsize $M$};
\node [centered, rotate=0.] at (.2,1.35) {\scriptsize $z_\zeta$};
\node [centered, rotate=0.] at (-.7,.5) {\scriptsize $a_\alpha$};
\node [centered, rotate=0.] at (.7,1.) {\scriptsize $b_\beta$};
\end{tikzpicture}
}

\newcommand{\BimoduleH}{
\begin{tikzpicture}[baseline={([yshift=-.8ex]current bounding box.center)},scale=.6]
\draw [->-] (3.2,5.6) -- (3.2,6.5);
\draw [->-] (3.2,6.5) -- (3.2,7.2);
\draw [->-] (3.2,7.2) -- (4.6,8.);
\draw [->-] (4.6,8.) -- (6.,7.2);
\draw [->-] (6.,7.2) -- (6.,5.6);
\draw [->-] (6.,5.6) -- (4.6,4.8);
\draw [->-] (4.6,4.8) -- (3.2,5.6);
\draw [decorate,decoration={snake,segment length=.12cm, amplitude=.04cm}, ->] (3.2,6.4) -- (4.2,6.4);
\node [centered, rotate=0.] at (2.7,6.) {\scriptsize $M_6$};
\node [centered, rotate=0.] at (2.7,6.9) {\scriptsize $M_0$};
\node [centered, rotate=0.] at (3.6,7.9) {\scriptsize $M_1$};
\node [centered, rotate=0.] at (5.6,7.9) {\scriptsize $M_2$};
\node [centered, rotate=0.] at (6.4,6.4) {\scriptsize $M_3$};
\node [centered, rotate=0.] at (5.6,4.9) {\scriptsize $M_4$};
\node [centered, rotate=0.] at (3.6,4.9) {\scriptsize $M_5$};
\node [centered, rotate=0.] at (4.5,6.4) {\scriptsize $M$};
\draw [->-] (2.6,7.6) -- (3.2,7.2);
\draw [->-] (4.6,8.8) -- (4.6,8.);
\draw [->-] (6.6,7.6) -- (6.,7.2);
\draw [->-] (6.6,5.2) -- (6.,5.6);
\draw [->-] (4.6,4.) -- (4.6,4.8);
\draw [->-] (2.6,5.2) -- (3.2,5.6);
\node [centered, rotate=0.] at (2.3,7.8) {\scriptsize $N_1$};
\node [centered, rotate=0.] at (4.6,9.) {\scriptsize $N_2$};
\node [centered, rotate=0.] at (6.9,7.8) {\scriptsize $N_3$};
\node [centered, rotate=0.] at (6.9,5.) {\scriptsize $N_4$};
\node [centered, rotate=0.] at (4.6,3.7) {\scriptsize $N_5$};
\node [centered, rotate=0.] at (2.3,5.) {\scriptsize $N_6$};
\end{tikzpicture}\qquad\Longrightarrow\qquad\frac{1}{d_\mathcal{A}^{9}}\sum_{x_i, y_i\in L_{M_i}}\sum_{e_i\in L_{N_i}}\sum_{p_\alpha,q_\beta\in L_M}\nonumber\\
\T\qquad
\begin{tikzpicture}[baseline={([yshift=-.8ex]current bounding box.center)},scale=1.5]
\draw [blue, ->-] (3.2,5.6) -- (3.2,5.8);
\draw [blue, ultra thick, snake it] (3.2,5.8) -- (3.2,6.1);
\draw [blue, ->-] (3.2,6.1) -- (3.2,6.4);
\node [centered, rotate=0.] at (3.05,5.7) {\scriptsize $x_6$};
\node [centered, rotate=0.] at (3.02,5.95) {\scriptsize $M_6$};
\node [centered, rotate=0.] at (3.05,6.25) {\scriptsize $y_6$};
\draw [red, thick, densely dashed] (3.2, 5.8) -- (2.65, 5.8);
\draw [red, thick, densely dashed] (3.2, 6.1) -- (4.1, 6.1);
\fill [red] (4.05, 6.1) circle (0.05);
\draw [blue, ->-] (3.2,6.4) -- (3.2,6.7);
\draw [blue, ultra thick, snake it] (3.2,6.7) -- (3.2,7.);
\draw [blue, ->-] (3.2,7.) -- (3.2,7.2);
\node [centered, rotate=0.] at (3.05,6.55) {\scriptsize $x_0$};
\node [centered, rotate=0.] at (3.02,6.85) {\scriptsize $M_0$};
\node [centered, rotate=0.] at (3.05,7.1) {\scriptsize $y_0$};
\draw [red, thick, densely dashed] (3.2, 6.7) -- (2.65, 6.7);
\draw [red, thick, densely dashed] (3.2, 7.) -- (3.8, 7.);
\fill [red] (3.8, 7.) circle (0.05);
\draw [blue, ->-] (3.2,7.2) -- (3.55,7.4);
\draw [blue, ultra thick, snake it] (3.55,7.4) -- (4.25,7.8);
\draw [blue, ->-] (4.25,7.8) -- (4.6,8.);
\node [centered, rotate=0.] at (3.45,7.2) {\scriptsize $x_1$};
\node [centered, rotate=0.] at (3.8,7.72) {\scriptsize $M_1$};
\node [centered, rotate=0.] at (4.35,8.) {\scriptsize $y_1$};
\draw [red, thick, densely dashed] (3.55, 7.4) -- (3.25, 7.9);
\draw [red, thick, densely dashed] (4.25, 7.8) -- (4.45, 7.45);
\fill [red] (4.47, 7.4) circle (0.05);
\draw [blue, ->-] (4.6,8.) -- (4.95,7.8);
\draw [blue, ultra thick, snake it] (4.95,7.8) -- (5.65,7.4);
\draw [blue, ->-] (5.65,7.4) -- (6.,7.2);
\node [centered, rotate=0.] at (5.9,7.4) {\scriptsize $y_2$};
\node [centered, rotate=0.] at (5.4,7.72) {\scriptsize $M_2$};
\node [centered, rotate=0.] at (4.77,7.8) {\scriptsize $x_2$};
\draw [red, thick, densely dashed] (4.95, 7.8) -- (5.25, 8.3);
\draw [red, thick, densely dashed] (5.65, 7.4) -- (5.38, 7.);
\fill [red] (5.39, 7.02) circle (0.05);
\draw [blue, ->-] (6.,7.2) -- (6.,6.7);
\draw [blue, ultra thick, snake it] (6.,6.7) -- (6.,6.1);
\draw [blue, ->-] (6.,6.1) -- (6.,5.6);
\draw [red, thick, densely dashed] (6., 6.7) -- (6.5, 6.7);
\draw [red, thick, densely dashed] (5.6, 6.1) -- (6., 6.1);
\fill [red] (5.56, 6.1) circle (0.05);
\node [centered, rotate=0.] at (5.85,6.9) {\scriptsize $x_3$};
\node [centered, rotate=0.] at (6.2,6.4) {\scriptsize $M_3$};
\node [centered, rotate=0.] at (6.15,5.85) {\scriptsize $y_3$};
\draw [blue, ->-] (6.,5.6) -- (5.65,5.4);
\draw [blue, ultra thick, snake it] (5.65,5.4) -- (4.95,5.);
\draw [blue, ->-] (4.95,5.) -- (4.6,4.8);
\node [centered, rotate=0.] at (5.95,5.35) {\scriptsize $x_4$};
\node [centered, rotate=0.] at (5.4,5.05) {\scriptsize $M_4$};
\node [centered, rotate=0.] at (4.9,4.78) {\scriptsize $y_4$};
\draw [red, thick, densely dashed] (5.65, 5.4) -- (5.95, 4.9);
\draw [red, thick, densely dashed] (4.95, 5.) -- (4.55, 5.7);
\fill [red] (4.58, 5.64) circle (0.05);
\draw [blue, ->-] (4.6,4.8) -- (4.25,5.);
\draw [blue, ultra thick, snake it] (4.25,5.) -- (3.55,5.4);
\draw [blue, ->-] (3.55,5.4) -- (3.2,5.6);
\node [centered, rotate=0.] at (3.48,5.6) {\scriptsize $y_5$};
\node [centered, rotate=0.] at (3.75,5.05) {\scriptsize $M_5$};
\node [centered, rotate=0.] at (4.5,5.03) {\scriptsize $x_5$};
\draw [red, thick, densely dashed] (4.25, 5.) -- (3.95, 4.5);
\draw [red, thick, densely dashed] (3.55, 5.4) -- (4.25, 5.8);
\fill [red] (4.25, 5.8) circle (0.05);
\draw [blue, ->-] (3.2,6.4) -- (3.6,6.4);
\draw [blue, ultra thick, snake it] (3.6,6.4) -- (4.,6.4);
\draw [blue, ->] (4.,6.4) -- (4.5,6.4);
\fill [blue] (3.2,6.4) circle (0.07);
\draw [red, thick, densely dashed] (3.6, 6.4) arc (180:0:1.);
\draw [red, thick, densely dashed] (4., 6.4) arc (-180:0:.8);
\node [centered, rotate=0.] at (3.4,6.5) {\scriptsize $p_\alpha$};
\node [centered, rotate=0.] at (3.8,6.53) {\scriptsize $M$};
\node [centered, rotate=0.] at (4.6,6.4) {\scriptsize $q_\beta$};
\node [centered, rotate=0.] at (3.4,6.8) {\color{red} $\times$};
\node [centered, rotate=0.] at (3.8,7.3) {\color{red} $\times$};
\node [centered, rotate=0.] at (4.9,7.5) {\color{red} $\times$};
\node [centered, rotate=0.] at (5.75,6.7) {\color{red} $\times$};
\node [centered, rotate=0.] at (5.4,5.6) {\color{red} $\times$};
\node [centered, rotate=0.] at (4.3,5.4) {\color{red} $\times$};
\node [centered, rotate=0.] at (3.6,5.8) {\color{red} $\times$};
\node [centered, rotate=0.] at (3.6,6.2) {\color{red} $\times$};

\draw [blue, ultra thick, snake it] (2.6,7.6) -- (2.9,7.4);
\draw [blue, ->-] (2.9,7.4) -- (3.2,7.2);
\fill [blue] (3.2,7.2) circle (0.07);
\draw [red, thick, densely dashed] (2.9,7.4) -- (2.68,7.07);
\node [centered, rotate=0.] at (2.5,7.7) {\scriptsize $N_1$};
\node [centered, rotate=0.] at (3.1,7.43) {\scriptsize $e_1$};
\draw [blue, ultra thick, snake it] (4.6,8.6) -- (4.6,8.3);
\draw [blue, ->-] (4.6,8.3) -- (4.6,8.);
\fill [blue] (4.6,8.) circle (0.07);
\draw [red, thick, densely dashed] (4.6,8.3) -- (4.2,8.3);
\node [centered, rotate=0.] at (4.6,8.7) {\scriptsize $N_2$};
\node [centered, rotate=0.] at (4.72,8.1) {\scriptsize $e_2$};
\draw [blue, ultra thick, snake it] (6.6,7.6) -- (6.3,7.4);
\draw [blue, ->-] (6.3,7.4) -- (6.,7.2);
\fill [blue] (6.,7.2) circle (0.07);
\draw [red, thick, densely dashed] (6.3,7.4) -- (6.08,7.73);
\node [centered, rotate=0.] at (6.7,7.7) {\scriptsize $N_3$};
\node [centered, rotate=0.] at (6.2,7.19) {\scriptsize $e_3$};
\draw [blue, ultra thick, snake it] (6.6,5.2) -- (6.3,5.4);
\draw [blue, ->-] (6.3,5.4) -- (6.,5.6);
\fill [blue] (6.,5.6) circle (0.07);
\draw [red, thick, densely dashed] (6.3,5.4) -- (6.7,5.73);
\node [centered, rotate=0.] at (6.8,5.1) {\scriptsize $N_4$};
\node [centered, rotate=0.] at (6.3,5.6) {\scriptsize $e_4$};
\draw [blue, ultra thick, snake it] (4.6,4.2) -- (4.6,4.5);
\draw [blue, ->-] (4.6,4.5) -- (4.6,4.8);
\fill [blue] (4.6,4.8) circle (0.07);
\draw [red, thick, densely dashed] (4.6,4.5) -- (5.,4.5);
\node [centered, rotate=0.] at (4.65,4.1) {\scriptsize $N_5$};
\node [centered, rotate=0.] at (4.38,4.67) {\scriptsize $e_5$};
\draw [blue, ultra thick, snake it] (2.6,5.2) -- (2.9,5.4);
\draw [blue, ->-] (2.9,5.4) -- (3.2,5.6);
\fill [blue] (3.2,5.6) circle (0.07);
\draw [red, thick, densely dashed] (2.9,5.4) -- (3.12,5.07);
\node [centered, rotate=0.] at (2.5,5.1) {\scriptsize $N_6$};
\node [centered, rotate=0.] at (3.12,5.38) {\scriptsize $e_6$};
\end{tikzpicture}
}

\newcommand{\BraidA}{
\begin{tikzpicture}[baseline={([yshift=-.8ex]current bounding box.center)},scale=.3]
\draw [->-] (3.,0) -- (0.,4.);
\draw [densely dashed] (0,0) -- (1.2,1.6);
\draw [densely dashed] (1.8,2.4) -- (3.,4.);
\node [centered, rotate=0.] at (-.25,-.4) {\scriptsize $0$};
\node [centered, rotate=0.] at (3.25,-.4) {\scriptsize $a$};
\node [centered, rotate=0.] at (3.25,4.4) {\scriptsize $0$};
\node [centered, rotate=0.] at (-.25,4.4) {\scriptsize $a$};
\end{tikzpicture}
}

\newcommand{\BraidB}{
\begin{tikzpicture}[baseline={([yshift=-.8ex]current bounding box.center)},scale=.3]
\draw [->-] (1.5,0) -- (1.5,4.);
\draw [densely dashed] (0,0) arc  [start angle = 180, end angle = 90, x radius = 1.5, y radius = 1.2];
\draw [densely dashed] (1.5,3.) arc  [start angle = -90, end angle = 0, x radius = 1.5, y radius = 1.2];
\node [centered, rotate=0.] at (0.,-.4) {\scriptsize $0$};
\node [centered, rotate=0.] at (1.5,-.35) {\scriptsize $a$};
\node [centered, rotate=0.] at (4.,4.4) {\scriptsize $0$};
\node [centered, rotate=0.] at (1.5,4.35) {\scriptsize $a$};
\end{tikzpicture}
}

\newcommand{\BraidC}{
\begin{tikzpicture}[baseline={([yshift=-.8ex]current bounding box.center)},scale=.3]
\draw [->-] (3,0) -- (1.8,1.6);
\draw [->-] (1.2,2.4) -- (0.,4.);
\draw [densely dashed] (0,0) -- (3.,4.);
\node [centered, rotate=0.] at (-.2,-.35) {\scriptsize $0$};
\node [centered, rotate=0.] at (3.2,-.35) {\scriptsize $a$};
\node [centered, rotate=0.] at (3.2,4.35) {\scriptsize $0$};
\node [centered, rotate=0.] at (-.2,4.35) {\scriptsize $a$};
\end{tikzpicture}
}

\newcommand{\BraidD}{
\begin{tikzpicture}[baseline={([yshift=-.8ex]current bounding box.center)},scale=.4]
\draw [blue, ->-] (3.,0) -- (0.,4.);
\draw [red] (0,0) -- (1.2,1.6);
\draw [red] (1.8,2.4) -- (3.,4.);
\node [centered, rotate=0.] at (-.25,-.4) {\scriptsize $\A$};
\node [centered, rotate=0.] at (3.25,-.4) {\scriptsize $M$};
\node [centered, rotate=0.] at (3.25,4.4) {\scriptsize $\A$};
\node [centered, rotate=0.] at (-.25,4.4) {\scriptsize $M$};
\end{tikzpicture}
}

\newcommand{\BraidE}{
\begin{tikzpicture}[baseline={([yshift=-.8ex]current bounding box.center)},scale=.4]
\draw [blue, ->-] (1.5,0) -- (1.5,1.2);
\draw [blue, ultra thick, snake it] (1.5,1.2) -- (1.5,2.8);
\draw [blue, ->-] (1.5,2.8) -- (1.5,4.);
\draw [red] (0,0) arc  [start angle = 180, end angle = 90, x radius = 1.5, y radius = 1.2];
\draw [red] (1.5,2.8) arc  [start angle = -90, end angle = 0, x radius = 1.5, y radius = 1.2];
\node [centered, rotate=0.] at (0.,-.4) {\scriptsize $\A$};
\node [centered, rotate=0.] at (1.5,-.35) {\scriptsize $M$};
\node [centered, rotate=0.] at (2.,2.) {\scriptsize $M$};
\node [centered, rotate=0.] at (3.2,4.4) {\scriptsize $\A$};
\node [centered, rotate=0.] at (1.5,4.35) {\scriptsize $M$};
\end{tikzpicture}
}

\newcommand{\BraidF}{
\begin{tikzpicture}[baseline={([yshift=-.8ex]current bounding box.center)},scale=.4]
\draw [blue, ->-] (3,0) -- (1.8,1.6);
\draw [blue, ->-] (1.2,2.4) -- (0.,4.);
\draw [red] (0,0) -- (3.,4.);
\node [centered, rotate=0.] at (-.2,-.35) {\scriptsize $\A$};
\node [centered, rotate=0.] at (3.2,-.35) {\scriptsize $M$};
\node [centered, rotate=0.] at (3.2,4.35) {\scriptsize $\A$};
\node [centered, rotate=0.] at (-.2,4.35) {\scriptsize $M$};
\end{tikzpicture}
}

\newcommand{\BraidG}{
\begin{tikzpicture}[baseline={([yshift=-.8ex]current bounding box.center)},scale=.5]
\draw [->-] (0, 6) -- (0, 7);
\draw [->-] (0, 9) -- (0, 10);
\draw [->-] (0, 7) arc (-90:90:1);
\draw [->-] (0, 7) arc (270:180:1);
\draw [->-] (-1, 8) arc (180:90:1);
\draw [dashed] (-1, 8) -- (0., 8);
\node [centered, rotate=0.] at (0,5.7) {\scriptsize $M_1$};
\node [centered, rotate=0.] at (0,10.3) {\scriptsize $M_4$};
\node [centered, rotate=0.] at (-1.8,8.) {\scriptsize $M_2$};
\node [centered, rotate=0.] at (1.8,8.) {\scriptsize $M_3$};
\node [centered, rotate=0.] at (.2,8) {\scriptsize $0$};
\draw [blue, ->-] (0, 0.) -- (0, 1);
\draw [blue, ->-] (0, 3.) -- (0, 4.);
\draw [red, decorate,decoration={snake,segment length=.12cm, amplitude=.04cm}] (-1, 2.) -- (0.1, 2.);
\draw [blue, ->-] (0, 1) arc (-90:-180:1);
\draw [blue, ->-] (0, 1) arc (-90:90:1);
\draw [blue, ->-] (-1, 2) arc (180:90:1);
\node [centered, rotate=0.] at (0.,-.3) {\scriptsize $M_1$};
\node [centered, rotate=0.] at (0.,4.3) {\scriptsize $M_4$};
\node [centered, rotate=0.] at (-1.2,1.) {\scriptsize $M_2$};
\node [centered, rotate=0.] at (-1.3,3.) {\scriptsize $M_2$};
\node [centered, rotate=0.] at (1.7,2.) {\scriptsize $M_3$};
\node [centered, rotate=0.] at (.35,2.) {\scriptsize $\A$};
\fill [blue] (0.,1.) circle (0.18);
\fill [blue] (0.,3.) circle (0.18);
\fill [blue] (-1.,2.) circle (0.18);
\draw [->-] (12, 6) -- (12, 7);
\draw [->-] (12, 9) -- (12, 10);
\draw [->-] (12, 7) arc (-90:90:1);
\draw [->-] (12, 7) arc (270:90:1);
\node [centered, rotate=0.] at (9.,8.) {\large $\T$};
\node [centered, rotate=0.] at (12.,5.7) {\scriptsize $M_1$};
\node [centered, rotate=0.] at (12.,10.3) {\scriptsize $M_4$};
\node [centered, rotate=0.] at (10.2,8.) {\scriptsize $M_2$};
\node [centered, rotate=0.] at (13.8,8.) {\scriptsize $M_3$};
\node [centered, rotate=0.] at (12.,8.) {\color{red} $\times$};
\draw [blue, ->-] (12, 0) -- (12, 1);
\draw [blue, ->-] (12, 3.) -- (12, 4.);
\draw [red] (11, 2.) arc (90:62:4);
\draw [blue, ->-] (12, 1) arc (-90:-180:1);
\draw [blue, ->-] (12, 1) arc (-90:-30:1);
\draw [blue, ultra thick, snake it] (12.866, 1.5) arc (-30:30:1);
\draw [blue, ->-] (12.866, 2.5) arc (30:90:1);
\draw [blue, ->-] (11, 2) arc (180:90:1);
\node [centered, rotate=0.] at (8.7,1.9) {\Large $\frac{1}{d_\A^2}$};
\node [centered, rotate=0.] at (9.6,2.) {\large $\T$};
\node [centered, rotate=0.] at (12.,-.2) {\scriptsize $M_1$};
\node [centered, rotate=0.] at (12.,4.2) {\scriptsize $M_4$};
\node [centered, rotate=0.] at (10.8,1.) {\scriptsize $M_2$};
\node [centered, rotate=0.] at (10.8,3.) {\scriptsize $M_2$};
\node [centered, rotate=0.] at (13.7,2.) {\scriptsize $M_3$};
\node [centered, rotate=0.] at (12.,2.3) {\color{red} $\times$};
\node [centered, rotate=0.] at (12,1.5) {\color{red} $\times$};
\fill [blue] (12.,1.) circle (0.18);
\fill [blue] (12.,3.) circle (0.18);
\fill [blue] (11.,2.) circle (0.18);
\draw [red, decorate,decoration={snake,segment length=.12cm, amplitude=.04cm}] (12.866, 2.5) arc [start angle = 150, end angle = 90, x radius = 1.5, y radius = 1.];
\node [centered, rotate=0.] at (13.5,3.5) {\scriptsize $\A$};
\draw [ultra thick, ->] (3, 8) -- (8, 8);
\draw [ultra thick, ->] (3, 2) -- (7.5, 2);
\draw [ultra thick, ->] (-2.5, 7) arc (130:230:3);
\draw [ultra thick, ->] (14.5, 7) arc (50:-50:3);
\node [centered, rotate=0.] at (5.5,8.5) {Contrast};
\node [centered, rotate=0.] at (-4.5, 5) {$\D$};
\node [centered, rotate=0.] at (16.5, 5) {$\D$};
\end{tikzpicture}
}

\newcommand{\ToricCodeA}{
\begin{tikzpicture}[baseline={([yshift=-.8ex]current bounding box.center)},scale=.45]

\draw [] (3.2,5.6) -- (3.2,6.8);
\draw [] (3.2,7.2) -- (4.6,8.);
\draw [] (4.6,8.) -- (6.,7.2);
\draw [] (6.,5.6) -- (4.6,4.8);
\draw [] (4.6,4.8) -- (3.2,5.6);
\draw [] (6.,7.2) -- (6.,5.6);
\draw [decorate,decoration={snake,segment length=.12cm, amplitude=.04cm}] (3.2,6.4) -- (4.4,6.4);
\draw [] (2.6,7.6) -- (3.2,7.2);
\draw [] (4.6,8.8) -- (4.6,8.);
\draw [] (6.6,7.6) -- (6.,7.2);
\draw [] (6.6,5.2) -- (6.,5.6);
\draw [] (4.6,4.) -- (4.6,4.8);
\draw [] (2.6,5.2) -- (3.2,5.6);
\draw [] (3.2,6.6) -- (3.2,7.2);
\node [centered, rotate=0.] at (2.3,7.8) {\scriptsize $e_1$};
\node [centered, rotate=0.] at (5.,8.8) {\scriptsize $e_2$};
\node [centered, rotate=0.] at (6.9,7.8) {\scriptsize $e_3$};
\node [centered, rotate=0.] at (6.9,5.) {\scriptsize $e_4$};
\node [centered, rotate=0.] at (5.,4.) {\scriptsize $e_5$};
\node [centered, rotate=0.] at (2.3,5.) {\scriptsize $e_6$};
\node [centered, rotate=0.] at (2.85,6.) {\scriptsize $i_6$};
\node [centered, rotate=0.] at (2.85,6.9) {\scriptsize $i_0$};
\node [centered, rotate=0.] at (3.6,7.95) {\scriptsize $i_1$};
\node [centered, rotate=0.] at (5.6,7.95) {\scriptsize $i_2$};
\node [centered, rotate=0.] at (6.4,6.4) {\scriptsize $i_3$};
\node [centered, rotate=0.] at (5.6,4.9) {\scriptsize $i_4$};
\node [centered, rotate=0.] at (3.7,4.9) {\scriptsize $i_5$};
\node [centered, rotate=0.] at (4.7,6.4) {\scriptsize $p$};
\end{tikzpicture}
}

\newcommand{\ToricCodeB}{
\begin{tikzpicture}[baseline={([yshift=-.8ex]current bounding box.center)},scale=.45]

\draw [] (3.2,5.6) -- (3.2,6.8);
\draw [] (3.2,7.2) -- (4.6,8.);
\draw [] (4.6,8.) -- (6.,7.2);
\draw [] (6.,5.6) -- (4.6,4.8);
\draw [] (4.6,4.8) -- (3.2,5.6);
\draw [] (6.,7.2) -- (6.,5.6);
\draw [decorate,decoration={snake,segment length=.12cm, amplitude=.04cm}] (3.2,6.4) -- (4.4,6.4);
\draw [] (2.6,7.6) -- (3.2,7.2);
\draw [] (4.6,8.8) -- (4.6,8.);
\draw [] (6.6,7.6) -- (6.,7.2);
\draw [] (6.6,5.2) -- (6.,5.6);
\draw [] (4.6,4.) -- (4.6,4.8);
\draw [] (2.6,5.2) -- (3.2,5.6);
\draw [] (3.2,6.6) -- (3.2,7.2);
\node [centered, rotate=0.] at (2.3,7.8) {\scriptsize $e_1$};
\node [centered, rotate=0.] at (5.,8.8) {\scriptsize $e_2$};
\node [centered, rotate=0.] at (6.9,7.8) {\scriptsize $e_3$};
\node [centered, rotate=0.] at (6.9,5.) {\scriptsize $e_4$};
\node [centered, rotate=0.] at (5.,4.) {\scriptsize $e_5$};
\node [centered, rotate=0.] at (2.3,5.) {\scriptsize $e_6$};
\node [centered, rotate=0.] at (2.6,6.) {\scriptsize $-i_6$};
\node [centered, rotate=0.] at (2.6,6.9) {\scriptsize $-i_0$};
\node [centered, rotate=0.] at (3.5,8.) {\scriptsize $-i_1$};
\node [centered, rotate=0.] at (5.6,8.) {\scriptsize $-i_2$};
\node [centered, rotate=0.] at (6.6,6.4) {\scriptsize $-i_3$};
\node [centered, rotate=0.] at (5.6,4.9) {\scriptsize $-i_4$};
\node [centered, rotate=0.] at (3.6,4.9) {\scriptsize $-i_5$};
\node [centered, rotate=0.] at (4.7,6.4) {\scriptsize $p$};
\end{tikzpicture}
}

\newcommand{\Edge}[1]{
\begin{tikzpicture}[baseline={([yshift=-.8ex]current bounding box.center)},scale=1.]
\draw [] (0, 0) -- (0, 1.5);
\node [right, rotate=0.] at (0.,.75) {\scriptsize ${#1}$};
\end{tikzpicture}
}

\newcommand{\Tail}[1]{
\begin{tikzpicture}[baseline={([yshift=-.8ex]current bounding box.center)},scale=1.]
\draw [decorate,decoration={snake,segment length=.12cm, amplitude=.04cm}] (0, 0) -- (0, 1.5);
\node [right, rotate=0.] at (0.,.75) {\scriptsize ${#1}$};
\end{tikzpicture}
}

\newcommand{\ToricCodeD}{
\begin{tikzpicture}[baseline={([yshift=-.8ex]current bounding box.center)},scale=1.]
\draw [] (0, 0) -- (0, 1.5);
\draw [red] (0, .4) -- (-.5, .4);
\draw [red] (0, 1.1) -- (.5, 1.1);

\node [centered, rotate=0.] at (.15,.15) {\scriptsize $x$};
\node [centered, rotate=0.] at (.15,.75) {\scriptsize $u$};
\node [centered, rotate=0.] at (.15,1.35) {\scriptsize $y$};
\node [centered, rotate=0.] at (-.6,.4) {\scriptsize $a$};
\node [centered, rotate=0.] at (.6,1.1) {\scriptsize $b$};
\end{tikzpicture}
}